\documentclass[fleqn,usenatbib]{mnras}

\usepackage{newtxtext,newtxmath}

\usepackage[T1]{fontenc}
\usepackage{ae,aecompl}
\usepackage{multirow}
\usepackage{tabularx}
\usepackage{supertabular}
\usepackage{longtable}
\usepackage[]{url}
\usepackage{graphicx}	
\usepackage{amsmath}	
\usepackage{amssymb}	

\def\deg{$^{\circ}$}
\usepackage{graphicx}	
\usepackage{amsmath}	
\usepackage{amssymb}	
\usepackage{multirow}
\usepackage[left]{lineno}
\usepackage{booktabs}
\usepackage{gensymb}
\usepackage{xcolor,color}
\usepackage[caption = false]{subfig}

\title[GRANDMA: the first six months of O3]{The first six months of the Advanced LIGO's and Advanced Virgo's third observing run with GRANDMA}

\author[GRANDMA consortium]{
S. Antier$^{1}$,
S. Agayeva$^{2}$, 
V. Aivazyan$^{3, 4}$,
S. Alishov$^{2}$,
E. Arbouch$^{5}$,
\newauthor
A. Baransky$^{6}$,
K. Barynova$^{7}$,
J. M. Bai$^{8}$,
S. Basa$^{9}$,
S. Beradze$^{3,  4}$,
\newauthor
E. Bertin$^{5}$,
J. Berthier$^{10}$,
M. Bla\v{z}ek$^{11}$, 
M. Bo\"er$^{12}$,
O. Burkhonov$^{13}$,
\newauthor
A. Burrell$^{15}$,
A. Cailleau$^{16}$,
B. Chabert$^{15, 17}$,
J. C. Chen$^{18}$,
N. Christensen$^{12}$,
\newauthor
A. Coleiro$^{1}$,
B. Cordier$^{19}$,
D. Corre$^{20}$,
M. W. Coughlin$^{21}$,
D. Coward$^{15}$,
\newauthor
H. Crisp$^{15}$,
C. Delattre$^{10}$, 
T. Dietrich$^{14}$, 
J. -G. Ducoin$^{20}$,
P. -A. Duverne$^{20}$,
\newauthor
G. Marchal-Duval$^{20}$,
B. Gendre$^{15}$,
L. Eymar$^{12}$,
P. Fock-Hang$^{10}$, 
X. Han$^{22}$, 
P. Hello$^{20}$,
\newauthor
E. J. Howell$^{15}$,
R. Inasaridze$^{3, 4}$,
N. Ismailov$^{2}$,
D. A. Kann$^{11}$,
G. Kapanadze$^{3, 4}$,
\newauthor
A. Klotz$^{16, 17}$,
N. Kochiashvili$^{3}$,
C. Lachaud$^{1}$, 
N. Leroy$^{20}$,
A. Le Van Su$^{9}$,
\newauthor
W. L. Lin$^{23}$,
W. X. Li$^{23}$,
P. Lognone$^{20}$,
R. Marron$^{1}$,
J. Mo$^{23}$,
J. Moore$^{15}$,
\newauthor
R. Natsvlishvili$^{3}$,
K. Noysena$^{12, 16}$,
S. Perrigault$^{10}$,
A. Peyrot$^{10}$,
D. Samadov$^{2}$,
\newauthor
T. Sadibekova$^{13, 19}$,
A. Simon$^{24}$,
C. Stachie$^{12}$,
J. P. Teng$^{10}$,
P. Thierry$^{10}$,
C. C. Th\"one$^{11}$, 
\newauthor
Y. Tillayev$^{13}$,
D. Turpin$^{22}$,
A. de Ugarte Postigo$^{11}$, 
F. Vachier$^{10}$,
M. Vardosanidze$^{3, 4}$,
\newauthor
V. Vasylenko$^{24}$,
Z. Vidadi$^{2}$,
X. F. Wang$^{23}$,
C. J. Wang$^{8}$,
J. Wei$^{22}$,
\newauthor
S. Y. Yan$^{23}$,
J. C. Zhang$^{23}$,
J. J. Zhang$^{8}$,
X. H. Zhang$^{23}$
}
\date{Accepted XXX. Received YYY; in original form ZZZ}
\pubyear{2019}

\begin{document}
\label{firstpage}
\pagerange{\pageref{firstpage}--\pageref{lastpage}}
\maketitle

\begin{abstract}
We present the Global Rapid Advanced Network Devoted to the Multi-messenger Addicts (GRANDMA). The network consists of 21 telescopes with both photometric and spectroscopic facilities. They are connected together thanks to a dedicated infrastructure. The network aims at coordinating the observations of large sky position estimates of transient events to enhance their follow-up and reduce the delay between the initial detection and the optical confirmation. The GRANDMA program mainly focuses on follow-up of gravitational-wave alerts to find and characterise the electromagnetic counterpart during the third observational campaign of the Advanced LIGO and Advanced Virgo detectors. But it allows for any follow-up of transient alerts involving neutrinos or gamma-ray bursts, even with poor spatial localisation. We present the different facilities, tools, and methods we developed for this network, and show its efficiency using observations of LIGO/Virgo S190425z, a binary neutron star merger candidate. We furthermore report on all GRANDMA follow-up observations performed during the first six months of the LIGO-Virgo observational campaign, and we derive constraints on the kilonova properties assuming that the events' locations were imaged by our telescopes.
\end{abstract}

\begin{keywords}
methods: observational -- Stars: neutron -- Gravitational waves: Individual: S190425z
\end{keywords}



\section{Introduction}

The first gravitational wave (GW) observation of a coalescing binary neutron star (BNS), GW170817, by the Advanced LIGO~\citep{TheLIGOScientific:2014jea} and Advanced Virgo~\citep{TheVirgo:2014hva} (aLIGO/Virgo) detectors, and the prompt joint observations by \textit{Fermi}-GBM and \textit{INTEGRAL} of the short GRB 170817A \citep{LSC_GW_GRB_2017ApJ,goldstein_ordinary_2017,savchenko_integral_2017}
established the association between short duration $\gamma$-Ray Bursts (sGRBs) and BNS mergers \citep[]{Eichler1989Nature}. The localisation to within 28\,deg$^2$, together with a luminosity distance measured by the GW signal of 40 Mpc \citep{LSC_BNS_2017PhRvL}  prompted a groundbreaking electromagnetic (EM) follow-up campaign that initiated the era of GW multi-messenger astronomy \citep{LSC_MM_2017ApJ}.
The independent observation of the macronova/kilonova (KN) of GW170817 \citep{LSC_MM_2017ApJ,Andreoni_2017PASA,Hallinan_2017Sci,Kasliwal_2017Sci,Troja_2017Natur} and its spectral evolution from blue to red helped to understand the merger process and allowed an identification of the host galaxy. 
The agreement between theoretical kilonova predictions (see \citealt{Me2017} and references therein) with the observational data revealed that BNS mergers are an important side for the production of heavy elements, e.g. \citep{JuBa2015, KaKa2019}. 
In addition, the joint GW/EM observation allowed a new measurement of the expansion rate of the universe \citep{2017Natur.551...85A,HoNa2018}, it conclusively showed BNS mergers can be engines for sGRBs~\citep{LSC_MM_2017ApJ}, it placed precise limits on the propagation speed of GWs~\citep{LSC_MM_2017ApJ}, it ruled out a number of alternative theories of gravity, e.g.~\citep{EzZu:2017,BaFe:017,CrVe:2017}, and
it placed new constraints on the equation of state of cold supranuclear matter, e.g.~\citep{BaBa2013,LSC_BNS_2017PhRvL,CoDi2018,RaPe2018,RaDa18,CoDi2018b,CaTe2019}. Late-time observations from X-rays to the radio bands monitored the afterglow \citep{margutti_binary_2018,troja_outflow_2018,lyman_optical_2018, ruan_brightening_2018,resmi_low_2018,Fong2019ApJ} before eventually showing that a successful jet was launched and that the early non-kilonova emission might be from a structured jet viewed $20-30${\deg} from the jet axis \citep{Mooley_2018,Howell2019MNRAS,van_Eerten_2018,Ghirlanda2019Sci}.

Despite all these advances and the general consistency between predictions and observational data, a number of things are still unknown. Most notably, there are a number of different kilonova models, see e.g.~\citep{Metzger:2019zeh} and references therein, which explain the observational data well, but predict different light curve evolutions for the first hours after the merger of the two compact objects~\citep{2018ApJ...855L..23A}. Thus, an early time observation within the first few hours after the merger might have provided important, additional information about the fast ejecta components, the composition of the ejected material, and the heating rate for the unbound material. Early time data could also rule out other contribution channels or kilonova-precursors powered by free neutron decay~\citep{Metzger:2014yda}.

The third aLIGO/Virgo observing run (O3), which started in April 2019, is now underway with expected BNS detection rates of between $1-16$ events a year \citep{Howell2019MNRAS} and the number of triggers for binary black-hole (BBHs) mergers being sent out as GW alerts of order 1/week\footnote{See https://www.gw-openscience.org/alerts/} and a reasonable chance of detecting the first black hole - neutron star merger \citep{pannaraleprospects2014,Bhattacharya_2018}. With the increased detection ranges\footnote{At the time of writing, the detection ranges for average orientated and located BNSs have reached around 140\,Mpc, 120\,Mpc and 50\, Mpc for LIGO Livingston, LIGO Hanford and Virgo respectively; see https://www.gw-openscience.org/detector\_status/ for the current status of the instruments.} and thus  increased detection rates comes an intense effort to observe and monitor any EM counterparts. This task is very challenging and as experienced during the second observing run \citep{2019arXiv190103310T}, surveying sky areas of the order of $100-1000$s of square degrees requires an important coordinated effort between telescopes spread across different locations on Earth. 

A world-wide network is thus required to optimise the EM follow-up of GW events. For this purpose, the Global Rapid Advanced Network Devoted to the Multi-messenger Addicts (GRANDMA) was created in April 2018, re-using existing facilities built in the 1960's and 1970's as well as newer robotic telescope systems. The aim of this paper is to present this collaboration and its network of telescopes. We list the components of the network in section~\ref{GRANDMAtel}. The observational strategy and the full GRANDMA system is presented in section~\ref{strategy}. In section~\ref{O3asum}, we present statistics on all GRANDMA follow-up observations obtained during the first six months of the LIGO-Virgo observing run O3, up until the one-month pause in October 2019. We present the GRANDMA follow-up of the S190425z alert to illustrate the versatility of the program from the rapid scanning of the GW sky localisation area to the analysis of the counterpart candidates.  We also derive constraints on the kilonova properties assuming that the events' locations were imaged by our telescopes for S190425z, the closest alert from the first half of the run. In the concluding section, we discuss possible improvements to telescope networks to face new challenges presented by time domain astronomy.
\section{GRANDMA consortium}
\label{GRANDMAtel}

\subsection{A network for GW follow-up of transients with large localisation uncertainties}

\begin{figure*}
\begin{center}
\includegraphics[scale=0.4]{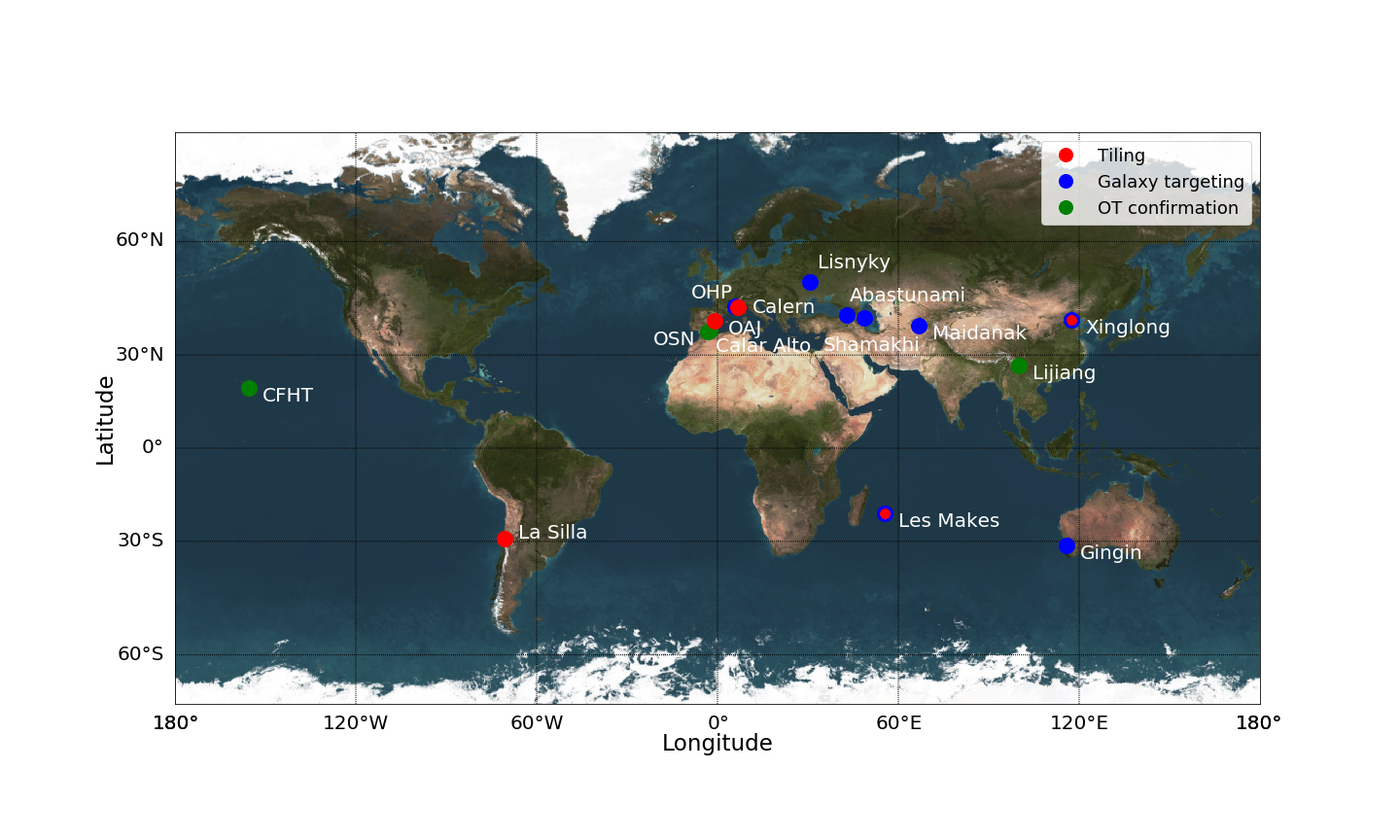}
\caption{Locations of the 15 observatories involved in the GRANDMA network. The colour encodes the observation strategy followed by telescopes at a given observatory: \textit{red} for tiling, \textit{blue} for targeting galaxies, \textit{green} for following-up candidates.}
\label{fig:GRANDMAnet2}
\end{center}
\end{figure*}

GRANDMA is a world-wide network of 21 telescopes with both photometric and spectroscopic facilities, with a large amount of time allocated for observing transient alerts as a telescope network (see Figure~\ref{fig:GRANDMAnet2}, Table~\ref{tab:GRANDMAtelphoto} and Table~\ref{tab:GRANDMAtelspectro}). The GRANDMA consortium includes 15 observatories, 24 institutions and groups from nine countries. Most of the allocated time available to GRANDMA is proprietary, i.e. warranted recurring time. 
Because of that, and the large amount of observing time,
the collaboration can cover thousand of square degrees within 24 hours which allows for a rapid scan of the sky localisation area of any GW alert triggered during aLIGO/Virgo O3. In this sense, GRANDMA has the capacity to catch the early stage of the visible components of the post-merger ejected material with rapid online colour identification. 
While GRANDMA has some shortcomings in performing very deep spectroscopic follow-ups that require large telescopes (for example 4 m class telescopes), 
it offers  unique worldwide  observational  capabilities complementary to those of other collaborations.


The GRANDMA network has access to four wide-field telescopes (field of view [FoV] $>\mathrm{1\,deg}^2$) located on three continents, with the TAROT network ($\approx18$ mag in 60~s) \citep{2008PASP..120.1298K,2018cosp...42E2475N} and the OAJ-T80 telescope ($\approx21$ mag in 180~s). GRANDMA also has access to ten remote and robotic telescopes with narrower fields-of-view, reaching at a minimum a limiting magnitude of 18. The two groups of telescopes are used to rapidly find counterpart candidates, either tiling the localisation area of the GW alerts, or targeting host-galaxy candidates. These 14 telescopes can use the full available night to observe GW sky localisation areas, depending only on weather conditions. The remaining telescopes of the collaboration are used to confirm an association and to perform dedicated follow-up observations, either photometrically or spectroscopically.

\begin{figure}
\begin{center}
\includegraphics[width=1.0\columnwidth]{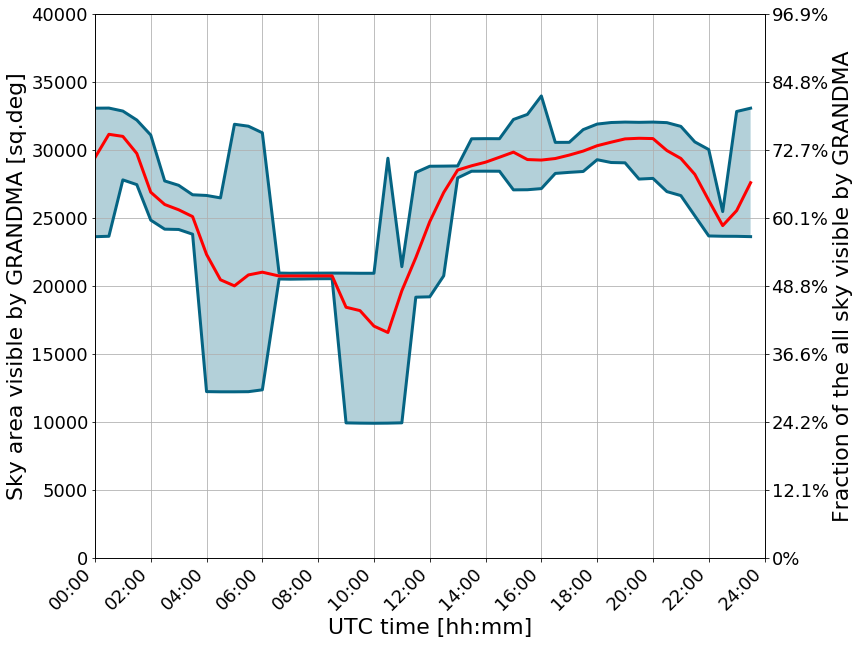}
\caption{The sky area visible during local nighttime (in square degrees and in percentage) to the GRANDMA network (15 observatories) as a function of time during a day. The red line represents the daily sky coverage averaged over a year while the blue area gives the minimum and maximum sky coverage reachable during a year. The visibility gap between 04:00 and 12:00 is due to the lack of observatories at western longitudes.}
\label{fig:GRANDMAnet}
\end{center}
\end{figure}

The prompt full coverage of the transient sky accessible by GRANDMA at a given time is presented in Figure~\ref{fig:GRANDMAnet}. In 24 hours, thanks to an extensive distribution on Earth and especially at eastern longitudes, the GRANDMA network is able to access over 60\% of sky (up to more than 72\%) during nights\footnote{At each observatory location, the night time is chosen to start when the local Sun elevation is below $-13^\circ$. The local sky visible area is defined with all the pointings having a local elevation of 25$^\circ$ above the local horizon with no restriction on the western and eastern hour angle accessible by each telescope.} to a limiting magnitude of 18 mag about 75\% of the time. Due to a lack of western observatories, the sky coverage is reduced to 40-50\% during the western nights. All the telescopes are connected to the ICARE (\textit{Interface and Communication for Addicts of the Rapid follow-up in multi-messenger Era}) which automatically provides dedicated observation plans to any telescope (described in section~\ref{strategy}) as soon as a GW alert is received. Finally, in the case of a confident detection of an electromagnetic counterpart to a GW event, GRANDMA has access to four 2m-class telescopes in three different time zones with photometric depth capabilities of $\approx22$ mag and spectroscopic depth capabilities of $\approx18.5$ mag, as shown in Table~\ref{tab:GRANDMAtelspectro}. All results of the observations and counterpart candidates are centralised in a unique database to optimise the follow-up and re-point telescopes to counterpart candidates.


\begin{table*}
\begin{tabular}{ccccccc}
Telescope      & Location        & Aperture & FOV            & Filters                                 & $3-\sigma$ limit     & Maximum Night slot\\
Name           &                 & (m)      &  (deg)     &                                         & (AB mag)             & (UTC) \\
\hline
\hline
TAROT/TCH      & La Silla Obs.   & 0.25 & $1.85\times1.85$   & Clear, $g^\prime r^\prime i^\prime $    &  18.0 in 60s (Clear)           & 06h-15h\\
CFHT/WIRCAM    &    CFH Obs.             &    3.6  &      $0.35\times0.35$             &        $JH$                                &      22.0 in 200s ($J$)       & 10h-16h\\
CFHT/MEGACAM   &     CFH Obs.            &     3.6  &      $1.0\times1.0$               &  $g^\prime r^\prime i^\prime z^\prime$                                       &         23.0 in 200s ($r^\prime$ )    &  10h-16h \\
Zadko          & Gingin Obs.      & 1.00 & $0.17\times0.12$ & Clear, $g^\prime r^\prime i^\prime I_C$ & 20.5 in 40s (Clear)  & 12h-22h\\
TNT            & Xinglong Obs.   & 0.80 &  $0.19\times0.19$  & $BVg^\prime r^\prime i^\prime$          &    19.0 in 300s ($R_C$)  & 12h-22h\\
Xinglong-2.16  & Xinglong Obs.   &  2.16    &      $0.15\times0.15$              &     $BVRI$                                    &      21.0 in 100s ($R_C$)       & 12h-22h \\
GMG-2.4        & Lijiang Obs.    &  2.4    &     $0.17\times0.17$                &         $BVRI$                                &     22.0 in 100s ($R_C$)        & 12h-22h\\
UBAI/NT-60     & Maidanak Obs.   & 0.60 & $0.18\times0.18$   & $BVR_CI_C$                              &    18.0 in 180s ($R_C$) & 14h-00h\\
UBAI/ST-60     & Maidanak Obs.   & 0.60 & $0.11\times0.11$   & $BVR_CI_C$                              &    18.0 in 180s ($R_C$) & 14h-00h\\
TAROT/TRE      & La Reunion      & 0.18 &  $4.2\times4.2$  & Clear                                   &     16.0 in 60s (Clear)        & 15h-01h\\
Les Makes/T60  & La Reunion.     & 0.60 &  $0.3\times0.3$  & Clear, $BVR_C$                          &   19.0 in 180s ($R_C$)       & 15h-01h\\
Abastumani/T70 & Abastumani Obs. & 0.70 & $0.5\times0.5$   & $BVR_CI_C$                              &    18.2 in 60s ($R_C$)   & 17h-03h\\
Abastumani/T48 & Abastumani Obs. & 0.48 & $0.33\times0.33$ & $UBVR_CI_C$                             &    15.0 in 60s ($R_C$)   & 17h-03h\\
ShAO/T60       & Shamakhy Obs.   & 0.60 & $0.28\times0.28$ & $BVR_CI_C$                              &    19.0 in 300s ($R_C$)  & 17h-03h \\
Lisnyky/AZT-8  & Kyiv Obs.       & 0.70 & $0.38\times0.38$ & $UBVR_CI_C$                             & 20.0 in 300s($R_C$) & 17h-03h\\
TAROT/TCA      & Calern Obs.     & 0.25 & $1.85\times1.85$ & Clear, $g^\prime r^\prime i^\prime$     &  18.0 in 60s (Clear)          & 20h-06h\\
IRIS           & OHP             & 0.5  &  $0.4\times0.4$  &   Clear,$u^\prime g^\prime r^\prime i^\prime z^\prime $ & 18.5 in 60s ($r^\prime$)   & 20h-06h\\
T120           & OHP             & 1.20 &   $0.3\times0.3$     &     $BVRI$                                    &   20.0 in 60s ($R$)           & 20h-06h\\
OAJ/T80        & Javalambre Obs. & 0.80 &   $1.4\times1.4$                 & $r^\prime$                              & 21.0 in 180s ($r^\prime$) & 20h-06h\\
OSN/T150       & Sierra Nevada Obs. & 1.50 &  $0.30\times0.22$                  & $BVR_CI_C$                              & 21.5 in 180s ($R_C$) & 20h-06h\\
CAHA/2.2m      & Calar Alto Obs. & 2.20 &  $0.27\diameter$                & $ u^\prime g^\prime r^\prime i^\prime z^\prime $                              & 23.7 in 100s ($r^\prime$) & 20h-06h\\

\hline
\end{tabular}
\caption{List of telescopes of the GRANDMA consortium and their photometric performance when using their standard setup.}
\label{tab:GRANDMAtelphoto}
\end{table*}

\begin{table*}
\begin{tabular}{ccccccc}
Telescope/Instrument & Location & Wavelength range & Spectral resolution $\lambda/\Delta\lambda$ &  Limiting mag\\
\hline
\hline
2.2m CAHA/CAFOS   & Calar Alto Obs. & 3200-7000/6300-11000 & 400 & 20 in 1h \\
ShAO/T2m & Shamakhy Obs. & $3800-7000$ & 2000 & 17 in 1h \\
Xinglong-2.16/BFOSC & Xinglong Obs. & $3600-9600$ & 1000 & 18 in 1h \\
GMG-2.4/YFOSC & Lijiang Obs. & $3400-9100$ & 2000 & 19 in 1h \\
\hline
\end{tabular}
\caption{List of telescopes of the GRANDMA consortium with spectroscopic capabilities.}
\label{tab:GRANDMAtelspectro}
\end{table*}

Since the beginning of April 2019 and until the one-month pause in October 2019, GRANDMA has already responded to 27/33 alerts sent by the LIGO-Virgo collaboration (see section~\ref{O3asum}), making it one of the most active networks following up GW alerts.
Presented below are descriptions of the telescopes participating in the GRANDMA network.

\subsection{Abastumani Astrophysical Observatory at Ilia State University (Iliauni, Georgia)}
The 70 cm Meniscus telescope (focal length of 2460 mm) of the Abastumani Astrophysical Observatory (AbAO, $41^{\circ}45^\prime17^{\prime\prime}N$, $42^{\circ}49^\prime20^{\prime\prime}E$) was mounted in 1955 at the altitude of 1610 m. A pl4240-ccd-camera-back-illum-63-5mm-shutter-grade-1 with $BVR_CI_C$ filters on the telescope provides a FoV of $30^\prime\times30^\prime$, with a readout of 10 s. The limiting magnitude in the $R_C$ filter with an exposure time of 1 min is 18.2 mag (AB system, 3-$\sigma$). The 48 cm Cassegrain telescope (at an altitude of 1605 m) with an equivalent focal length of 7715 mm is equipped with an Apogee Alta KAF-16801E CCD together with $UBVR_CI_C$ filters, and has FoV of $20^\prime\times20^\prime$. The limiting magnitude in the $R_C$ filter for an exposure of 1 min is 15 mag (AB system, 3-$\sigma$).

\subsection{HETH group at IAA}
The HETH ({\bf H}igh {\bf E}nergy {\bf T}ransients and their {\bf H}osts) group, a research group at the Instituto de Astrof\'isica de Andaluc\'ia (IAA) in Granada, Spain, contributes with three different observing programmes to the GRANDMA consortium. All the telescopes allotted time have been obtained in competitive calls for observing time for Semesters 2019A and 2019B.  

The Javalambre Auxiliary Survey Telescope (JAST/T80) is an 83cm, F4.5 survey telescope located at the Javalambre observatory\footnote{https://oajweb.cefca.es} in Teruel, Arag\'on, Spain, at an altitude of 1082 m at 40$^\circ$02\arcmin33.7\arcsec N 1$^\circ$00\arcmin59.6\arcsec W. The telescope is equipped with the T80-cam installed at the Cassegrain focus, with a FoV of $1.4^\circ\times1.4^\circ$ and 12 SDSS and narrow-band filters. The camera has a $9.2\textnormal{k}\times9.2$k CCD with 10$\mu$m pixels, which yields a pixel scale of 0\farcs55/pixel.
This telescope is used for tiling observations (proposals \#~1800140,1800150,1900160, PI Kann) to cover parts or the full error box during the first night, taking up to 50 pointings of 180s in the $r^\prime$ band with a limiting magnitude of 21 per tile. A second part of the program allows to follow-up a counterpart (e.g. a KN) in the $i^\prime$ band for up to $\sim2$ weeks reaching a limiting magnitude of $i^\prime>22$ mag in a 600s exposure.

The second facility is the Observatorio de Sierra Nevada\footnote{https://www.osn.iaa.csic.es} (OSN), managed by IAA and located at an altitude of 2896 m in the Sierra Nevada ski resort in Granada, Spain, at 37$^\circ$03\arcmin46.4\arcsec N 3$^\circ$23\arcmin09.8\arcsec W. The 150cm telescope at OSN is equipped with an Andor iKon-LDZ936N $2048\times2048$ pixel camera in the Nasmyth focus and has a FOV of $7\farcm92\times7\farcm92$. This camera is used to detect possible counterparts in candidate hosts, for which we use the Johnson-Cousins $R_C$ filter (2019/1 PI Izzo, 2019/2 PI Blazek). An additional run to follow-up a possible KN in $BVR_CI_C$ filters is also included, reaching $\sim I_C=21$ mag in a 540s exposure.

Finally, we use several instruments at the Centro Astron\'omico Hispano Andaluz (CAHA) observatory\footnote{https://www.caha.es} at the Sierra de Los Filabres, Spain, at an altitude of 2196 m, at 37$^\circ$13\arcmin23.4\arcsec N 2$^\circ$32\arcmin45.6\arcsec W, also managed by the IAA. We use the Bonn University Simultaneous Camera (BUSCA), mounted at the Cassegrain focus of the 2.2 m telescope, for follow-up of GW counterparts or KNe in different filters (2019F-15, 2019H-22, PI Kann). The Calar Alto Faint Object Spectrograph (CAFOS), also mounted on the 2.2 m telescope, is used for imaging and spectroscopic follow-up of sufficiently bright candidate counterparts (2019F-16, 2019H-23, PI Kann). CAFOS is an imager and spectrograph which we use with the B400 grism, which has a limiting magnitude of $\sim21$ mag in a 1h exposure.

\subsection{Observatoire de Haute-Provence}
IRiS (Initiation to Research in Astronomy for Schools) is a robotic telescope installed in 2013 on the site of the Observatoire de Haute-Provence, France in order to allow for discoveries in the transient sky, while offering the opportunity to learn modern observational techniques. The telescope is a Ritchey-Chr\'etien type telescope with a diameter of 50 cm mounted on a German-type mount and equipped with a focal plane composed of a camera equipped with an E2V 42-40 sensor ($2\mathrm{k}\times2\mathrm{k}$), a filter wheel and filters (Clear, SDSS $u^\prime g^\prime r^\prime i^\prime z^\prime$, as well as narrow bands centred on the H$\alpha$, CH4 and OIII lines). The entire telescope and its equipment can be fully controlled remotely in order to automatically follow-up  interesting transients or perform a series of targeted exposures of known galaxies to find optical counterparts of GRBs, GWs and neutrino events.

In the case of optical counterparts deemed significant, we can use the T120 telescope, also located at Observatoire de Haute-Provence. It is a 120 cm diameter telescope equipped with an Andor Ikon L 936 camera with a $2048\times2048$ CCD with a FoV of $13\farcm1\times13\farcm1$ and with $BVR_CI_C$ filters.

\subsection{Les Makes observatory}
The 60cm telescope at Les Makes Observatory (R\'eunion Island) is a
Ritchey-Chr\'etien F/8 reflector. A SBIG ST11K CCD camera
with Astronomik LRGB filters provides a
FoV of $26^\prime\times17^\prime$. A typical exposure uses a $3\times3$ binning, providing a spatial sampling of 1\farcs16/pixel. The telescope and camera are controlled by the AudeLA software using the RobObs script. This combination is adapted for galaxy targeting. 

\subsection{Lisnyky observatory}
The 70cm F/4 telescope at the observational station Lisnyky of Taras Shevchenko National University of Kyiv, Ukraine, is equipped with a FLI PL4710 back-illuminated CCD and $UBVRI$ Bessel filters. For faint objects we use a mode with $2\times2$ binning, which gives a scale of 1\farcs96/pixel. The FoV of the instrument is $16^\prime\times16^\prime$. The limiting magnitude of a 300~s exposure image is 20 mag under good sky conditions. It is possible to reach 21.5-22 mag with 1800 s exposures.

\subsection{Shamakhy Astrophysical Observatory of Azerbaijan}

The 60cm telescope at Shamakhy Astrophysical Observatory (ShAO, $40^{\circ}46^\prime20^{\prime\prime}N$, $48^{\circ}35^\prime04^{\prime\prime}E$)  (original equivalent focal length was 7.5 m) Schmidt-Cassegrain system was mounted in 1975 at an altitude of 1500 m. For use with GRANDMA collaboration observations, this telescope has been modified, with the equivalent focal length being reduced to 4.7 m. It is equipped with a FLI CCD camera with $4\textnormal{k}\times4\textnormal{k}$ pixels with size 9 $\micron$ and standard $BVR_CI_C$ filters, which provides a $17^\prime\times17^\prime$ FoV with a readout time of 6 seconds. The limiting magnitude in the $R_C$ filter of a 300 s exposure image is 19 mag with precision 3-$\sigma$ under good sky conditions.

The 2~m telescope at Shamakhy Astrophysical Observatory (Coude, Cassegrain and direct focuses) was mounted in 1967 at an altitude of 1435 m. This telescope has been equipped with various spectrographs for spectral observations. For faint objects we can use a $2\times2$ Kanberra prism spectrograph. Typical observations in the visual range cover $3700-8000$ {\AA} up to 17 mag with a resolution of $R=2000$.

\subsection{TAROT network}
\label{TAROT}
TAROT stands for T\'elescope \`a Action Rapide pour les Objets Transitoires (Rapid Action Telescope for Transient Objects). Two TAROT 25\,cm aperture telescopes
\citep{2008PASP..120.1298K} are installed at Calern in France (hereafter TCA) 
and La Silla in Chile (hereafter TCH).
Initially designed to study optical counterparts of GRBs these telescopes
are now able to react to GW alerts. Their FoV is $1.86^\circ\times1.86^\circ$. 
TAROT R\'eunion (hereafter TRE) is an 18\,cm aperture telescope which has been installed in 2016 
at Les Makes observatory \citep{Klotz2019ExA}.
TAROT R\'eunion is located
at $21.1988^{\circ}$\,S, $55.4102^{\circ}$\,E, and an elevation of 991\,m above sea level.
Its camera is a FLI 16803 that covers 
a FoV of $4.2^\circ\times4.2^\circ$
to be efficient when tiling the large skymaps delivered
by GW interferometer alerts.

\subsection{Tsinghua-NAOC Telescope}
The Tsinghua-NAOC (National Astronomical Observatories of China) Telescope, abbreviated as TNT, is an 80 cm Cassegrain reflecting telescope located at Xinglong Observatory($117^{\circ}34^\prime39^{\prime\prime}E$, $40^{\circ}23^\prime40^{\prime\prime}N$, elevation: $\sim900m$).
It was designed and manufactured by APM-Teleskope in Germany and is under joint operation of Tsinghua University and NAOC since 2004. The main scientific goal of this telescope is to monitor and obtain photometric data of multiple transient phenomena such as supernovae, active galactic nuclei, and gamma-ray bursts, etc.

This telescope is equipped with a $1340\times1300$ CCD camera providing a FoV of $11\farcm5\times 11\farcm2$. The filter system used on the TNT is standard Johnson $UBV$ and Cousins $R_CI_C$ system \citep{BessellPASP}. The limiting magnitudes of TNT for a S/N of 100 with a 5 min exposure are 18.8 mag in $V$ band and 18.7 mag in $R_C$ band \citep{HuangRAA}.

\subsection{Zadko collaboration}
The Zadko Telescope \citep{cow10, cow17} is a 1~m Cassegrain telescope (with aperture f/4), located in Western Australia at coordinates $115^\circ42^\prime49^{\prime\prime} E$ $31^\circ21^\prime24^{\prime\prime} S$. It has been designed by DFM Engineering Ltd, and donated to the University of Western Australia by resource company Claire Energy. The telescope has been active since the first of April 2009, with various upgrades done when needed \citep[see][for details]{cow17}. For the O3 run, a new QHY 163M camera with a field of view of $17\arcmin\times12\arcmin$ has been installed, allowing for reaching R $\sim$ 21 mag in about 90 seconds.

The telescope is controlled by the same software as the TAROT network, Robotic Observatory Software (ROS), aimed at observing the transient sky as quickly as possible when something occurs in the Universe. Like any ROS-compliant instrument, the Zadko Telescope is unsupervised with various pipelines in charge of listening to several alert channels and to schedule, perform, process, and calibrate the observations. In the special case of GRANDMA, Zadko bypassed some of them as GRANDMA offers its own tools (see below).

In addition to the transient sky, the Zadko telescope is also used to perform outreach and teaching activities, and to study minor bodies in our solar system. Lastly, the Zadko telescope also collaborates with other entities (such as NASA) to perform on-demand observations for full-time coverage, due to its unique location: this is the only one-meter-class professional telescope located in Western Australia, which is located at the antipodes of the US mainland.

\subsection{UBAI collaboration}
The Ulugh Beg Astronomical Institute (UBAI) recently joined the GRANDMA project with two Zeiss-600 0.6m Cassegrain telescopes (with an aperture f/12.5), Northern (NT-60) and Southern (ST-60). Both are located at high-altitude Maidanak Observatory ($66^{\circ}54'$ E, $38^{\circ}40'$ N, 2600m above sea level, \href{http://www.maidanak.uz/}{http://www.maidanak.uz}) in the South-East of Uzbekistan at a distance of about 120 km south of Samarkand. The site provides about $60\%$ of the available yearly dark time ($\sim1800$ hours) with median seeing 0\farcs69 at $0.5\mu$m \citep{2000A&AS..145..293E}. The maximum amount of clear time corresponds to the summer-autumn period.

Operational since the 1980s, the telescopes are presently equipped with FLI $1024\times1024$ CCD cameras and $BVRI$ Bessel filters, the FoV is $10\farcm8\times10\farcm8$ for the NT-60 and $6\farcm5\times6\farcm5$ for the ST-60. The limiting magnitude in the $R$ filter is 18 mag with a 3 min exposure.

\subsection{Collaboration with other groups}

GRANDMA can follow-up counterpart candidates with two Chinese consortia that have synoptic surveys. First, the Tsing\-hua Ma-huateng Telescopes (TMTS) are composed by $4\times40$ cm tubes on a single mount, located at Xinglong Observatory in Hebei Province, China. Each tube has a field of view of about 4.4 square degrees and resolution of 1\farcs8/pixel. It can reach 19 mag ($5\sigma$) in 60 seconds. Every night, the TMTS can scan about 8000 square degrees of the sky to search for optical transients such as supernovae and flare stars.

Secondly, GRANDMA also collaborates with SVOM, a Sino-French Collaboration preparing a mission dedicated to the study of the multi-messenger transient sky with a core programme devoted to GRB science \citep{Wei16}. Currently, the SVOM ground segment used for O3 is mainly located in China at the Jilin and Xinglong Observatories. The full network has multiple photometric capabilities with:
\begin{enumerate}
    \item A 1.2m Chinese Ground Follow-up Telescope (C-GFT at Jilin) equipped with a $4\mathrm{k}\times4\mathrm{k}$ CCD camera observing with no filter ($r_{lim}\sim20$ with 100 s exposure). The FoV of this telescope is $1.5^\circ\times1.5^\circ$ (1\farcs35/pixel angular resolution).
    \item Three Ground Wide-field Angle Cameras (GWAC at Xinglong) telescopes, each equipped with four ``JFOV'' $4\mathrm{k}\times4\mathrm{k}$ CCD E2V camera ($\diameter=18$ cm) and one ``FFOV'' $3\mathrm{k}\times3\mathrm{k}$ CCD camera ($\diameter=3.5$ cm). One mount has a FoV of 25$^\circ\times25^\circ$ ($\sim500$ square degrees) and typically a limiting unfiltered magnitude of 16 is reached with an exposure of 10 seconds (up to 18 at best by stacking hundreds of single images). The GWACs perform a systematic survey of the sky every night with online analysis of any discovered transients \citep{Turpin19}.
    \item Two 60 cm telescopes (GWAC/F60A and GWAC/ F60B at Xinglong) equipped with a $2\mathrm{k}\times2\mathrm{k}$ and $1\mathrm{k}\times1\mathrm{k}$ CCD cameras using $UBVR_C I_C$ Johnson-Cousins filters. With a FoV of $\sim 20' \times 20'$, they have a typical 1$''$ angular resolution and a limiting magnitude of $R_{lim}\sim19$ mag in a 120 s exposure.
    \item A 30~cm telescope (GWAC/F30 at Xinglong) is also operated with $UBVR_C I_C$ Johnson-Cousins filters with a quite wide FoV of $1.8^\circ\times1.8^\circ$ and $R_{lim}\sim17$ mag in a 100 s exposure.
\end{enumerate}

Besides, in case of significant optical candidates during a GW alert, GRANDMA can trigger both photometric and spectroscopic observations on the two Chinese 2-m class telescopes through the GRANDMA collaborators. 
\begin{enumerate}
    \item The 2.16-m telescope, located at the Xinglong Observatory, has Cassegrain and Coude focuses, and it is simple to change from one to the other (only needing one minute). It is equipped with the BFOSC (Beijing-Faint Object Spectrograph and Camera), OMR Low Resolution Spectrograph and HRS (High Resolution Fiber-Fed Spectrograph) as shown in Table~\ref{tab:GRANDMAtelspectro}. 
    \item The GMG 2.4-m telescope is located at the Lijiang station of Yunnan Observatories, and is similar to the Liverpool and Nordic Optical telescopes. It is equipped with the YFOSC (Yunnan Faint Object Spectrograph and Camera), and with a back-illuminated $2048\times4096$ E2V42-90 CCD. Using YFOSC, the limiting magnitude in photometry varies from 22 to 23 mag depending on the exposure time and the seeing. For low/medium spectral resolution observations, we expect a magnitude limit of about 19 mag depending on features in the spectrum.
\end{enumerate}


GRANDMA obtained 10 hours per semester of target-of-opportunity time for Semesters 2019 A and B on the Canada-France-Hawaii telescope (CFHT) equipped with WIRCAM and MegaCam, on the Big Island, Hawaii, United States, which can reach a 24 mag limit. We will use both of them for following-up counterpart candidates of GW alerts in the optical and near infrared band ($g^\prime r^\prime i^\prime z^\prime$ for MegaCam, $JH$ for WIRCAM) and characterise the colour evolution of the kilonovae. In addition, MegaCam can participate in the GW skymap observations and to find directly the optical counterparts by targetting galaxies inside the GW volume.

GRANDMA is also connected to gamma-ray counterpart searches of GWs. First, we have developed our own offline detection pipeline to detect GRBs called F-WBSB (\textit{Fermi}/GBM-like search with Wild Binary Segmentation for the detection of Bursts) presented in \citealt{Antier2019}. In case of GW alerts, we extract archival data over the full day both in the \textit{Fermi}/GBM and \textit{INTEGRAL} SPI/ACS instruments. We consider a gamma-ray transient as interesting in the [-10 min, 1 hour] time window around the GW trigger. The pipeline gives an independent approach from standard online and offline analysis \citep{2015ApJS..217....8B,2019ApJ...871...90B}, which helps to confirm sub-threshold triggers found by other groups with unique algorithms based on binary segmentation \citep{Fryzlewicz14}.

\subsection{Citizen science programme}

GRANDMA has also developed its own citizen science programme called \textit{kilonova-catcher}\footnote{\href{https://grandma-kilonovacatcher.lal.in2p3.fr/}{https://grandma-kilonovacatcher.lal.in2p3.fr/}}: Amateur astronomers having any observational capabilities are invited to participate in the follow-up of GW alerts. The participants are welcomed to register on the \href{https://grandma-kilonovacatcher.lal.in2p3.fr/}{GRANDMA platform} and to provide their telescope characteristics and location.

In case of a GW alert, induced by the merging of two compact objects involving at least one neutron star at a maximal distance of 200 Mpc, GRANDMA will compute a list of galaxies consistent with the distance range of the GW event as well as its sky localisation given by the probability skymap (see section~\ref{obsgwscan}). According to the telescope information given by each \textit{kilonova-catcher} user, a short list of promising GW host galaxies candidates will then be automatically assigned to a telescope user using Slack alert messages, emails and also displayed on the GRANDMA amateur platform on the telescope user page. The \textit{kilonova-catcher} users are then encouraged on a best-effort basis to point their instruments towards those galaxies centred in their FoV and to adapt their exposure time to reach at least $17-18$ mag. This limiting magnitude is based on the scenario of the kilonova AT 2017gfo detected in association with GW170817 \citep{LSC_MM_2017ApJ}. The images, in the FITS format, are then stored using an uploading application on the GRANDMA platform. If the electromagnetic counterpart is found during the global follow-up of any GW event, we will look back at the archival \textit{kilonova-catcher} data to find if the field was observed by some amateur telescopes. If so, then the images will be reduced and the observations will be integrated into the full scientific analysis. Some amateur astronomers have acquired significant experience in searching for optical transient sources, particularly in the supernova domain. Their experience will be also profitable to potentially send promising optical transient candidates to a committee of GRANDMA scientists who will analyse them and take the decision to trigger deeper follow-up observations with the professional instruments.

Presently, GRANDMA reaches a participation of about 30 amateur astronomers across the world. 16 telescopes representing 60\% of the participants are located in France, the others mostly in Europe, especially in Spain with four telescopes. Other locations cover some regions in Canada, Chile, Mexico, Germany, Luxembourg, Malta, Mauritania, Australia and New Zealand. The diversity of telescopes varies from 80 mm aperture to 600 mm with an average FoV of $40^\prime$. Their expected limiting magnitude for a one minute exposure is about 17 mag with a resolution of 1\farcs05/pixel. A few telescope are remotely operated, especially the largest ones, but most of the amateur astronomers directly control their telescope, sitting to the side of the pillar of their instruments. Nowadays, the amateur community has well-advanced skills in optical data-reduction techniques thanks to extensive tutorials given by many astronomical associations and specialised websites. Therefore, the non-professional world is perfectly able to produce data that are scientifically exploitable.
The instantaneous coverage in different optical bands of the GW probability skymap by multiple telescopes will result in the unique opportunity to catch the early stage of the visible components of the post-merger ejected material.

\section{Observation strategy during the LIGO/Virgo O3 observational campaign}
\label{strategy}


\subsection{Organisation of the follow-up procedure}

In order to perform follow-up of alerts and possible associated transients, a system of shifts has been set up, with a team of four follow-up advocates (FA) or ``shifters'' designed to cover a entire week. Each day of a given week is divided in four six-hour long shifts, each under the supervision of one FA. In principle each FA is assigned the same daily slot during the week. In case of an alert, the first responsibility of the FA is to check that the observation plan has been computed, sent to the network of observatories and has been well-received by them. The FA has then to follow all notices (GCNs) related to the alerts currently in hand. 

If an optical transient is found during the campaign (by GRANDMA or by others) and the first observations can be consistent with a kilonova or GRB afterglow signal, the FA may ask to follow it up by dedicated telescopes of the network. This may require the interruption of current observations for some telescopes and the modification of the observation plan. When a counterpart is formally identified (by the GRANDMA network or by others), the FA may also ask for observations by spectroscopic facilities of the network (the CAHA 2.2m telescope for instance) or through ToO programs we have access to. 

Lastly, the FA is in charge of reporting everything that happened during their shift, either internally through a system of logbooks, or toward the whole community using an automatic writing tool generating the GCNs related to GRANDMA observations and possible discoveries.

\subsection{GRANDMA infrastructure}

Like any project aimed toward robotic astronomy and studying the transient sky, the GRANDMA project relies heavily on a central database and cloud applications, developed for that purpose. We also constructed a dedicated web interface 
for helping the FA in their tasks.

At the core of the project lies a MySQL database centralising all informations about the alerts and the observations done by any of the telescopes of the collaboration. That database is populated automatically by various scripts, provided by the collaboration, and running at the telescope sites, or by programs running on our main server. It is accessed by some scripts which need specific information about the collaboration (i.e. our tool dedicated to produce GCNs sent by the collaboration) and the ongoing observations (the position of the tiles already observed for instance).

Each time a new event is detected by the aLIGO/Virgo collaboration and released to the public via a GCN, our server automatically retrieves and analyses the sky map of the error box of the event. It then produces a set of observation plans optimised for each telescope, following the method presented in the next section. That observation plan is then propagated to the observers. All of these mechanisms widely use the IVOA~\footnote{http://www.ivoa.net/} methods for sharing data and information.

A web interface to the database allows at any moment to follow the ongoing procedure and adjust the observation strategies if needed. As a matter of consequence, each new detection of a GW counterpart candidate can be inserted into the database and followed-up by small FoV instruments and/or spectroscopic instruments. That interface provides the only possible human supervision of the infrastructure, and is made for the benefit of the FAs.

\subsection{Observation strategy for GW alerts}
\label{obsgwscan}

The beginning of O3 has provided a wide variety of GW candidates for follow-up, both in terms of the merging objects and the sky localisation areas requiring examination.
Due to the large sky localisations and differences in telescope configurations and sensitivities, optimised techniques for follow-up efforts are important. The GRANDMA observation strategy uses different approaches depending on the FoV of the telescopes. The core of the observation plans relies on \texttt{gwemopt} \citep{CoTo2018}, open-source software designed to optimise the scheduling of target-of-opportunity observations using either synoptic or galaxy-targeted based searches.
Briefly, for synoptic survey instruments ($\gtrsim1$ deg$^{2}$), it divides the skymap into a grid of ``tiles'' the size and shape of the FoV of each telescope.
For each tile, the available segments for observation are computed, including accounting for their rising and setting, in addition to their distance from the moon.
Observations are scheduled by computing a weight based on the three-dimensional probability distribution from the GW alert \citep{Singer:2016eax}; for nearby sources ($\lesssim300$ Mpc), which is galaxy-weighted, following the distribution of galaxies and using their properties such as $B$-band luminosity.
For narrow-FoV telescopes ($\lesssim1$ deg$^{2}$) with nearby alerts ($\lesssim300$ Mpc), galaxy-targeted observations are performed, where one galaxy at a time is imaged within the three-dimensional probability distribution as suggested by \citet{2013ApJ...767..124N,2016ApJ...820..136G}.

Each observation is processed by the reduction pipelines to find transient candidates. 
In general, these images are compared to previous observations, either images or databases of known objects (as traces of cosmics, known variable stars,
minor planets and known supernovae), to determine whether the objects are new with respect to the GW transient.
For this reason, survey instruments with recent pre-existing limits are at an advantage in this endeavor.
In addition, transients should have at least two detections separated in time to mitigate the follow-up of asteroids. 

Each team of the GRANDMA consortium can use their own detection pipeline; a subset have low-latency reduction processes (the TAROT network as an example, \citealt{2018cosp...42E2475N}). However, the GRANDMA consortium provides to individual teams a fast pipeline for detection of potential new transients, called \texttt{gmadet}, which is publicly available through the GitHub platform\footnote{http://github.com/dcorre/gmadet}. The issue of new transient detection in the telescope images is complex and has several approaches. The error regions given by the aLIGO/Virgo interferometers are usually very large, significantly larger than any single observed image. An easy solution would be if every telescope in our network had previous images of the observed field from different periods. Those templates can be used for image subtraction, and new potential transients can be discovered afterward by standard photometric techniques on the difference images. This approach is especially useful for the cases when the optical counterpart would be hidden visually inside the light of its host galaxy. However, telescopes in the GRANDMA network are not all-sky surveys and do not have a significant amount of templates for such an image subtraction. The situation is improved for those GW events with estimated distances inside the range of the GLADE catalogue of galaxies \citep{2018gladecatalog}. Those galaxies can be examined manually, and we can perform photometry for crowded fields (e.g., Daophot) for the detection of optical candidates. However, most of the events are outside of the GLADE catalogue range. Therefore the optical candidates can be practically anywhere in the telescope images. 

In GRANDMA, there are various differences in the data reduction methods, including calibration and the filter systems employed. This can lead to differences in light curves that are not due to physical effects. Thus, the collaboration is working on a homogeneous procedure to calibrate the images such that they have photometric consistency. In order to harmonise the detection process among the telescopes, which is crucial to affirm source discoveries with independent observations, we developed a common algorithm written in the Python programming language and using several external astronomical packages and modules to perform astrometry and photometry. The astrometric calibration is performed using the Astrometry.net package \citep{2010astrometrynet}, the source detection is performed using either IRAF \citep{1986iraf} or Sextractor \citep{sextractor} depending on the user preference. After stellar detection we perform a cross-check of all detected sources with several astronomical catalogues - PanSTARRS, USNO-B1, Gaia and GSC using the CDS X-match service. The final stage consists of photometry of the unidentified sources, and then a subsequent upload of the potential transients into the GRANDMA internal database. Sub-images centered on these potential transients are displayed on the web interface and compared to the same fields from the PanSTARRS all-sky survey using Aladin Lite~\footnote{https://aladin.u-strasbg.fr/AladinLite/}. There is an on-going effort to automatically discard artefacts among the transient candidates using a machine learning algorithm.

Follow-up of the remaining optical transients with dedicated telescopes is performed to classify them; this includes both photometric and spectroscopic follow-up. 
While classifying supernovae and M-dwarf flares spectroscopically is relatively straightforward, photometrically, one needs to rely on the rapid timescales over which kilonovae evolve \citep{Me2017}.
Comparison to GW170817/AT 2017gfo shows that the evolution in luminosity is expected to be $\Delta r \sim 1$\,mag per day, and slightly slower at later times; for this reason, transients can be ruled out by measurements of their luminosity evolution. 
In general, typical supernovae near peak evolve much more slowly ($\Delta r \sim 0.1$\,mag per day), and therefore can be rejected by tracking their photometric evolution.
In this way, the network can follow-up objects both photometrically and spectroscopically to both classify and exclude sources. 
In parallel, the GRANDMA consortium is developing a similar approach to harmonise spectroscopic reduction with generic data pre-processing and calibration.

\section{GRANDMA electromagnetic follow-up campaign of O3 run A}
\label{O3asum}
\subsection{Summary of the 6 months of the campaign}

The GRANDMA consortium followed up 27/33 alerts during the first six months of the O3 campaign, as shown in Table~\ref{tab:GRANDMArecapO3BBH} for the binary black hole merger candidates, and Table~\ref{tab:GRANDMArecapO3BNS} for the systems containing at least one neutron star based on LIGO-Virgo low latency results (see the \href{https://emfollow.docs.ligo.org/userguide/}{LIGO-Virgo userguide} for more information~\footnote{https://emfollow.docs.ligo.org/userguide/}). All the sky localization of the alerts can be shown in Appendix~\ref{skycoveragelink}.

We first note that none of our follow-ups have led to a complete coverage of any error box. This can be explained by the fact that Sun and Moon constraints do not allow us to perform full coverage (see the example of S190924h). In addition, the luminosity distances of the alerts sent by LIGO-Virgo are mostly above 150 Mpc (the closest binary neutron star candidate was estimated to lie at $156\pm41$ Mpc), where the galaxy catalog we are using to perform galactic tiling is no longer complete.

The TAROT telescopes fully automatically responded to any alert. This helped us to track down several issues in the database and web interface, improving the entire GRANDMA system. From April to September 2019, the delay between the GW trigger time and the first observation by the TAROT network decreased from several hours to dozens of minutes. While our first answer to an alert was delayed by more of twelve hours, in September 2019, the ICARE infrastructure was able to receive the GW alert, and produce the observation plan and send it to all GRANDMA telescopes within $10-15$ minutes. Then, the automatic TAROT pipeline (CADOR)~\citep{BourezLaas2008CADORAT} was able to schedule the first observations to start within $5-15$ minutes at one of the observatories. For some other telescopes, the insertion into the schedule needs human validation, which can take up to one hour. We are working on reducing that time delay. The most rapid follow-up was for S190515ak with a delay of 16 minutes between the GW trigger and the first observation of TAROT-Calern.

Lastly, the coverage of GRANDMA improved during the first six months from 100 deg$^2$ to 300 deg$^2$ with a maximum coverage (for S190910d) of 540 deg$^2$ in 68~hours thanks to a more efficient coordination of the telescopes.

We also highlight the participation of the OAJ/T80 to the S190426c alert, a binary neutron star merger candidate; these observations covered 11 \% of the sky localization probability at 19.6 mag in the $r^\prime$ band. We also triggered various telescopes with the S190425z alert discussed in the following section~\ref{190425z}. 
Finally, ShAO/T60, Lisnyky/AZT-8, UBAI/NT-60 were triggered for the BNS candidate S190818y alert, using a galaxy targeting strategy, for the purpose of demonstrating a generic photometric detection pipeline on all the observations.
The follow-up activity increase also received help from a successful campaign with the amateur astronomers program for the S190901ap binary neutron star merger candidate alert, estimated at a luminosity distance of $241 \pm 79$\,Mpc. For that event, we reached a coverage of 90\% of the credible region of 14753 deg$^2$, pointing at some interesting galaxies contained in the GW volume \citep{gcn25688}.

Beside, our independent search for finding gamma-ray bursts in the Fermi-GBM data \citep{WBSpaper} in the period of around the GW trigger time (-5s, 1h) did not lead to any significant GRB detection.

Despite all our efforts, no significant counterparts were found by the GRANDMA consortium for any alert (in agreement with observations by other groups).

\begin{table*}
\caption[Summary of the GRANDMA observations]{Summary of the GRANDMA observations during the first six months for BBH candidates. Observations are not necessarily continuous during the time interval.
\href{https://gracedb.ligo.org/superevents/S190829u/view/}{S190829u} was retracted by the LIGO-Virgo collaboration due to a data quality issue.  90\% c.r. corresponds to the 90 \% credible region of the latest sky localisation area sent by LIGO-Virgo ($^a$, when only ``the bayestar sky localisation'' is available), $\delta \, t$ to the delay with respect to the GW trigger, $\Delta \, \mathrm{T}$ to the duration of the observations, Prob (\%), Area ${\mathrm{deg^2}}$ to the coverage of GRANDMA compared to the latest revision of the sky localisation area in percentage and in squares degrees. From 2019 May 19$^{th}$ to 2019 June 2$^{nd}$, TRE and TCH were under maintenance and there were bad observational conditions at the TCA site. \href{https://gracedb.ligo.org/superevents/S190924h/view/}{S190924h} was not observed due to moon constraints. The $\delta \, t$ is dependant on the delay of the first sky localisation area sent by the LVC.}
\label{tab:GRANDMArecapO3BBH}
\begin{tabular}{ccccc|cccccc}
Alert & Time & Type & Dist & 90\% c.r. & Telescope       & $\delta \, t$ & $\Delta \, \mathrm{T}$ & Lim. mag &  Prob & Area   \\
 & (UTC) &  & (Mpc) & (deg$^2$) &  & (h) & (h) &  &  (\%) & (${\mathrm{deg^2}}$)   \\
\hline
\multirow{4}{*}{\href{https://gracedb.ligo.org/superevents/S190930s/view/}{S190930s}} & \multirow{4}{*}{13:35:41} & \multirow{4}{*}{MG (95\%)} & \multirow{4}{*}{$709 \pm 191$} & \multirow{4}{*}{1748} & GRANDMA & 1.9  & 57.0  & 17-18  & 19.0  & 260  \\
 &  &  &  &   & TCA  & 4.4 &  55.2  &  18 &  5.1  & 68  \\
   &  &    & &  & TCH  & 12.7 & 46.0  & 18  & 3.0   & 38   \\
      &  &    & &  & TRE & 1.9 & 31.4    & 17  &  18.0   & 242   \\
\hline
\href{https://gracedb.ligo.org/superevents/S190924h/view/}{S190924h} & 02:19:258 & MG (99\%) & $548 \pm 112$ & 303 & \multicolumn{6}{c}{No observations} \\
\hline
\multirow{2}{*}{\href{https://gracedb.ligo.org/superevents/S190915ak/view/}{S190915ak}} & \multirow{2}{*}{23:57:02} & \multirow{2}{*}{BBH (99\%)} & \multirow{2}{*}{$ 1584 \pm 381$} &  \multirow{2}{*}{318} & GRANDMA & 0.3 & 50.7 & 18  &  36.1  & 80 \\
 &  &  &  &   & TCA  & 0.3  & 50.7  & 18 &  36.1  & 80  \\
\hline
\multirow{4}{*}{\href{https://gracedb.ligo.org/superevents/S190828l/view/}{S190828l}} & \multirow{4}{*}{06:55:09}  & \multirow{4}{*}{BBH (99\%)} & \multirow{4}{*}{$ 1528\pm 387$} & \multirow{4}{*}{359} & GRANDMA & 1.2 &  49.7 &  17-18  & 12.6  & 294   \\
  &  &  &  &   & TCA  & 12.6  & 31.2 & 18  & 0.3 & 67   \\
   &  &  &  &  & TCH  & 1.2 & 49.7 & 18  & 11.5  &  69  \\
  &  &  &  &   & TRE  & 14.2 & 23.7 & 17  & 0.8 & 157  \\
\hline
\multirow{4}{*}{\href{https://gracedb.ligo.org/superevents/S190828j/view/}{S190828j}} & \multirow{4}{*}{06:34:05}  &\multirow{4}{*}{ BBH (99\%)} &  \multirow{4}{*}{$1946 \pm 388$}  & \multirow{4}{*}{228}  & GRANDMA & 0.5  & 47.6  & 17-18   & 6.1  & 210   \\
  &  &  &  &   & TCA  & 12.5  & 31.3 & 18  & 2.7  &  67  \\
   &  &  &  &   & TCH  & 0.5 & 47.6 & 18  & 0.04  & 21    \\
  &  &  &  &   & TRE  & 15.6  & 23.4 & 17  & 3.4 & 122   \\
\hline
\multirow{4}{*}{\href{https://gracedb.ligo.org/superevents/S190728q/view/}{S190728q}} & \multirow{4}{*}{06:45:10} & \multirow{4}{*}{MG (52\%)} & \multirow{4}{*}{$ 874 \pm 171$} & \multirow{4}{*}{104} & GRANDMA & 0.5  & 50.2  & 17-18  & 88.3  & 180  \\
  &  &  &  &   & TCA  & 13.2  & 29.6 & 18 & 18.8  & 26   \\
   &  &  &  &   & TCH  & 0.5  & 50.2 & 18 & 23.2  & 56    \\
  &  &  &  &   & TRE  & 10.5 & 22.6 & 17 & 83.3 &   121 \\
\hline
\multirow{3}{*}{\href{https://gracedb.ligo.org/superevents/S190727h/view/}{S190727h}} & \multirow{3}{*}{06:03:33} & \multirow{3}{*}{BBH (92\%)} & \multirow{3}{*}{$ 2839 \pm 655$} & \multirow{3}{*}{151} & GRANDMA & 1.5  & 163.8  &  17-18 & 49.0  & 138   \\
   &  &  &  &   & TCH  & 1.6 & 163.8 & 18 & 48.4 & 68  \\
  &  &  &  &   & TRE  & 16.7 & 17.2 &  17 & 0.7  & 70   \\
\hline
\multirow{4}{*}{\href{https://gracedb.ligo.org/superevents/S190720a/view/}{S190720a}} & \multirow{4}{*}{00:08:36} & \multirow{4}{*}{BBH (99\%)} & \multirow{4}{*}{$ 869 \pm 283$} & \multirow{4}{*}{443} & GRANDMA & 0.3  &  30.8 & 17-18   & 11.0  & 161  \\
  &  &  &  &   & TCA  & 0.6 & 26.8 & 18 & 5.0 &  60  \\
   &  &  &  &   & TCH  & 1.7   & 29.4 & 18 & 4.0  &  50  \\
  &  &  &  &   & TRE  & 0.3 & 22.7 & 17 & 4.0 &  71  \\
\hline
\multirow{4}{*}{\href{https://gracedb.ligo.org/superevents/S190707q/view/}{S190707q}} & \multirow{4}{*}{09:33:26} & \multirow{4}{*}{BBH (99\%)} & \multirow{4}{*}{$ 781 \pm 211$} & \multirow{4}{*}{921} & GRANDMA & 5.7  & 14.1   & 17-18  & 14.6  & 128  \\
  &  &  &  &   & TCA  & 10.7 & 6.1 & 18 & 2.4  & 21    \\
   &  &  &  &   & TCH  & 13.4 & 6.3 & 18 & 4.0  & 25   \\
  &  &  &  &   & TRE  & 5.7 & 4.8 & 17 & 9.5 & 88   \\
\hline
\multirow{4}{*}{\href{ttps://gracedb.ligo.org/superevents/S190706ai/view/}{S190706ai}} & \multirow{4}{*}{22:26:41} & \multirow{4}{*}{BBH (99\%)} & \multirow{4}{*}{$ 5263 \pm 1402$} & \multirow{4}{*}{826} & GRANDMA & 0.5 &  22.9 & 17-18   & 16.0 & 167 \\
  &  &  &  &   & TCA  & 0.6 & 22.7 & 18 & 0.27  & 22   \\
   &  &  &  &   & TCH  & 0.5 & 9.1& 18 & 3.9  & 24   \\
  &  &  &  &   & TRE  & 3.2  & 2.0 & 17  & 11.8 &  121  \\
\hline
\multirow{2}{*}{\href{https://gracedb.ligo.org/superevents/S190701ah/view/}{S190701ah}} & \multirow{2}{*}{20:33:06} & \multirow{2}{*}{BBH (93\%)} & \multirow{2}{*}{$ 1849 \pm 446$} & \multirow{2}{*}{49} & GRANDMA & 2.7  & 2.9  & 17  & 41.7  & 71  \\
  &  &  &  &   & TRE  & 2.7 & 2.9 & 17  & 41.7  & 71    \\
\hline
\multirow{3}{*}{\href{https://gracedb.ligo.org/superevents/S190630ag/view/}{S190630ag}} & \multirow{3}{*}{18:52:05} & \multirow{3}{*}{BBH (94\%)} & \multirow{3}{*}{$ 926 \pm 259$} & \multirow{3}{*}{1483} & GRANDMA & 4.7 & 18.9  & 17-18   & 11.4 & 92  \\
 &  &  &  &   & TCA  & 4.7 & 3.0 & 18 & 8.9  & 21  \\
  &  &  &  &   & TRE  & 21.0 & 2.5 & 17  & 2.5  & 71   \\
\hline
\href{https://gracedb.ligo.org/superevents/S190602aq/view/}{S190602aq} & 17:59:27 & BBH (99\%) & $ 797 \pm 238$ & 1172 & \multicolumn{6}{c}{No observations} \\
\hline
\href{https://gracedb.ligo.org/superevents/S190521r/view/}{S190521r} & 07:43:59 & BBH (99\%) & $ 1136 \pm 279$ & 488 & \multicolumn{6}{c}{No observations} \\
\hline
\href{https://gracedb.ligo.org/superevents/S190521g/view/}{S190521g} & 03:02:29 & BBH (97\%) & $ 3931 \pm 953$ & 765 & \multicolumn{6}{c}{No observations} \\
\hline
\href{https://gracedb.ligo.org/superevents/S190519bj/view/}{S190519bj} & 15:35:44 & BBH (96\%) & $ 3154 \pm 791$ & 967 &\multicolumn{6}{c}{No observations}  \\
\hline
\href{https://gracedb.ligo.org/superevents/S190517h/view/}{S190517h}$^a$ & 05:51:01 & BBH (98\%) & $ 2950 \pm 1038$ & 939 & GRANDMA & 4.2 & 17.3  & 18 & 8.9 & 11 \\
  &  &  &  &   & TCH  & 4.2 & 17.3 & 18 & 8.9 & 11 \\
\hline
\multirow{3}{*}{\href{https://gracedb.ligo.org/superevents/S190513bm/view/}{S190513bm}$^a$}  & \multirow{3}{*}{20:54:28} & \multirow{3}{*}{BBH (94\%)} & \multirow{3}{*}{$ 1987 \pm 501$} & \multirow{3}{*}{691} & GRANDMA & 0.7 & 12.9   &  18 & 23.5 & 45  \\
 &  &  &  &   & TCA    & 0.7 & 5.6 & 18 & 7.0  & 24 \\
  &  &  &  &   & TCH  & 6.0  & 7.5 & 18 & 16.7  &  21   \\
\hline
\multirow{3}{*}{\href{https://gracedb.ligo.org/superevents/S190512at/view/}{S190512at}} & \multirow{3}{*}{18:07:14} & \multirow{3}{*}{BBH (99\%)} & \multirow{3}{*}{$ 1388 \pm 322$} & \multirow{3}{*}{252} & GRANDMA & 13.8  & 18.1  & 18  & 30.3  & 42  \\
 &  &  &  &   & TCA  &  25.5 & 5.8  & 18 & 7.2  & 22    \\
  &  &  &  &   & TCH  & 13.8 & 18.1 & 18 & 24 & 24\\
\hline
\multirow{2}{*}{\href{https://gracedb.ligo.org/superevents/S190503bf/view/}{S190503bf}$^a$} & \multirow{2}{*}{18:54:04} & \multirow{2}{*}{BBH (96\%)} & \multirow{2}{*}{$421 \pm 105$} & \multirow{2}{*}{448} & GRANDMA & 4.3 & 4.7  & 18   & 17.1 & 24  \\
 &  &  &  &   & TCH  & 4.3 & 4.7 &  18 &   17.1 &  24  \\
\hline
\multirow{2}{*}{\href{https://gracedb.ligo.org/superevents/S190421ar/view/}{S190421ar}} & \multirow{2}{*}{21:38:56} & \multirow{2}{*}{BBH (97\%)} & \multirow{2}{*}{$1628 \pm 535$} & \multirow{2}{*}{1444} & GRANDMA & 19.7 & 8.2  & 17  &  18.0 & 124  \\
 &  &  &  &   & TRE  & 19.7 & 8.2 & 17  &  18.0 & 124   \\
\hline
\multirow{3}{*}{\href{https://gracedb.ligo.org/superevents/S190412m/view/}{S190412m}$^a$} & \multirow{3}{*}{05:30:44} & \multirow{3}{*}{BBH (100\%)} & \multirow{3}{*}{$812 \pm 194$} & \multirow{3}{*}{156} & GRANDMA & 10.1 & 12.2  & 17-18   & 75.9 &  125\\
 &  &  &  &   & TCA  & 14.1 & 8.2 & 18 & 36.3 &  24 \\
 &  &  &  &   & TRE  & 10.1 & 7.7 & 17  & 73.4 & 123 \\
\hline
\href{https://gracedb.ligo.org/superevents/S190408an/view/}{S190408an}$^a$ & 18:18:02 & BBH (100\%) & $1473 \pm 358$ & 387 & \multicolumn{6}{c}{No observations}  \\
\hline
\end{tabular}
\end{table*}

\begin{table*}
\caption{Summary of the GRANDMA observations during the first six months for binary neutron-star or neutron-star black-hole merger candidates, using the latest versions of the LALInference sky localisations ($^a$, when only ``the bayestar sky localisation'' is available). Observations are not necessarily continuous during the time interval. \href{https://gracedb.ligo.org/superevents/S190518bb/view/}{S190518bb}, \href{https://gracedb.ligo.org/superevents/S190524q/view/}{S190524q}
\href{https://gracedb.ligo.org/superevents/S190808ae/view/}{S190808ae},
\href{https://gracedb.ligo.org/superevents/S190816ic/view/}{S190816i}, \href{https://gracedb.ligo.org/superevents/S190822c/view/}{S190822c} were retracted by the LIGO-Virgo collaboration due to data quality issues in the detectors.  90\% c.r. corresponds to the 90 \% credible region of the latest sky localisation area sent by LIGO-Virgo, $\delta \, t$ to the delay with respect to the GW trigger, $\Delta \, \mathrm{T}$ to the duration of the observations, Prob (\%), Area ${\mathrm{deg^2}}$ to the coverage of GRANDMA compared to the latest revision of the sky localisation area in percentage and in squares degrees. $\delta \, t$ is dependant on the delay of the first sky localisation area sent by the LIGO-Virgo collaboration, this generally took about few hours at the beginning of the campaign.}
\label{tab:GRANDMArecapO3BNS}
\begin{tabular}{ccccc|cccccc}
Alert & Time & Type & Dist & 90\% c.r. & Telescope       & $\delta \, t$ & $\Delta \, \mathrm{T}$ & Lim. mag &  Prob & Area   \\
 & (UTC) &  & (Mpc) & (deg$^2$) &  & (h) & (h) &  &  (\%) & (${\mathrm{deg^2}}$)   \\\hline
\multirow{4}{*}{\href{https://gracedb.ligo.org/superevents/S190930t/view/}{S190930t}$^a$} & \multirow{4}{*}{14:34:07} & \multirow{4}{*}{NSBH (74\%)} & \multirow{4}{*}{$108 \pm 38$} & \multirow{4}{*}{24220} & GRANDMA & 1.0  & 80.5 & 17-18  & 1.3  & 250  \\
 &  &  &  &   & TCA  & 3.5 & 80.0   & 18  &  0.7  & 135   \\
   &  &    & &  & TCH  &  11.9 & 71.5 & 18  &  0.6  & 110   \\
      &  &    & &  & TRE  & 1.0  & 80.5 & 17  &  0.4  &  94  \\
\hline
\multirow{4}{*}{\href{https://gracedb.ligo.org/superevents/S190923y/view/}{S190923y}$^a$} & \multirow{4}{*}{12:55:59} & \multirow{4}{*}{NSBH (68\%)} & \multirow{4}{*}{$438 \pm 133$} & \multirow{4}{*}{2107} & GRANDMA & 3.6 & 55.6  & 17-18   & 25.4  & 436  \\
 &  &  &  &   & TCA  & 5.5 & 46.3  & 18 &  2.2  & 64  \\
   &  &    & &  & TCH  & 10.7 & 32.7 & 18  & 3.5   &  64  \\
      &  &    & &  & TRE  & 3.6 & 55.6   & 17  &  20.0  &  312  \\
\hline
\multirow{3}{*}{\href{https://gracedb.ligo.org/superevents/S190910h/view/}{S190910h}} & \multirow{3}{*}{08:29:58} & \multirow{3}{*}{BNS (61\%)} & \multirow{3}{*}{$230 \pm 88$} & \multirow{3}{*}{24264} & GRANDMA & 10.5 & 129.1 & 18  & 1.1   & 337 \\
 &  &  &  &   & TCA  & 10.5 & 129.1  & 18 &  0.5  & 175  \\
   &  &    & &  & TCH  & 25.0 & 113.1 &  18 &  0.6  &  161  \\
\hline
\multirow{4}{*}{\href{https://gracedb.ligo.org/superevents/S190910d/view/}{S190910d}} & \multirow{4}{*}{01:26:19} & \multirow{4}{*}{NSBH (98\%)} & \multirow{4}{*}{$632 \pm 186$} & \multirow{4}{*}{2482} & GRANDMA & 1.0 & 67.6 & 17-18  & 37   & 540  \\
 &  &  &  &   & TCA  & 1.0 & 67.6  & 18 & 4   & 103  \\
   &  &    & &  & TCH  & 3.5 & 52.6   & 18 & 11   & 158  \\
 &  &  &  &   & TRE  & 18.3 & 44.2  & 17 & 32   & 360 \\
\hline
\multirow{4}{*}{\href{https://gracedb.ligo.org/superevents/S190901ap/view/}{S190901ap}} & \multirow{4}{*}{23:31:01} & \multirow{4}{*}{BNS (86\%)} & \multirow{4}{*}{$ 241 \pm 79$} & \multirow{4}{*}{14753} & GRANDMA & 0.4  & 58.6  & 17-18  & 9.1 & 354 \\
  &  &  &  &   & TCA  & 1.25 & 57.4 & 18 & 2.5  & 100   \\
   &  &  &  &   & TCH  & 4.5 & 46.6 &  18& 3.0 & 112   \\
  &  &  &  &   & TRE  & 0.4  & 48.6 & 17 & 7.2 & 279  \\
\hline
\multirow{5}{*}{\href{https://gracedb.ligo.org/superevents/S190814bv/view/}{S190814bv}} & \multirow{5}{*}{21:10:39} & \multirow{5}{*}{NSBH (99\%)} & \multirow{5}{*}{$ 267 \pm 52$} & \multirow{5}{*}{23} & GRANDMA & 0.5  & 22.5  &  17-23  & 90.2  & 161  \\
  &  &  &  &   & TCA  & 0.65 & 13.5 & 18  & 21.9  & 32   \\
  &  &  &  &   & TRE  & 0.5  & 22.2  &17& 89.0 & 139  \\
    &  &  &  &   & LesMakes  &  1.3 & 3.6 & 18.5 & 6.1  &  2  \\
    &  &  &  &   & CFHT  &  232.0 & 0.5 & 23.0 & $<1$  &  1  \\
\hline
\multirow{5}{*}{\href{https://gracedb.ligo.org/superevents/S190718y/view/}{S190718y}$^a$} & \multirow{5}{*}{14:35:12} & \multirow{5}{*}{BNS (2\%)} & \multirow{5}{*}{$ 227 \pm 165$} & \multirow{5}{*}{7246} & GRANDMA & 3.9  & 54.6 & 17-19 & 48.8  &  120   \\
 &  &  &  &   & TRE  & 7.4 & 51.1 & 17 & 48.8  & 120  \\
  &  &  &  &   & Shao/T60  & 4.6 & 1.3 & 16.5 & $<1$   & $<1$  \\
    &  &  &  &   & UBAI/T60  & 3.9  & 1.5 & 18.0  & $<1$  & $<1$  \\
    &  &  &  &   & Lisnyky/AZT-8  & 5.7 & 2.8 & 19.2 & $<1$  & $<1$  \\
\hline
\multirow{3}{*}{\href{https://gracedb.ligo.org/superevents/S190510g/view/}{S190510g}} & \multirow{3}{*}{02:59:39} & \multirow{3}{*}{BNS (42\%)} & \multirow{3}{*}{$ 227 \pm 92$} & \multirow{3}{*}{1166} & GRANDMA & 2.1 & 69.3  &18 & 45.0  & 51   \\
 &  &  &  &   & TCA  & 16.7 & 54.7 & 18 & 0.3 & 24    \\
  &  &  &  &   & TCH  & 2.1 & 68.3 & 18 & 44.7 & 31    \\
\hline
\multirow{5}{*}{\href{https://gracedb.ligo.org/superevents/S190426c/view/}{S190426c}} & \multirow{5}{*}{15:21:55} & \multirow{5}{*}{BNS (24\%)} & \multirow{5}{*}{$377 \pm 100$} & \multirow{5}{*}{1131} & GRANDMA & 6.3 & 60.5  & 17-20  & 28.6 & 152  \\
 &  &  &  &   & TCA  & 7.4  & 52.9 & 18 & 9.7 & 23    \\
 &  &  &  &   & TCH  & 15.4  & 51.4 & 18 & 4.3 & 25   \\
 &  &  &  &   & TRE  & 7.4 & 50.9 & 17  & 14.0  & 105  \\
  &  &  &  &   & OAJ  & 6.3  & 4.9 & 19.6 & 11.4 & 44    \\
\hline
\multirow{9}{*}{\href{https://gracedb.ligo.org/superevents/S190425z/view/}{S190425z}} & \multirow{9}{*}{08:18:05} & \multirow{9}{*}{BNS (99\%)} & \multirow{9}{*}{$156 \pm 41$} & \multirow{9}{*}{7461} & GRANDMA & 6.7 & 63.4  & 17-23   & 2.8  & 135 \\
 &  &  &  &   & TCA & 38.1 & 29.1  & 18.0 & 0.9 & 18.0 \\
 &  &  &  &   & TCH & 22.4 & 47.7  & 18.0 &  1.0  & 25 \\
 &  &  &  &   & TRE & 6.7 & 33.6 & 17.0   & 2.4 & 124 \\
 &  &  &  &   & Zadko & 6.7 & 28.0  & 16.0  & $<1$ & $<1$ \\
 &  &  &  &   & Abastumani/T70 & 14.2 & 2.5 & 16.5 & $<1$ &  $<1$ \\
 &  &  &  &   & Lisnyky/AZT-8 & 14.8 & 2.2 & 19.2  & $<1$ & $<1$ \\
 &  &  &  &   & Les Makes/T60 & 7.0 & 8.5  & 19.2  & $<1$ & $<1$ \\
 &  &  &  &   & GMG/2.4 & 9.3 & 3.0  & 19.0  & $<1$ & $<1$ \\
 &  &  &  &   & CAHA/2.2 & 81.6 & 1.7  & 23.0  & $<1$ & $<1$ \\

\hline
\end{tabular}
\end{table*}

\subsection{An example of follow-up: the binary neutron star merger candidate S190425z}
\label{190425z}

\subsubsection{S190425z: the first binary neutron star merger candidate of O3}

The GW event S190425z occurred on 2019-04-25 at 08:18:05 UTC \citep{LVC2019GCN24168}. At that moment, both Advanced LIGO-Livingston and Virgo were observing, while Advanced LIGO-Hanford had been down for 30 minutes. The lack of a detection by LIGO-Hanford resulted in a very large localisation region, with the initial 90\% credible region of the skymap for S190425z spanning $\sim10200$ deg$^2$. A refined analysis reduced this region to $\sim7500$ deg$^2$.

The false-alarm rate of S190425z, an estimate of its significance, was $\approx$ one per 70000 years (see definitions in the \href{https://emfollow.docs.ligo.org/userguide/}{the LIGO-Virgo userguide}). The candidate also had a large probability to be a binary neutron star merger and to have a remnant (both probabilities $>99\%$). Finally, the distance, inferred from the GW signal was estimated to be $156\pm41$ Mpc. 

Due to these properties, an intense follow-up campaign occurred in the month following the event, with more than 100 notices sent by various teams via the GCN network\footnote{https://gcn.gsfc.nasa.gov/other/GW190425z.gcn3}. Wide-field optical tiling collaborations like ZTF, ATLAS, Pan-STARRS and Gaia discovered multiple optical counterpart candidates \citep{Kasliwal2019GCN24191,McBrien2019GCN24197,Smith2019GCN24210,Smith2019GCN24262,Kostrzewa-Rutkowska2019GCN24345,Kostrzewa-Rutkowska2019GCN24354,Kostrzewa-Rutkowska2019GCN24366}, but all turned out to be supernovae or other sources unrelated to S190425z (\citealt{Pavana2019GCN24200,Perley2019GCN24204,Buckley2019GCN24205,Izzo2019GCN24208,Wiersema2019GCN24209,Castro-Tirado2019GCN24214,Short2019GCN24215} and \citealt{Dichiara2019GCN24220,Morokuma2019GCN24230,Jencson2019GCN24233,Carini2019GCN24252,Chang2019GCN24260,McCully2019GCN24295,Nicholl2019GCN24217,Nicholl2019GCN24321,Dimitriadis2019GCN24358}). At high energies, no contemporaneous emission was detected by any satellites \citep{Martin-Carillo2019GCN24169,Minaev2019GCN24170,Sugizaki2019GCN24177,Savchenko2019GCN24178,Casentini2019GCN24180,Chelovekov2019GCN24181,Sakamoto2019GCN24184,Piano2019GCN24186,Xiao2019GCN24213,Axelsson2019GCN24266,Svinkin2019GCN24417}. Neutrinos and cosmic rays were also not detected \citep{Icecube2019GCN24176,Alvarez-Muniz2019GCN24240}.

Several factors led to difficulties in finding the counterpart. First of all, the highly unconstrained localisation led to $\sim$\,50,000 galaxies within the three-dimensional estimated volume \citep{Dalya2019GCN24171,Cook2019GCN24232}. It is challenging to observe even most of these galaxies several times in order to pinpoint a transient event. Secondly, the large distance (four times further than GW170817) implied that a kilonova similar to AT 2017gfo/GW170817 would appear two to three magnitudes fainter. Observing such a faint object implies an increase in exposure time required for each image (for a given telescope), leading to it being highly challenging to succeed in a detection during the first twelve hours. Therefore, it is not surprising that no significant candidate counterparts were discovered. 

\subsubsection{GRANDMA follow-up observations of S190425z}

The LIGO Scientific Collaboration and the Virgo Collaboration sent a GCN notice 
42\,min after the trigger. Observation plans with both ``tiling'' for the TAROT network and ``galaxy targeting'' strategies for the other telescopes were computed automatically as described in Section~\ref{strategy}. These plans were sent to the telescopes 15\,min after the initial GCN reception. 

A total of seven GRANDMA telescopes responded to this alert. The observations started in the following order: Zadko (2019-04-25 14:56:19 UTC, 6.7\,hr post GW trigger time), TAROT/TRE (2019-04-25 14:56:19 UTC, 6.7\,hr post GW trigger time), Les Makes/T60 (2019-04-25 15:18:10 UTC, 7.0\,hr post GW trigger time), Lisnyky/AZT-8 (2019-04-25 22:32:38 UTC, 14.2\,hr post GW trigger time), Abastumani/T70 (2019-04-25 23:08:01 UTC, 14.8\,hr post GW trigger time), TAROT/TCH (2019-04-26  06:40:5 UTC, 22.4\,hr post GW trigger time), and TAROT/TCA (2019-04-26 22:25:38 UTC, 1.6\,days post GW trigger time).


\begin{figure*}
\begin{center}
\includegraphics[scale=0.24]{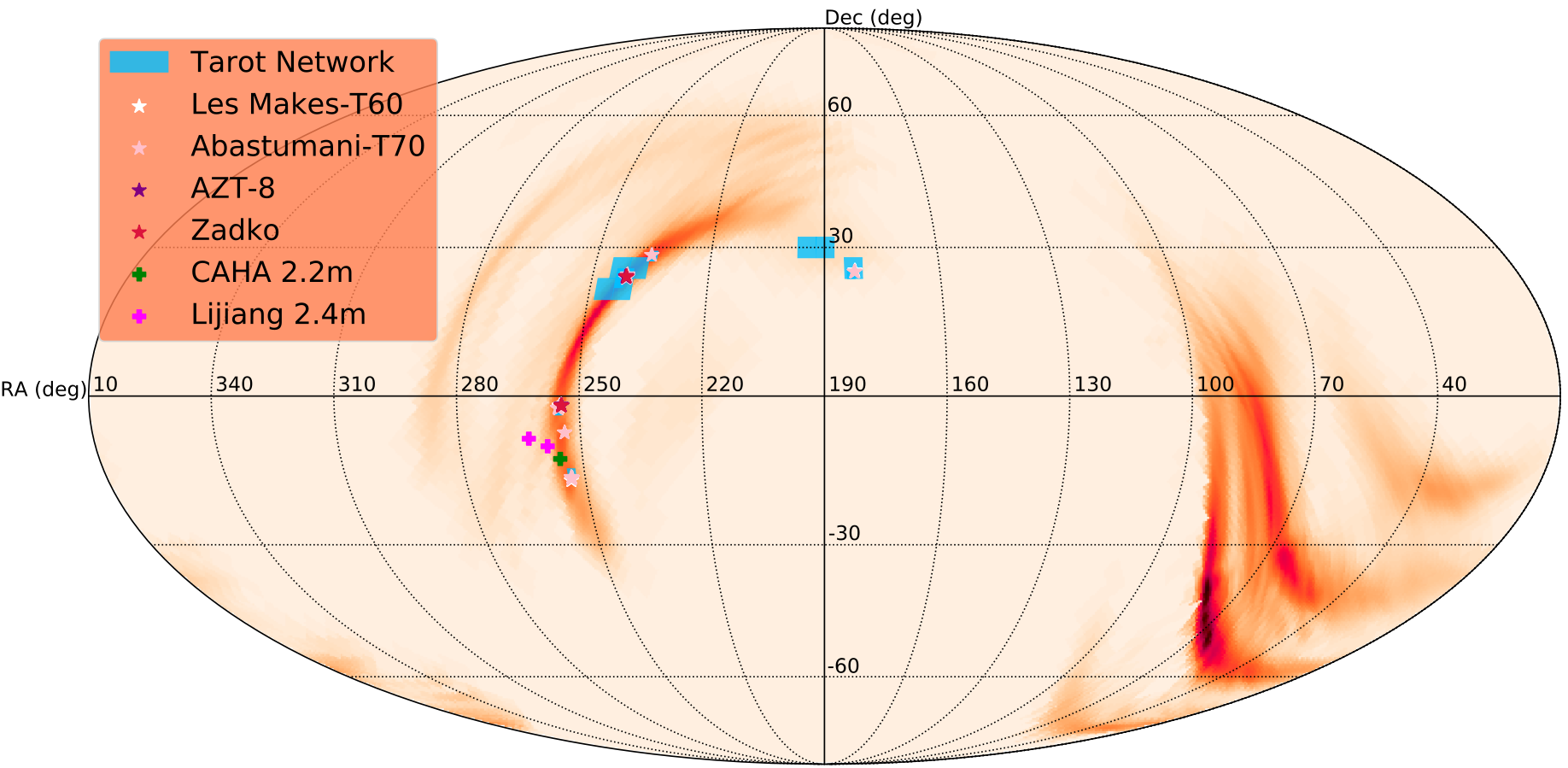}
\caption{GRANDMA follow-up of the GW candidate event S190425z, a binary neutron-star merger candidate. Blue tiled areas represent observational tiles obtained by the TAROT network. In red, the LALInference sky localisation area of S190425z is shown. Stars represent galaxy-targeted fields obtained by multiple observatories (see legend). Plus signs represent direct observations of candidate counterparts (not discovered by GRANDMA).}
\label{fig:GRANDMAS190425zfull}
\end{center}
\end{figure*}

\begin{figure}
\begin{center}
\includegraphics[scale=0.21]{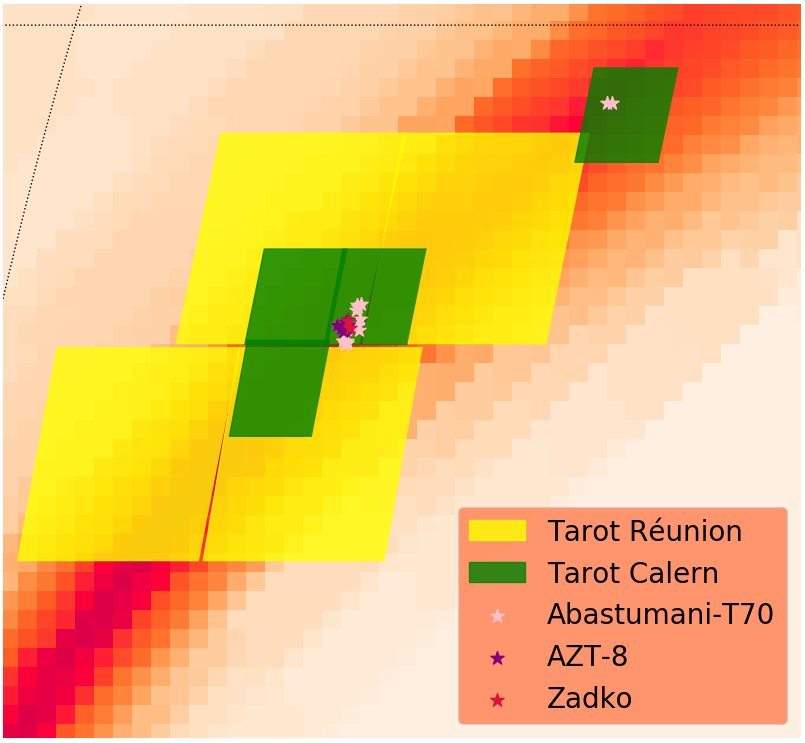}
\caption{Zoom into Fig. \ref{fig:GRANDMAS190425zfull} to show details of the observation tiles and the galaxy-targeted follow-up. While the fields for each of the TAROT telescopes do not (or barely) overlap, a more efficient system of assigning tiles to prevent overlap across multiple telescopes \citep{Networkpaper} had not been implemented here yet.}
\label{fig:GRANDMAS190425zzoom}
\end{center}
\end{figure}

The observations were previously reported in \citet{Blazek2019GCN24227}, \citet{Howell2019GCN24256} and \citet{Gendre2019GCN24763}. The joint coverage of the alert is shown in Figure~\ref{fig:GRANDMAS190425zfull} and Figure~\ref{fig:GRANDMAS190425zzoom}; the data are presented in Table~\ref{tab:GRANDMArecapO3BNS}. The TAROT network observed seven fields with multiple epochs per telescope corresponding to a few percent of the initial Bayestar sky localisation coverage (see Table~\ref{tab:TAROT190425z} in the Appendix). The Zadko telescope and camera were in test mode, with its sensitivity strongly decreased compared to its normal state; in total, images of 70 galaxies compatible with the GW distance were obtained in the five fields observed. The Les Makes/T60 recorded five images of 300\,s each: 52 galaxies compatible with the GW distance were observed. Abastumani-T70 observed eleven fields containing 41 galaxies in the $R$ band with a partly cloudy sky. Lisnyky/AZT-8 observed seven tiles with images of 26 potential host galaxies in the $R$ and $B$ bands. The full list of galaxies observed by Zadko, Les Makes/T60, Lisnyky/AZT-8, Abastumani/T70 is available in Table~\ref{tab:galaxytarget_S190425z} of Appendix A.

Each telescope team was in charge of the reduction of its own data, and each team used diverse techniques to search for new transients. The teams mostly compared their data with deeper archival imaging of the fields by hand, for example PanSTARRS, SDSS, or, if neither of those surveys cover this sky position, DSS. They also checked the Minor Planet Center database\footnote{\url{https://minorplanetcenter.net/cgi-bin/checkmp.cgi}} to see whether some spurious ``transient'' was found to be a passing asteroid and therefore not related to the event. In addition, the TAROT network used a different method to find transients described in~\citet{2018cosp...42E2475N}. There were no reported counterpart candidates. The upper limit for the kilonova or orphan afterglow emission at $5\sigma$ in the sky localisation area observed by GRANDMA were $Clear<16$ mag for Zadko, $Clear<17$ mag for the TAROT network,   $R_C<19.2$ for Les Makes/T60, $R_C,B<19.2$ for Lisnyky/AZT-8, and $R_C<16.5$ for Abastumani/T70. 

The automatic observations of the event by the TAROT network favoured multi-epoch tiling observations in order to catch the rise of the potential kilonova emission. The localisation was more than a factor of 100 larger than that of GW170817, which was used as our study case \citep{LSC_BNS_2017PhRvL}. Thus, the mapping of the skymap was far from optimal, with temporal redundancy not only in the same telescope but also in the full network. Also, the strategy employed for galaxies allowed overlap of fields. In this sense, S190425z helped to vet the full infrastructure and demonstrated our capacity to observe GW alerts with a high temporal cadence, albeit optimised for a smaller localisation. 

There are a variety of lessons learned from this first binary neutron-star candidate. First of all, coordination of multi-colour observations: the earliest observations of AT 2017gfo associated with GW170817 revealed that it was a very blue source (e.g. \citealt{Evans2017Science}), in contrast to the typical red colours of SNe, especially ones beyond peak. Secondly, we desire to better use the small FoV instruments by preventing overlapping tiles by slightly shifting the fields imaged, in addition to incorporating automatic interruption for observing counterpart candidates. Finally, the time needed to process the images was underestimated, and the difference of pipelines between each group led to some uncertainties on the coherence of the data analysis. This motivates a universal data reduction.

\subsubsection{External counterpart candidates found by ZTF and \textit{Swift} UVOT}

We also observed some of the candidate counterparts reported by other groups.
We observed the candidates ZTF19aarykkb and ZTF19aarzaod reported by the Zwicky Transient Facility \citep{Kasliwal2019GCN24191} with LJT. Multiple images were obtained in the $g^\prime$ and $r^\prime$ bands \citep{2019GCN.24267....1L}, at two epochs for ZTF19aarykkb, as well as a single $r^\prime$ epoch of ZTF19aarzaod. Images from the PanSTARRS survey were used for template subtraction, and stars from this catalogue for calibration. Both sources are clearly detected, and the derived magnitudes are comparable to those derived by other teams \citep{Tan2019GCN24193,Hiramatsu2019GCN24194,Bhalerao2019GCN24201,Burke2019GCN24206,Kilpatrick2019GCN24212,Castro-Tirado2019GCN24214,Sun2019GCN24234}. However, both events were ruled out as a possible counterpart, as they were spectroscopically identified as type II SNe \citep{Pavana2019GCN24200,Perley2019GCN24204,Buckley2019GCN24205,Izzo2019GCN24208,Wiersema2019GCN24209,Nicholl2019GCN24211,Castro-Tirado2019GCN24214,Dichiara2019GCN24220,Chang2019GCN24260}.

More or less at the same time, \textit{Swift} UVOT detected a bright, blue, fast-evolving optical transient \citep{Breeveld2019GCN24296}. We triggered the 2.2\,m CAHA telescope to obtain photometric and spectroscopic follow-up observations. However, before the observations had commenced, another group reported that the source had since strongly faded \citep{Kong2019GCN24301}. For this reason, we cancelled the spectroscopic observations, requesting only photometric data, and detected the source in $R_CI_C$ but not in $UBV$ \citep{Kann2019GCN24459}. It has been interpreted since as the UV-bright flare of a late dwarf star \citep{Lipunov2019GCN24326,Bloom2019GCN24337} and is therefore not related to S190425z.

\subsection{Constraining kilonova properties}

Following the presentation of the targeted follow-up observations by the GRANDMA network, we want to seek answers to the questions: can we derive constraints on the observed GW transients from our GRANDMA observations, and which kind of EM transients would have been missed by our observations? For this purpose, we consider as an example the Lisnyky/AZT-8's observation of the galaxy HyperLEDA 1109773, which was part of the S190425z follow-up campaign. HyperLEDA 1109773 is about 136.9 Mpc away and Lisnyky/AZT-8 reached a limiting magnitude of 19.2 mag in the $R_C$ and $B$ bands.
From the missing transient observation, we know that only compact binary mergers in HyperLEDA 1109773 would be in agreement with our observations, for which the maximum $R_C$-band magnitude would be below 19.2 mag. For a quantitative interpretation of this statement, we follow~Coughlin M., Dietrich T. et al. (in prep.), i.e. we use the kilonova model of~\citep{KaMe2017} in combination with a Gaussian process regression based interpolation \citep{DoFa2017} to create a surrogate model for arbitrary lightcurve and ejecta properties, see e.g.~\citep{CoDi2018b,CoDi2018} for details. Our analysis shows that only extremely high ejecta masses, 
$\sim 0.5 M_\odot$ or larger, are slightly disfavoured by our 
observations, for comparison the ejecta mass of GW170817 is estimated to be $\sim 0.05 M_\odot$. 
Thus, from our observations alone, kilonovae associated with smaller ejecta masses might have been missed. 
Such considerations provide a natural limit for target identification for GW alerts with large distances. However, if aperture times are increased then the GRANDMA network can be successfully employed for target characterisation of already observed transients. This is of particular importance for the astronomical community, since typically a large number of potential transients are obtained from large sky surveys and, as discussed, the GRANDMA network provides a perfect tool to follow these up.

\section{Discussion: towards network optimisation}

There has been a significant effort to optimise single-telescope observations (for example \citealt{2016ApJ...820..136G,CoTo2018,2017ApJ...838..108R,2016A&A...592A..82G}) for large target-of-opportunity sky areas in the context of time-domain astronomy.
There are a variety of considerations when it comes to how to use a single system.
For example, the difference in techniques between synoptic or galaxy-targeted based searches is not always clear.
While some telescopes have FoVs such that one or the other is clearly preferred, for systems with FoV $\approx0.1$ square degrees, the answer is less obvious.
One may also consider the difference in sensitivities (and potentially filters) between systems, as well as the ability to react promptly to an alert and the time allocated to observe targets of opportunity.
The use of a small-aperture system in the optical is likely to be significantly different than that of a large-aperture system in the near infrared, even if both rely on galaxy-targeting to search for counterparts.

A network level optimisation is much more complicated.
For example, the differences in sensitivity and FoV potentially lead to significantly different search strategies within the same network.
Moreover, there are often differences between the observatories on the policy regarding how to respond to different types of alerts, for example, binary neutron stars or binary black holes.
A network that has both synoptic and galaxy-targeted systems may make choices as to whether covering overlapping parts of the sky is appropriate.
The unequal distribution of the telescopes on the Earth should also be taken into account, with some regions having a higher concentration of telescopes and some regions only being covered by a single telescope. 
A centralised scheduler supervising all the telescopes will ideally be the best option, which can respond in real time to the availability of telescopes for the observations, accounting for GW sky localisation regions and their updates with candidates.
In reality, it is difficult to create such a scheduler due to a variety of reasons, including the political aspects of various programmes, target-of-opportunity approvals required every semester, detection-pipeline reactions to transients and the diversity of independent schedulers.

The ideal answer to this likely depends on whether the systems have similar sensitivities to transients, as well as if it is possible to schedule coordinated observations between the systems.
This includes accounting for the use of different filters or creating significant temporal separation between observations.
These multi-epoch observations of the same transients can then be useful to provide a measurement of luminosity evolution or colour information on a given source.
Beyond simply searching for new objects, there are also considerations of how they can be usefully followed-up.
Synoptic systems such as ZTF and MASTER as well as the space satellite \textit{Swift} have been reporting dozens of new transients, e.g. for S190425z \citep{2019arXiv190712645C,Breeveld2019GCN24296,Kasliwal2019GCN24191,Lipunov2019GCN24167}, and in general, each object requires either photometric or spectroscopic follow-up to classify it.

In addition to the dedicated follow-up systems, perhaps large-aperture systems dedicated to deep photometry and spectroscopic classification (or both), many of the smaller aperture systems, which begin the search for new candidates, may be more useful following up objects found either within or outside of the network. 

\section{Conclusion}

The GRANDMA consortium presented in this paper, which is a global network of 21 telescopes with both photometric and spectroscopic facilities, located in 15 different observatories, aims to face the challenges of time-domain astronomy; in particular, it is set up for the case of large uncertainties in the localisation of the transient phenomena as presented in \cite{Networkpaper}. 
The network morphology has 24-hour coverage and large availability of the telescopes, which helps to rapidly scan the GW sky localisation area to $\sim18$th magnitude. 
The tools have been set up for monitoring the full network include the joint agreement between the teams, the web interface, the uniform distribution of the observation plan, and the 24~hr follow-up advocates on duty to revise observation plans in case of interesting candidates and updated sky localisation areas.

Since the beginning of the third observational campaign organised by aLIGO/Virgo, O3, the GRANDMA programme has demonstrated its capacity to perform follow-up by observing 27/33 GW events.
 This includes scheduling the observations for rapidly scanning the GW localisation area with the help of the amateur astronomer community (as for S190901ap), low-latency detection pipelines for detecting transients, and vetting and characterising potential counterpart candidates.

One area of improvement going forward is to understand optimal coordination of a network of this size. \cite{Networkpaper} present basic strategies for telescope networks, including ways to tile localisations with telescopes of different FoV and depth. Still to be understood is, for example, how an individual telescope might prioritise observations within its assigned fields, especially those overlapping with other systems.
In addition, it will be important to understand how to incorporate the fact that certain systems may not be available for a particular follow-up campaign, such as due to limited target-of-opportunity time, engineering or other downtime, or similar.
There is ongoing development to create updated dynamical observation plans. In addition to joint observations, GRANDMA is also focusing on the development of generic data analysis (to be deployed by January 2020) and on the central multi-messenger plateform ICARE to improve the full process as well as integrating new telescopes into its programme.

\section*{Acknowledgements}
Parts of this research were conducted by the Australian Research Council Centre of Excellence for Gravitational Wave Discovery (OzGrav), through project number CE170100004. EJH acknowledges support from a Australian Research Council DECRA Fellowship (DE170100891).
AdUP and CCT acknowledge support from Ram\'on y Cajal fellowships RyC-2012-09975 and RyC-2012-09984 and the Spanish Ministry of Economy and Competitiveness through project AYA2017-89384-P. DAK acknowledges support from the Spanish research project AYA2017-89384-P. MB acknowledges funding as ``personal tecnico de apoyo'' under fellowship number PTA2016-13192-I. MC is supported by the David and Ellen Lee Postdoctoral Fellowship at the California Institute of Technology. SA is supported by the CNES Postdoctoral Fellowship at Laboratoire AstroParticule et Cosmologie. 
SA, AC, CL and RM acknowledge the financial support of the UnivEarthS Labex program at Sorbonne Paris Cit\'e (ANR-10-LABX-0023 and ANR-11-IDEX-0005-02). 
SA and NL acknowledge the financial support of the Programme National Hautes Energies (PNHE). DT acknowledges the financial support of the Chinese Academy of Sciences (CAS) PIFI post-doctoral fellowship program (program C). UBAI acknowledges support from the Ministry of Innovative Development through projects FA-Atech-2018-392 and VA-FA-F-2-010. 
IRiS has been carried out thanks to the support of the OCEVU Labex (ANR-11-LABX-0060) and the A*MIDEX project (ANR-11-IDEX-0001-02) funded by the ``Investissements d'Avenir'' French government program. IRiS and T120 thank all the Observatoire de Haute-Provence staff for the permanent support. SB, NK, RN and MV acknowledge Shota Rustaveli National Science Foundation (SRNSF grant No 218070). TAROT has been built with the support of the Institut National des Sciences de l'Univers, CNRS,France. TAROT is funded by the CNES and thanks the help of the technical staff of the Observatoire de Haute Provence, OSU-Pytheas.

\bibliographystyle{mnras}
\bibliography{references}

\appendix

\section{Sky Coverage Links}
\label{skycoveragelink}

\begin{table*}
\caption{Sky coverage observations of the GRANDMA consortium during the first six months.}
\begin{tabular}{ccc}
\hline
GW trigger & GRANDMA sky localization coverage Link & GCN report \\
\hline
S190930t & https://grandma-owncloud.lal.in2p3.fr/index.php/s/qEuqLP4MFC9aAst & - \\
S190930s & https://grandma-owncloud.lal.in2p3.fr/index.php/s/L9giV01FNOqOe7P & - \\
S190923y  & https://grandma-owncloud.lal.in2p3.fr/index.php/s/u1xnw9MGczJm3R8 & \cite{TurpinGCN25847} \\
S190915ak & https://grandma-owncloud.lal.in2p3.fr/index.php/s/55vKDfmLJzbHyBn & \cite{KannGCN25781} \\
S190910h & https://grandma-owncloud.lal.in2p3.fr/index.php/s/fJdB1OS0Fj3Pg3t & \citet{gcn25780}  \\
S190910d & https://grandma-owncloud.lal.in2p3.fr/index.php/s/yhviEDd1e633wJH & \citet{gcn25749} \\
S190901ap & https://grandma-owncloud.lal.in2p3.fr/index.php/s/F2xWVjaJdniTkWk & \citet{gcn25666,gcn25688} \\
S190828l & https://grandma-owncloud.lal.in2p3.fr/index.php/s/NmDYNp1AGY5BynT & \citet{gcn25593}\\
S190828j & https://grandma-owncloud.lal.in2p3.fr/index.php/s/tG92ASJWw5C1tWV & \citet{gcn25594}\\
S190814bv & https://grandma-owncloud.lal.in2p3.fr/index.php/s/IKRVU8DGxKtoIPK & \cite{gcn25338}, \citet{gcn25599} \\
S190728q & https://grandma-owncloud.lal.in2p3.fr/index.php/s/g55dL4wKghD3V5k & - \\
S190727h & https://grandma-owncloud.lal.in2p3.fr/index.php/s/ai1VdEgtcW48tbZ & - \\
S190720a & https://grandma-owncloud.lal.in2p3.fr/index.php/s/VTOjRQxVDU7Ze1g & - \\
S190718y  & https://grandma-owncloud.lal.in2p3.fr/index.php/s/MHpMFRhy4TGVahI & - \\
S190707q & https://grandma-owncloud.lal.in2p3.fr/index.php/s/FeZLc99kDMZaMbM & - \\
S190706ai & https://grandma-owncloud.lal.in2p3.fr/index.php/s/5yS9dQQ26FQ8FKz & - \\
S190701ah & https://grandma-owncloud.lal.in2p3.fr/index.php/s/ImQadVEpsVit1h6 & - \\
S190630ag & https://grandma-owncloud.lal.in2p3.fr/index.php/s/pfuQHsUBbJaFtSP & - \\
S190517h & https://grandma-owncloud.lal.in2p3.fr/index.php/s/j8EDTsfDQUDq1D4 & \citet{gcn24631} \\
S190513bm & https://grandma-owncloud.lal.in2p3.fr/index.php/s/tCCiIMIiYpnNYEX & \citet{gcn24549} \\
S190512at & https://grandma-owncloud.lal.in2p3.fr/index.php/s/BSbYZ2O5ZnHwna8 & - \\
S190510g & https://grandma-owncloud.lal.in2p3.fr/index.php/s/ipgBWxLiyL1nDfJ & - \\
S190503bf & https://grandma-owncloud.lal.in2p3.fr/index.php/s/6MdspJUbp0n8uaq & \citet{gcn24431} \\
S190426c & https://grandma-owncloud.lal.in2p3.fr/index.php/s/FeFP4EWVgi2ZE3p & \citet{gcn24327} \\
\multirow{2}{*}{S190425z} & \multirow{2}{*}{https://grandma-owncloud.lal.in2p3.fr/index.php/s/tZTxGSoCEMc4fwG} & \citet{Gendre2019GCN24763,Kann2019GCN24459} \\
 &  & \citet{Howell2019GCN24256,Blazek2019GCN24227} \\
S190421ar & https://grandma-owncloud.lal.in2p3.fr/index.php/s/08hcngYuxwONGWf & \citet{gcn24162} \\
S190412m & https://grandma-owncloud.lal.in2p3.fr/index.php/s/gyf5HfDVD7GS5GD & \citet{gcn24126} \\
\hline
\end{tabular}
\end{table*}

\begin{figure*}
\subfloat[S190930t]{\includegraphics[width = 3.5in]{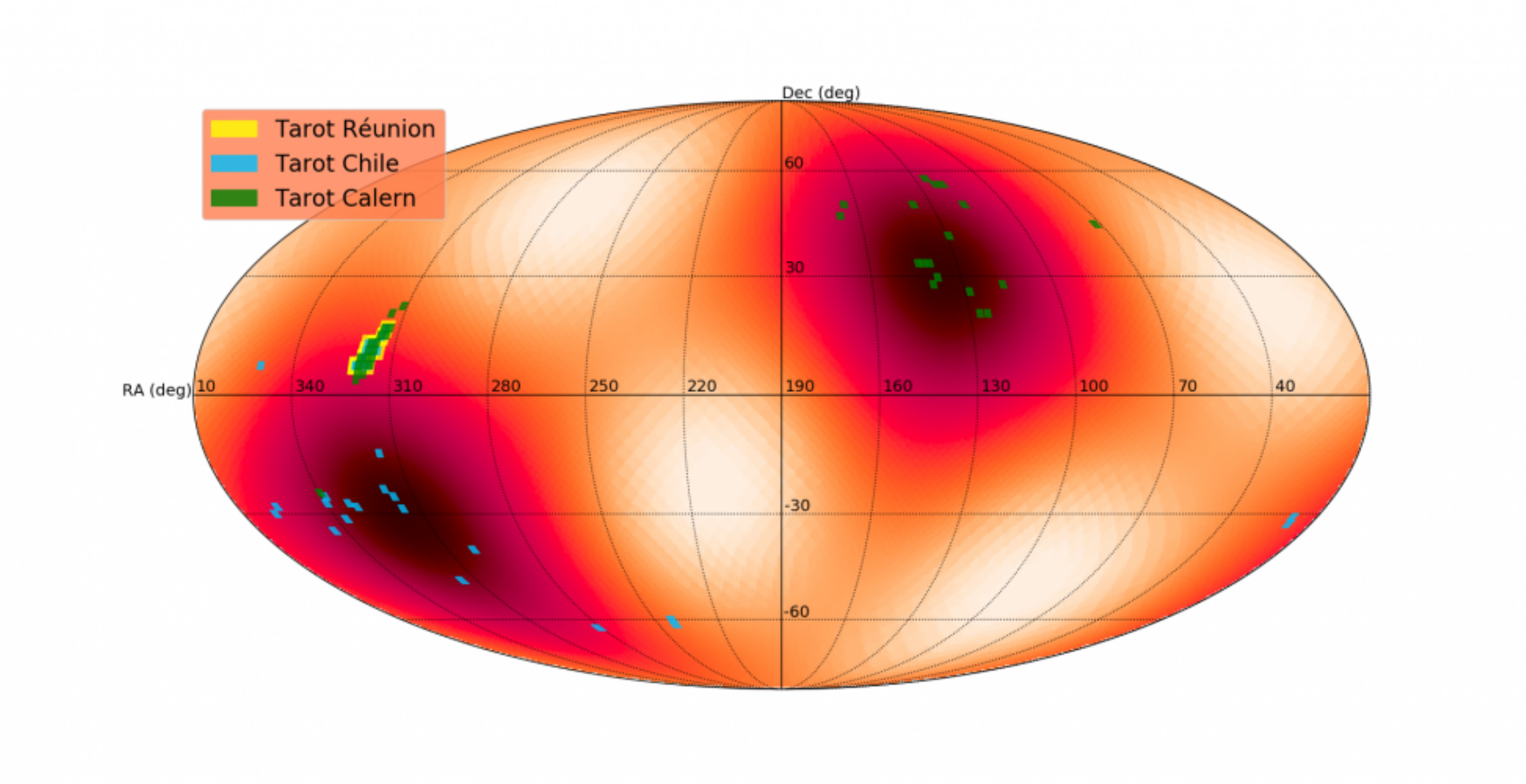}}
\subfloat[S190930s]{\includegraphics[width = 3.5in]{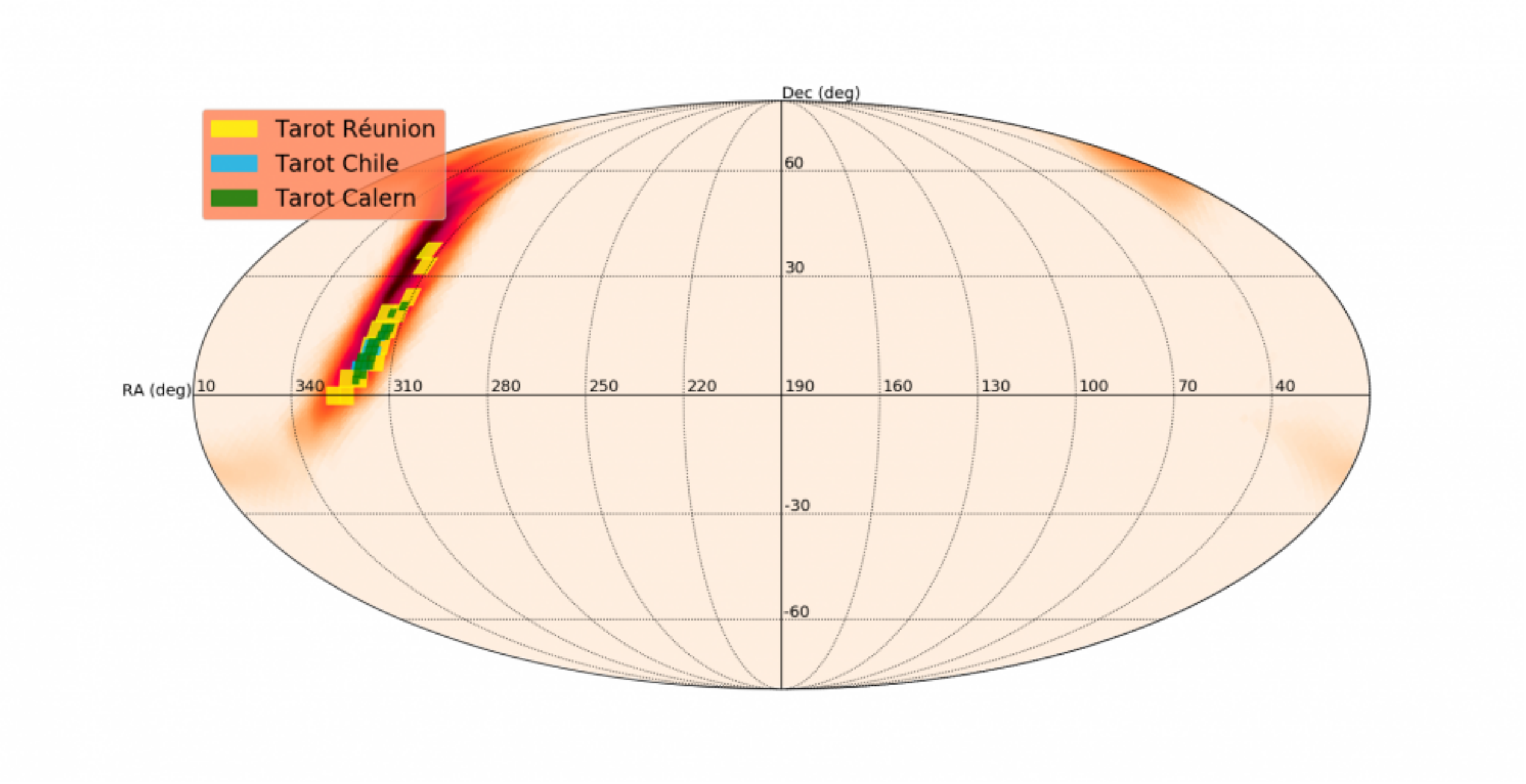}}\\
\subfloat[S190923y]{\includegraphics[width = 3.5in]{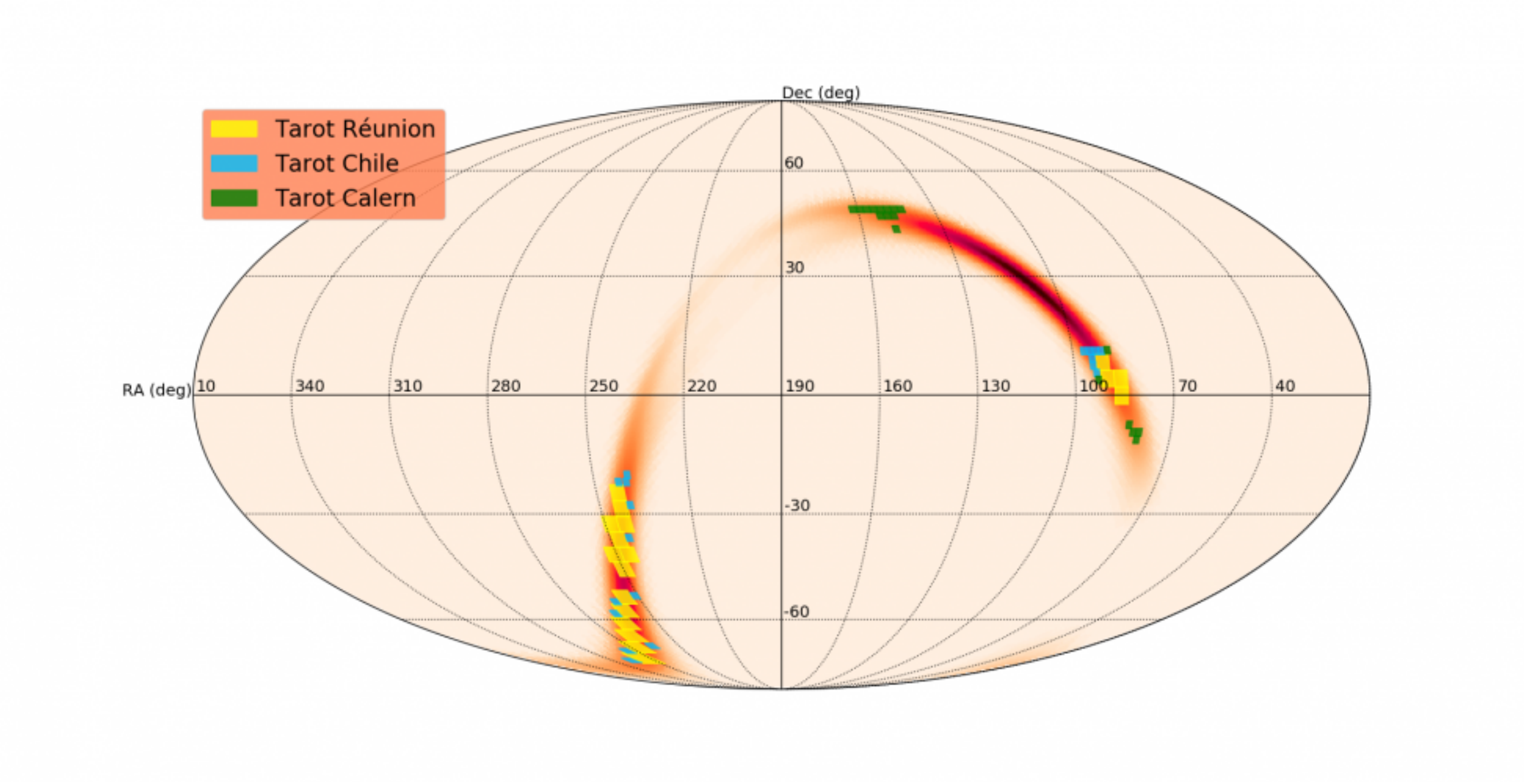}}
\subfloat[S190915ak]{\includegraphics[width = 3.5in]{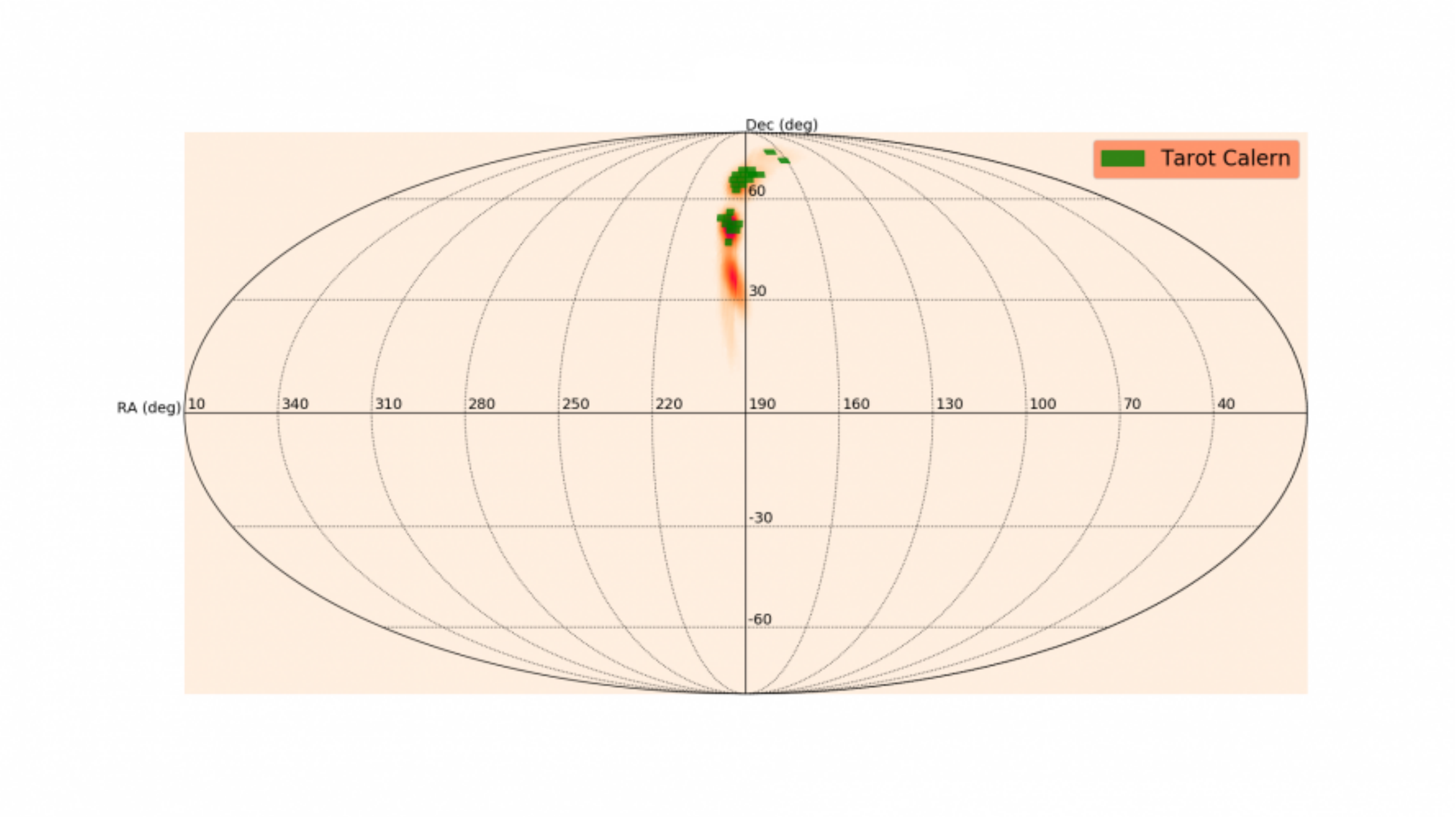}}\\
\subfloat[S190910h]{\includegraphics[width = 3.5in]{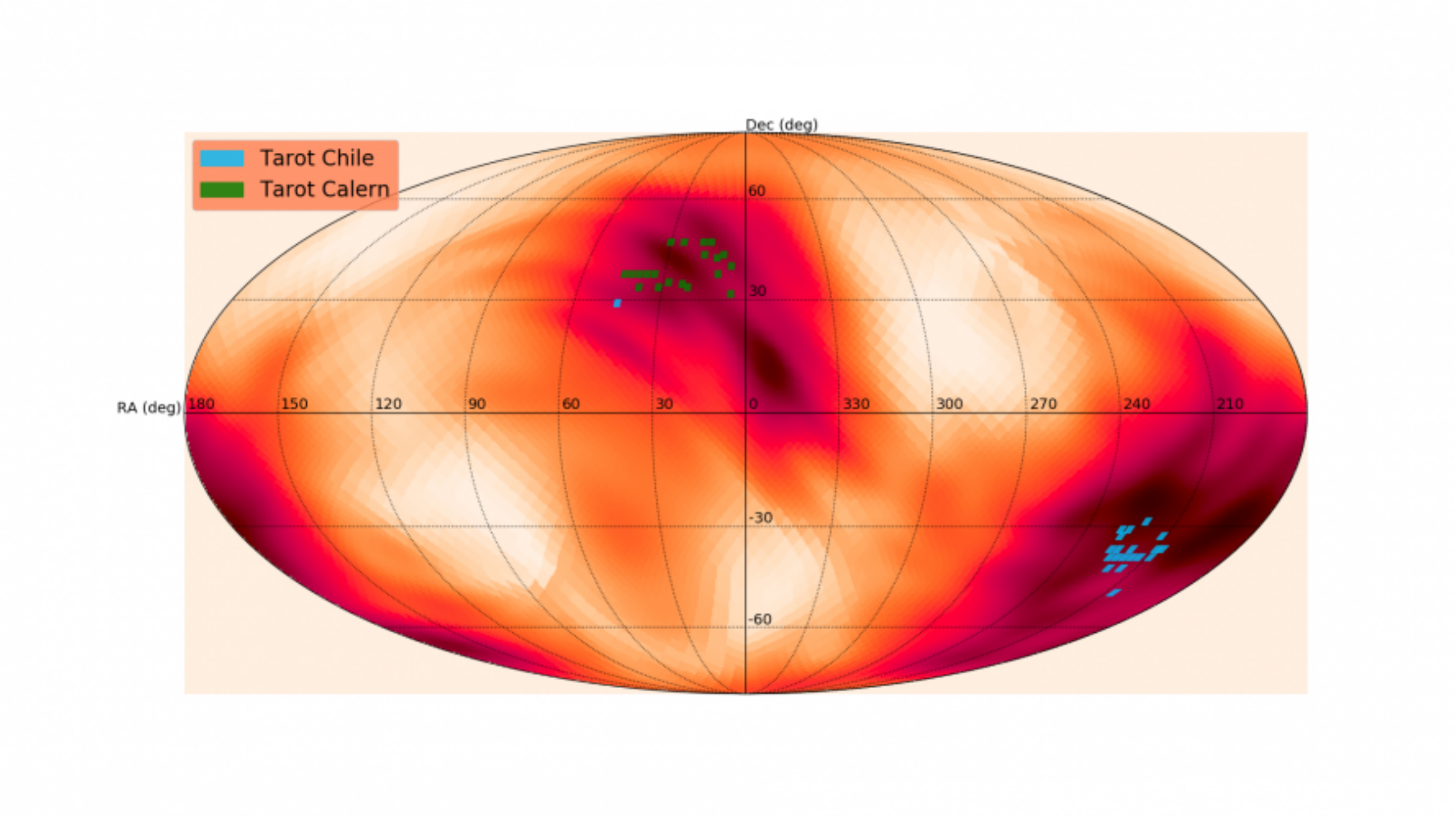}}
\subfloat[S190910d]{\includegraphics[width = 3.5in]{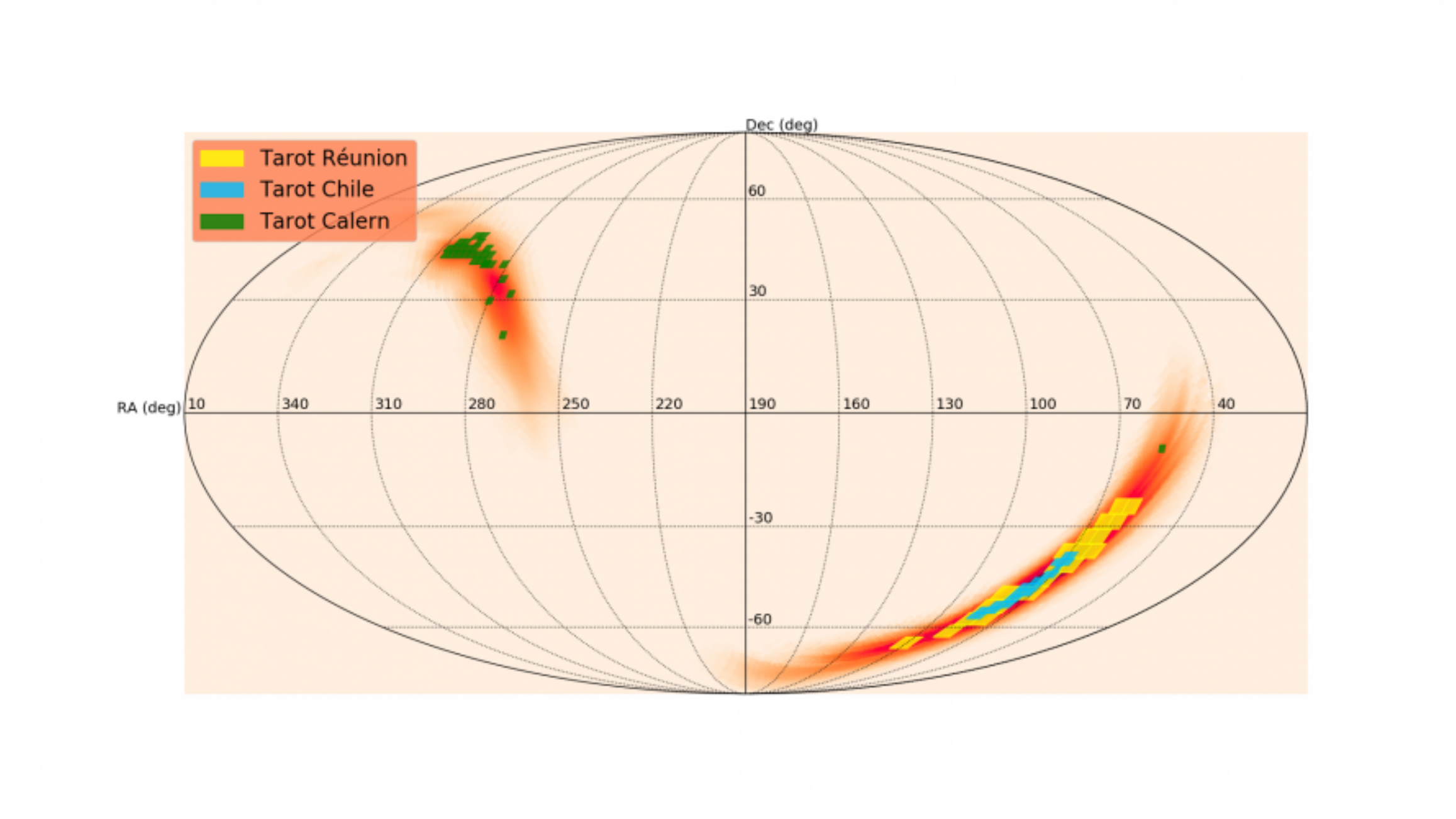}}\\
\subfloat[S190901ap]{\includegraphics[width = 3.5in]{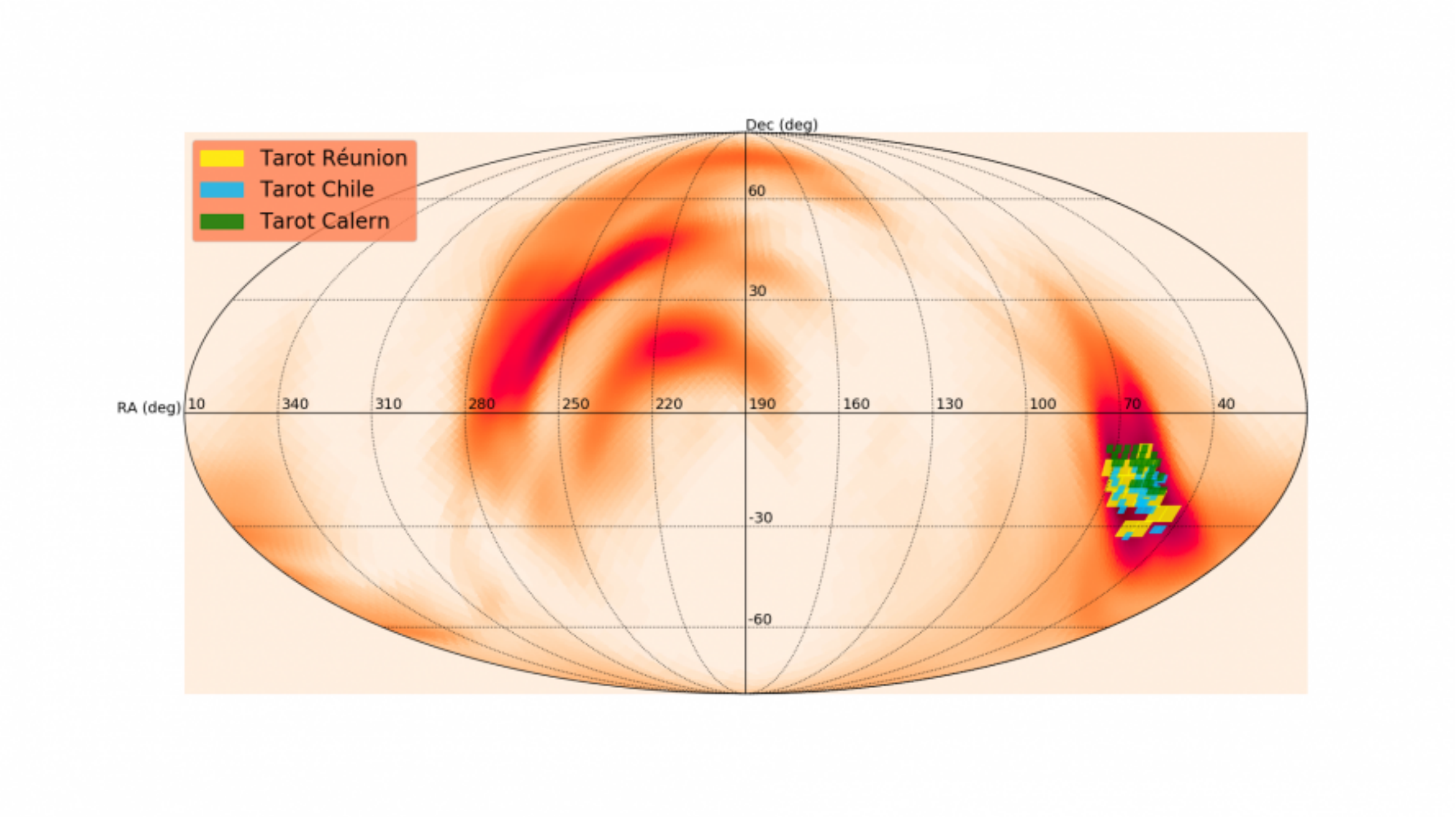}}
\subfloat[S190828l]{\includegraphics[width = 3.5in]{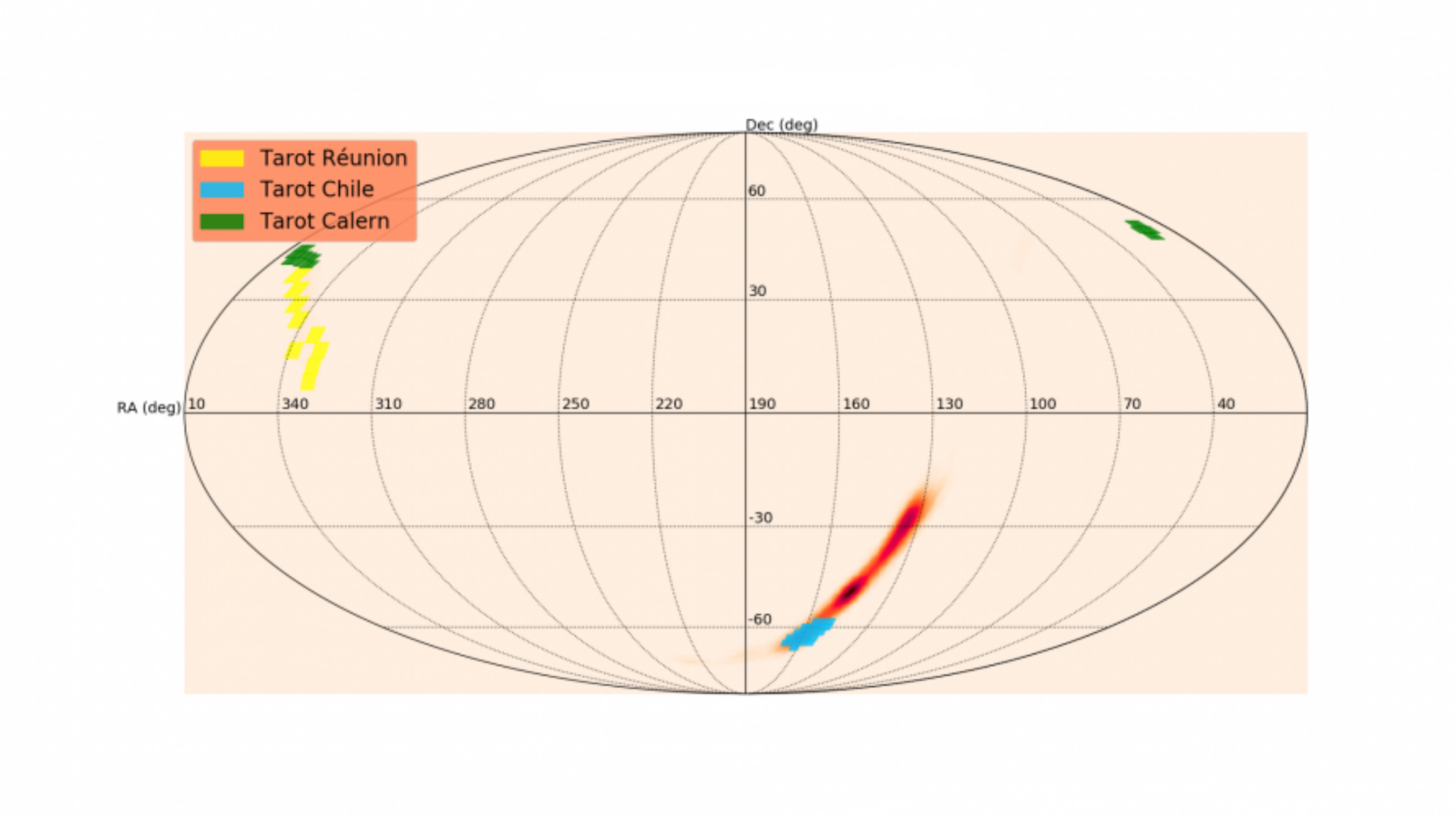}}\\
\caption{Sky coverage observations of the GRANDMA consortium during the first six months.}
\label{followup_plots}
\end{figure*}

\begin{figure*}
\subfloat[S190828j]{\includegraphics[width = 3.5in]{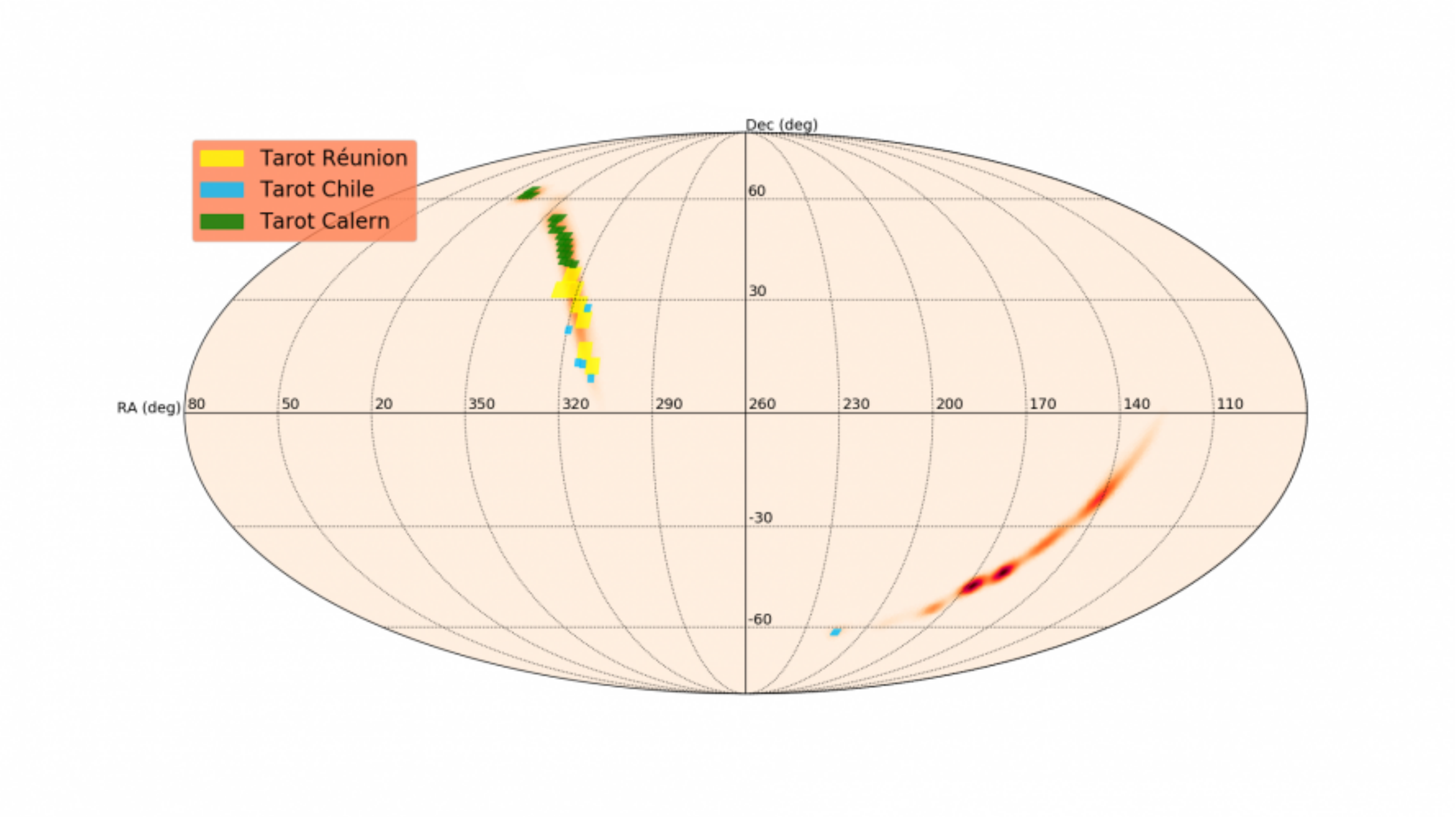}}
\subfloat[S190814bv]{\includegraphics[width = 3.5in]{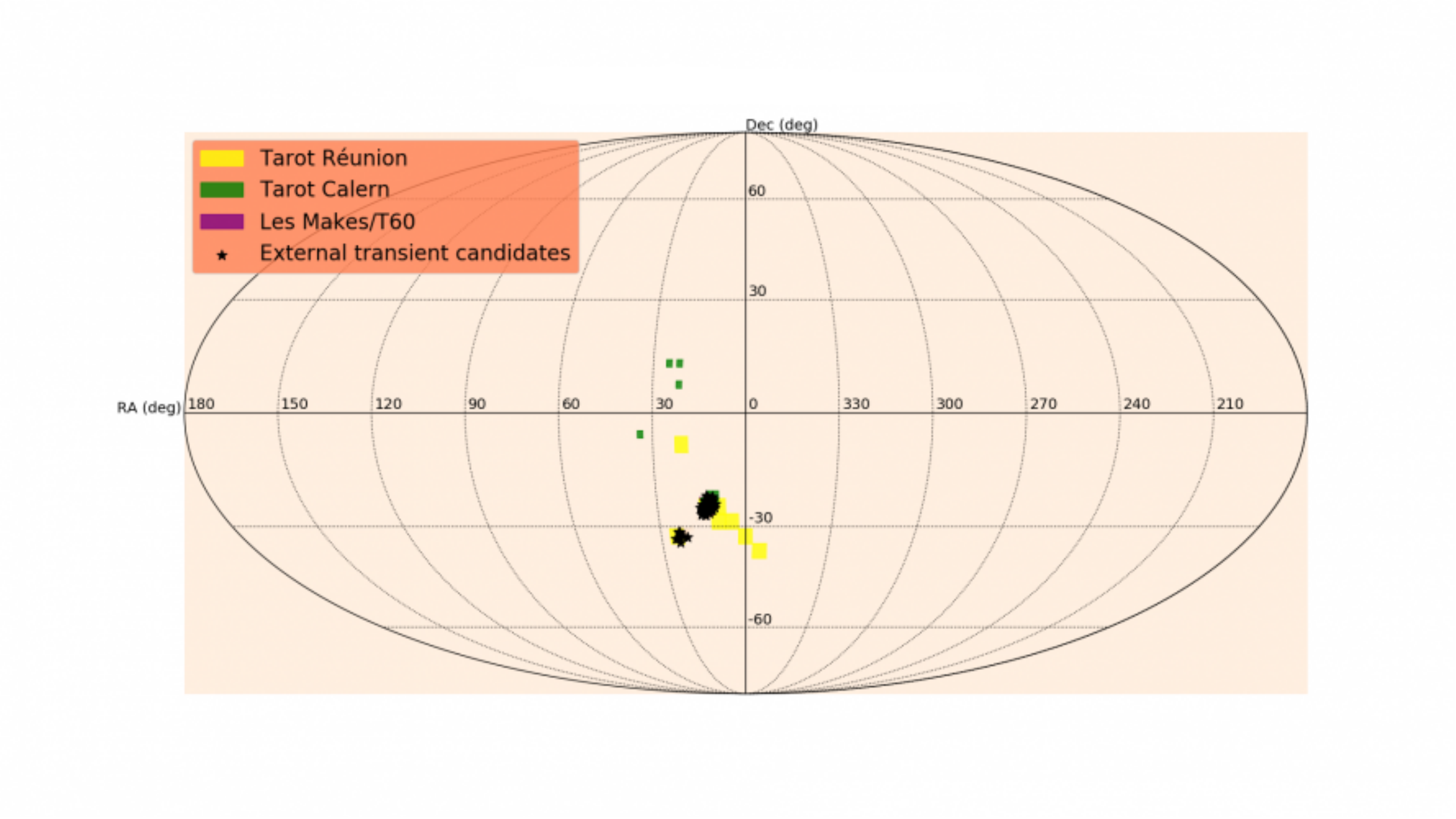}}\\
\subfloat[S190728q]{\includegraphics[width = 3.5in]{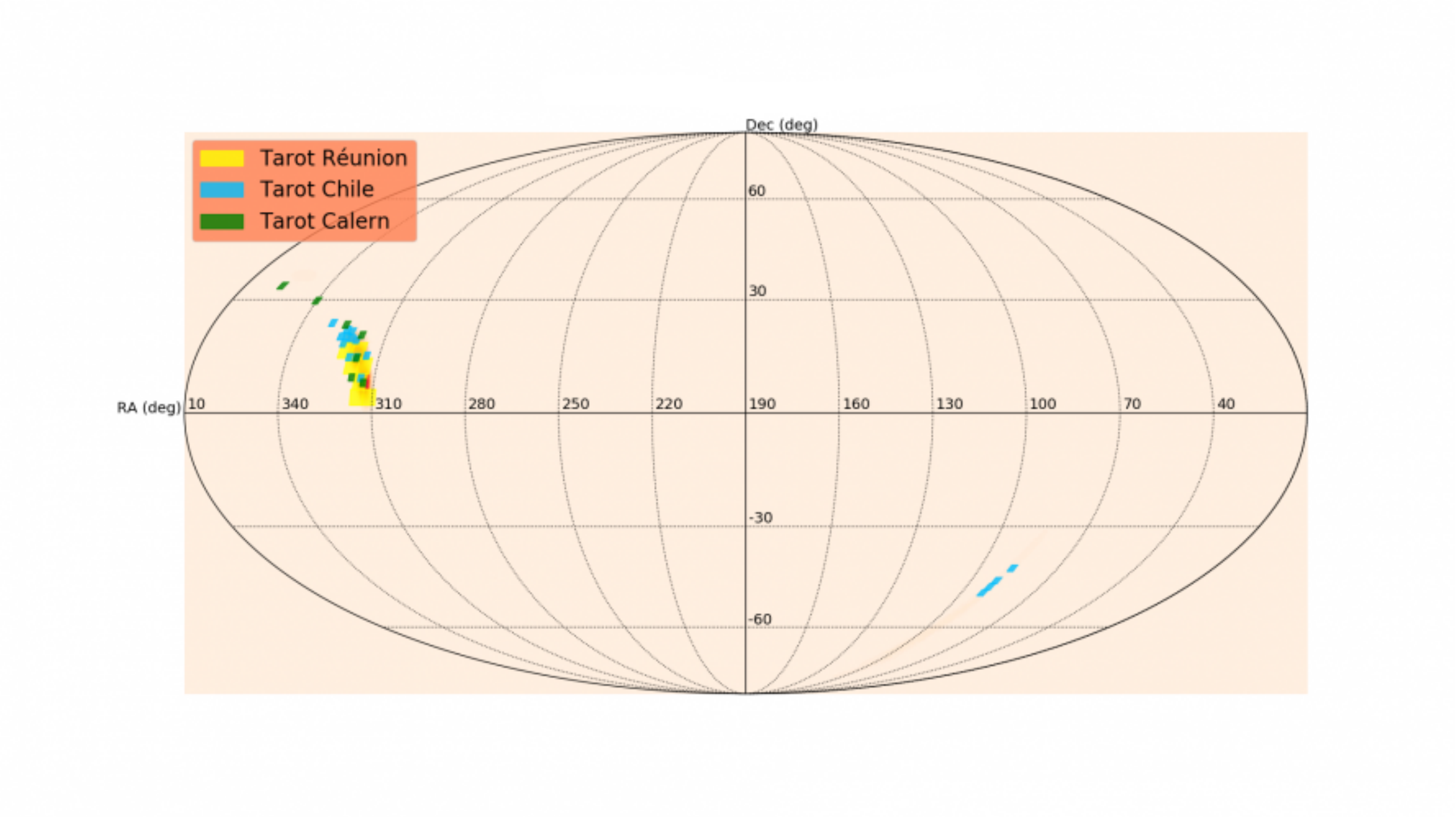}}
\subfloat[S190727h]{\includegraphics[width = 3.5in]{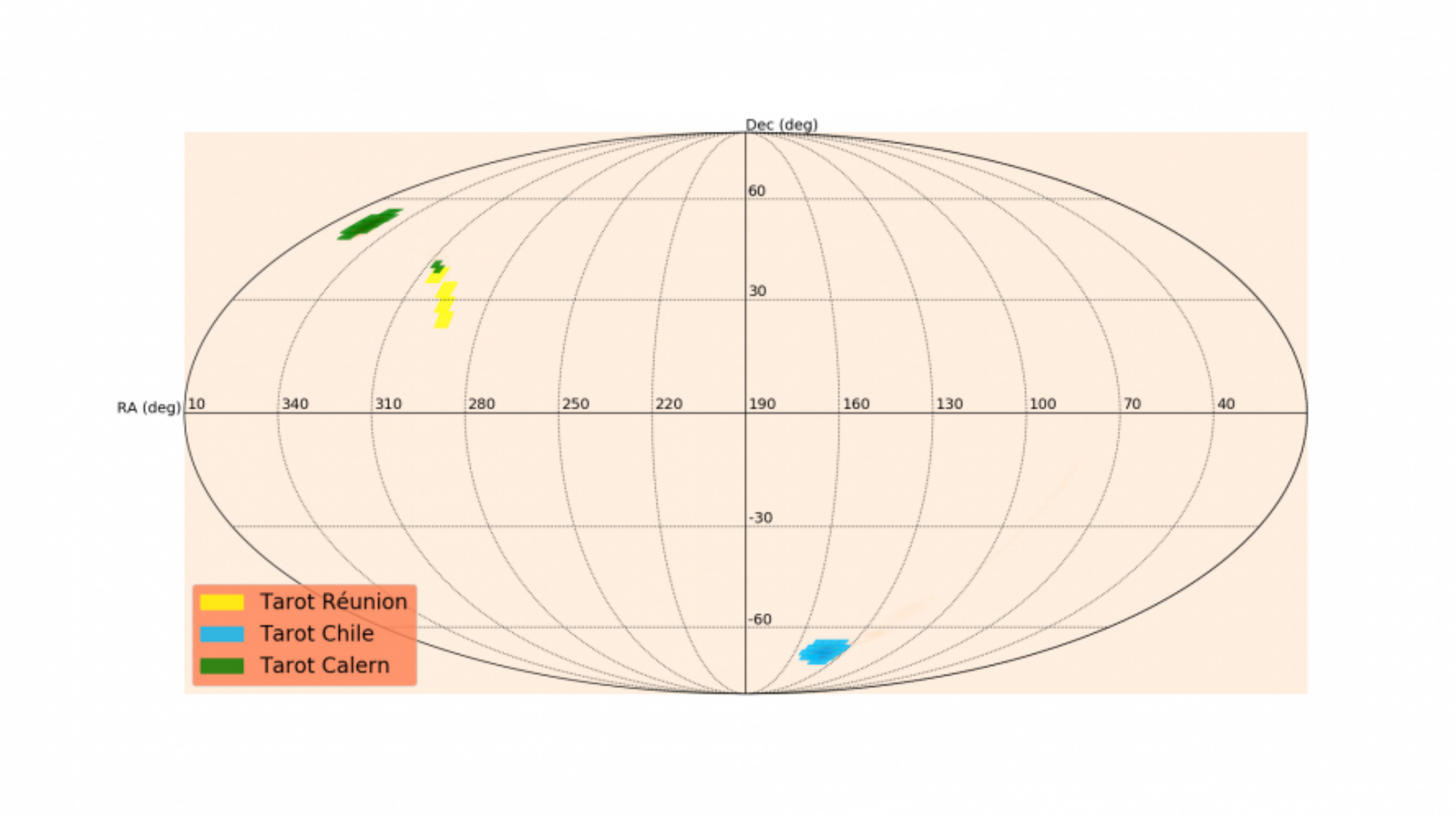}}\\
\subfloat[S190720a]{\includegraphics[width = 3.5in]{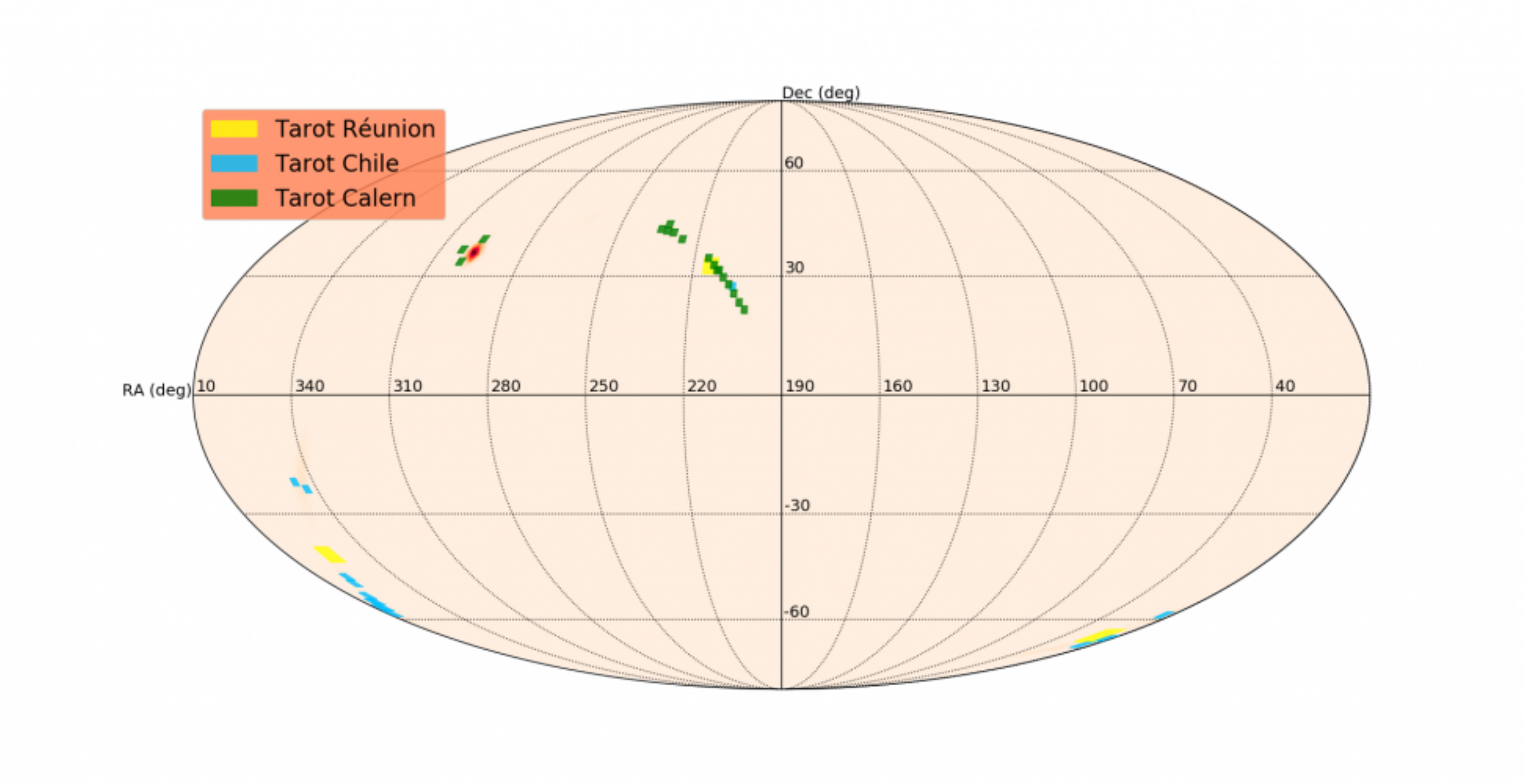}}
\subfloat[S190718y]{\includegraphics[width = 3.5in]{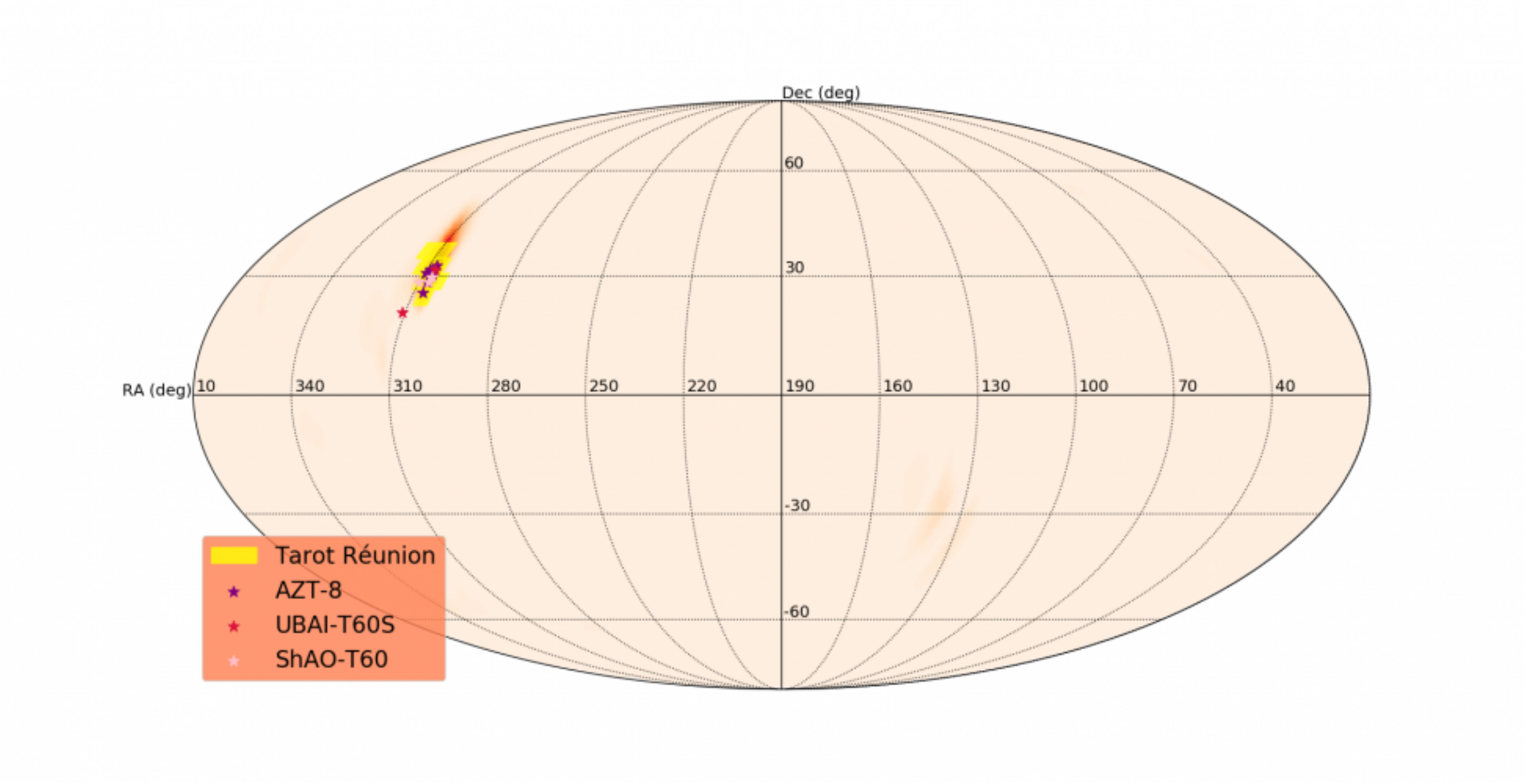}}\\
\subfloat[S190707q]{\includegraphics[width = 3.5in]{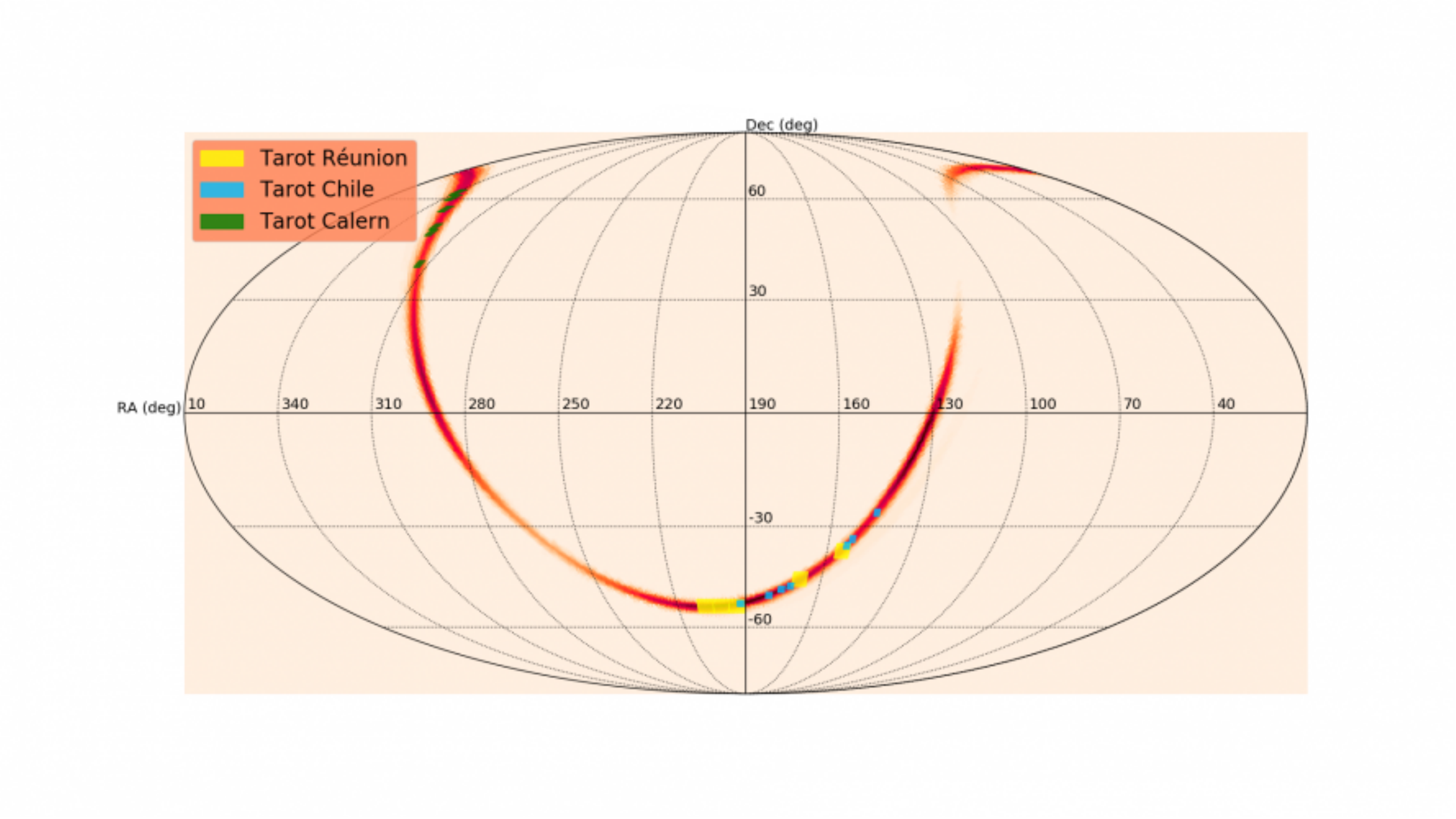}}
\subfloat[S190706ai]{\includegraphics[width = 3.5in]{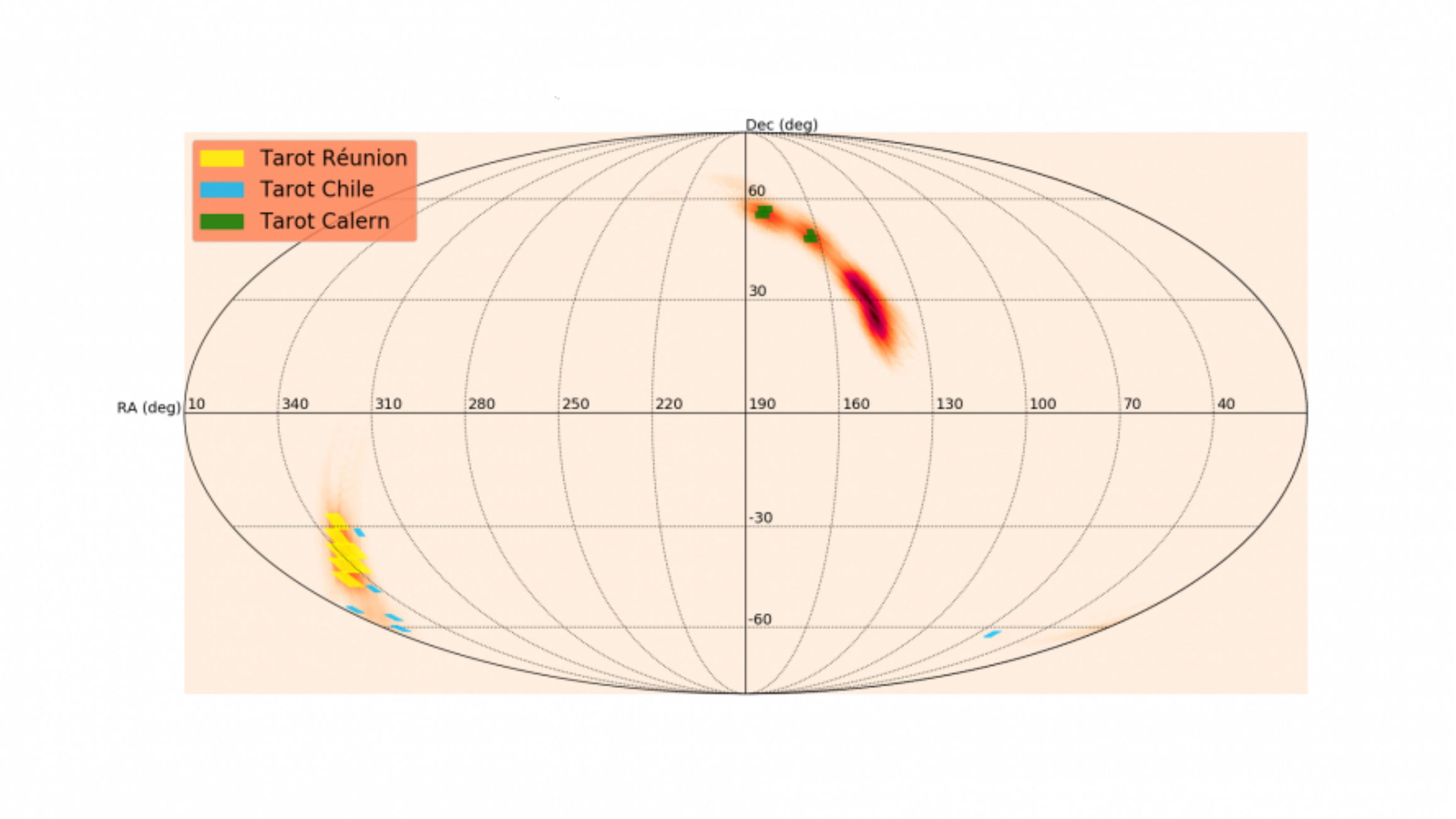}}\\
\contcaption{Sky coverage observations of the GRANDMA consortium during the first six months.}
\end{figure*}

\begin{figure*}
\subfloat[S190701ah]{\includegraphics[width = 3.5in]{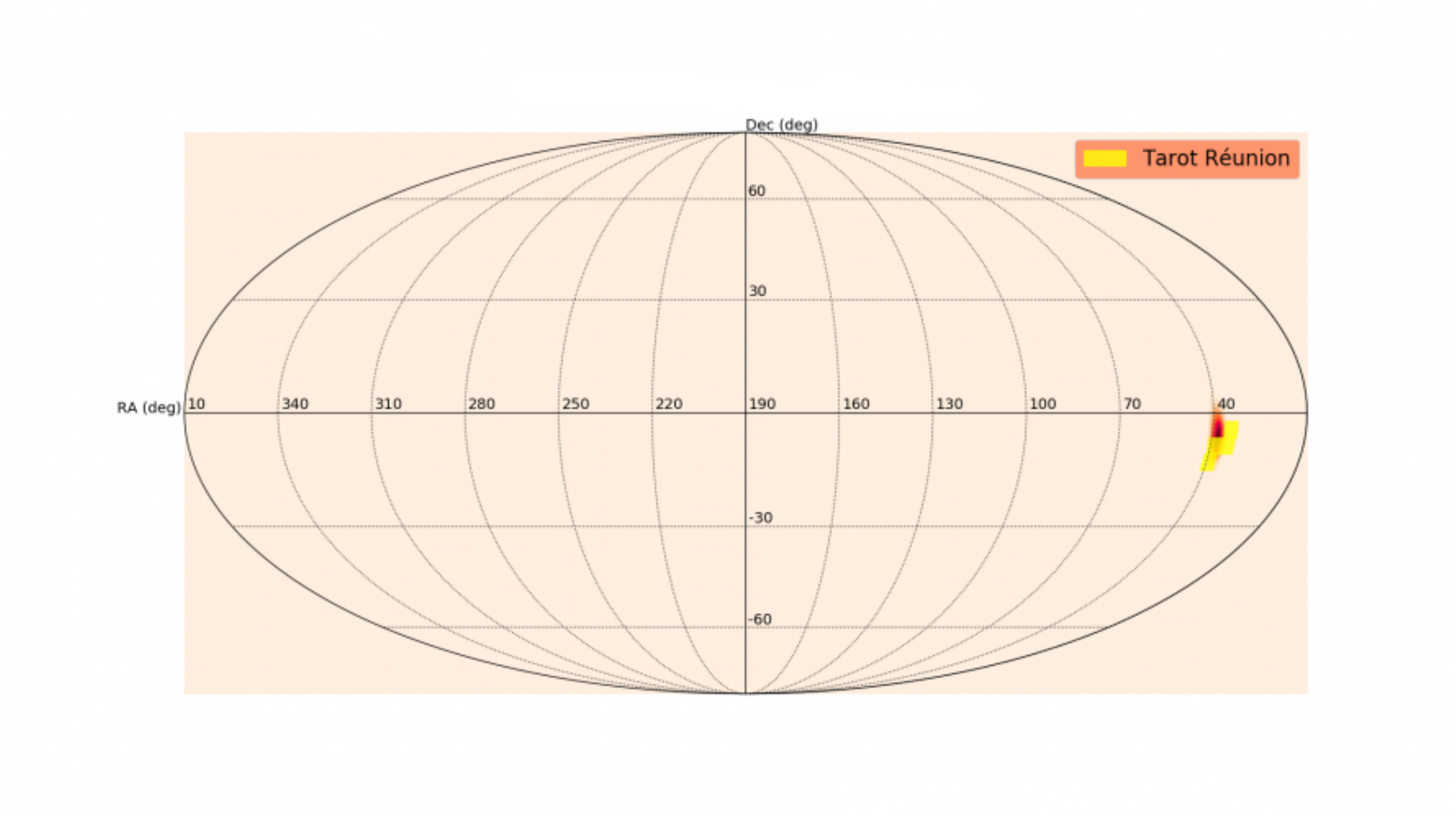}}
\subfloat[S190630ag]{\includegraphics[width = 3.5in]{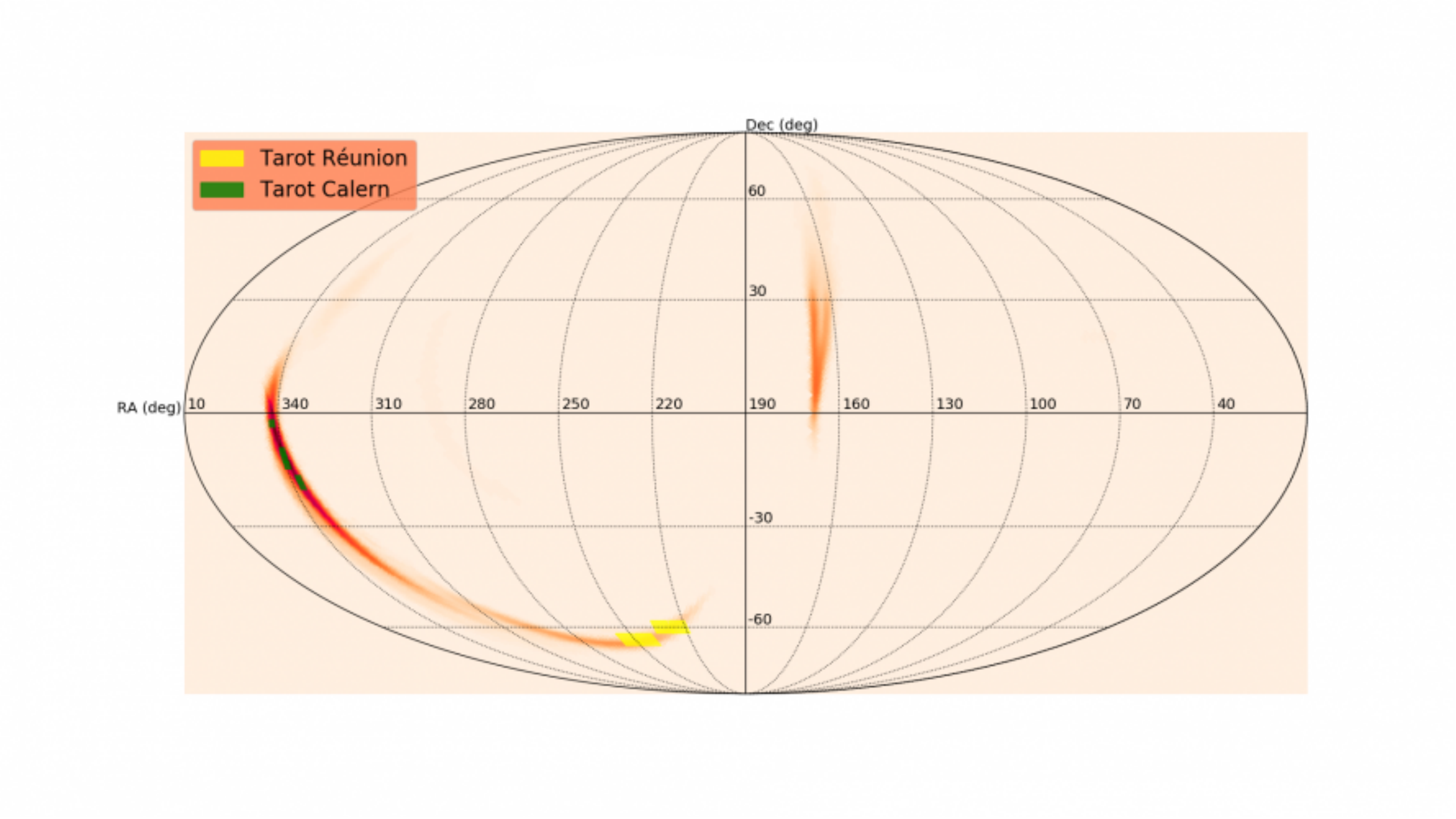}}\\
\subfloat[S190517h]{\includegraphics[width = 3.5in]{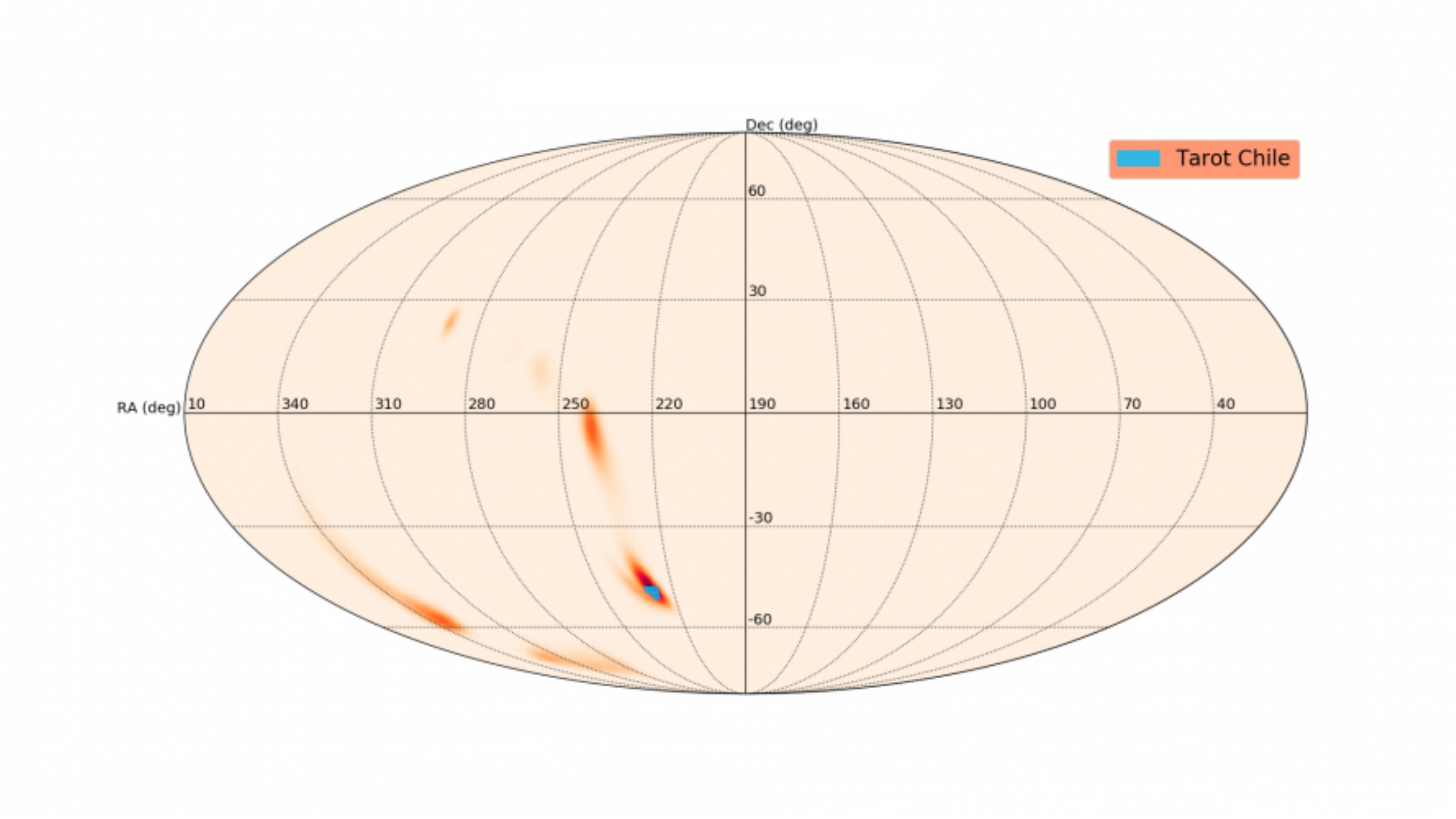}}
\subfloat[S190513bm]{\includegraphics[width = 3.5in]{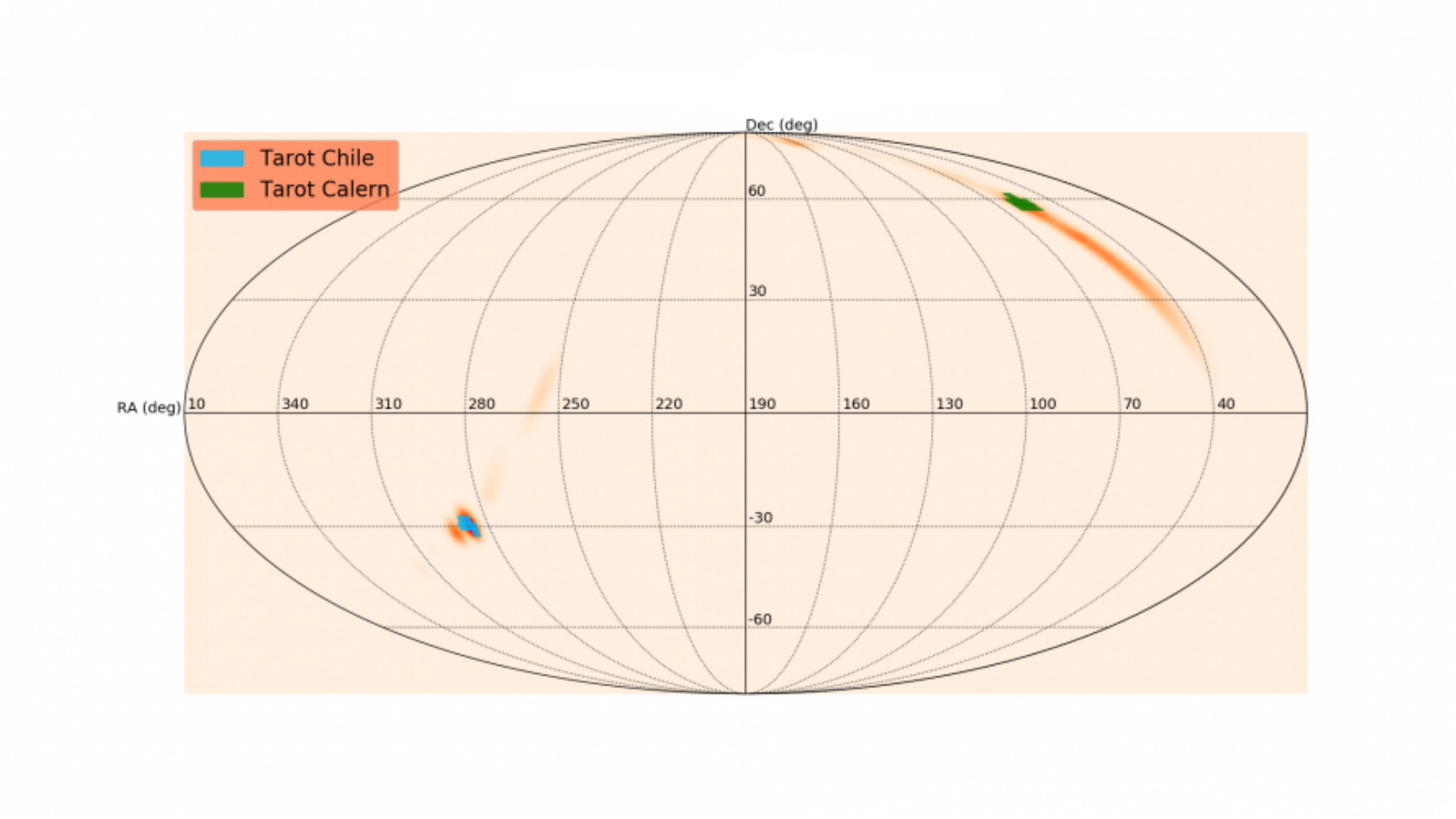}}\\
\subfloat[S190512at]{\includegraphics[width = 3.5in]{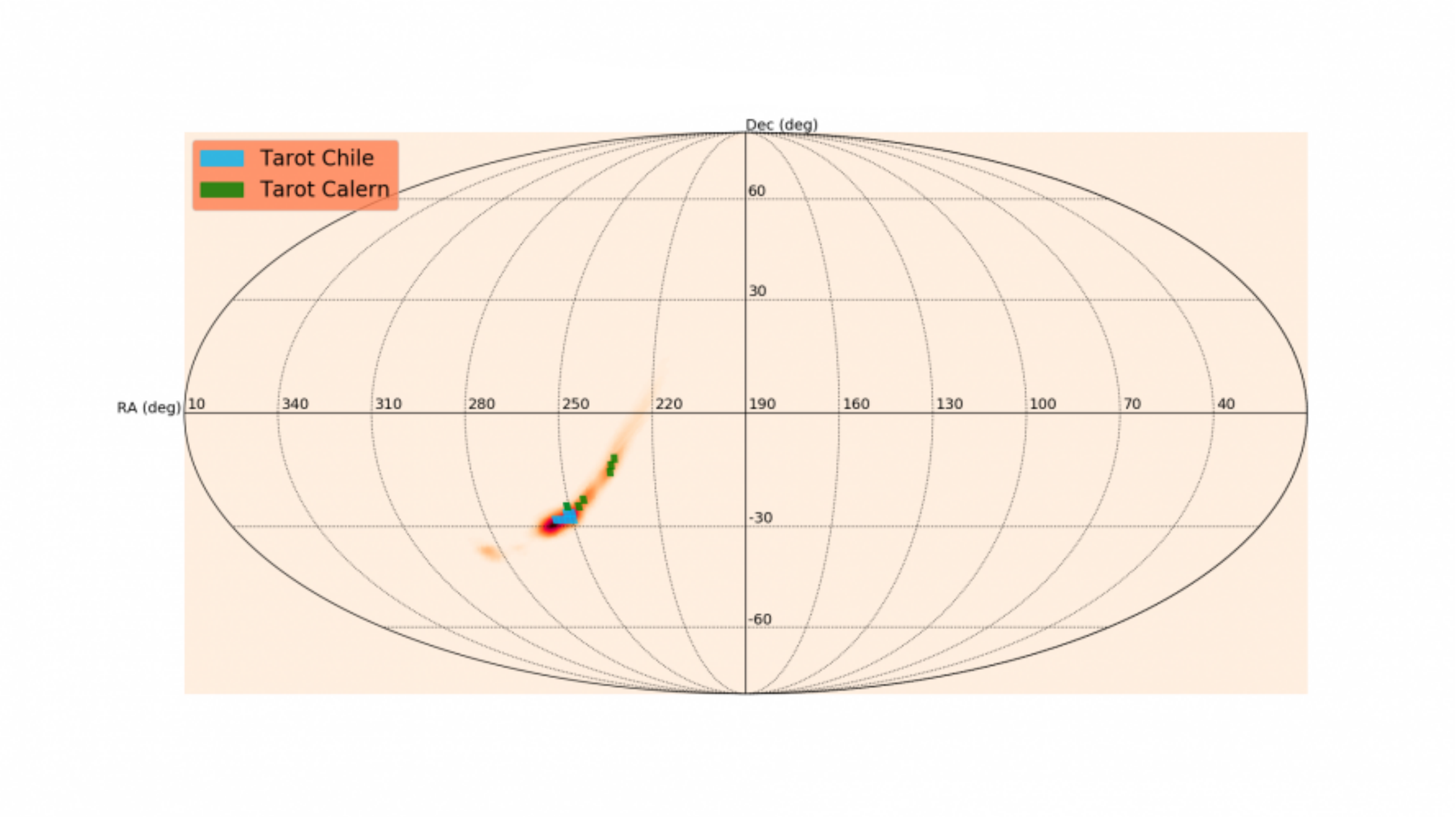}}
\subfloat[S190510g]{\includegraphics[width = 3.5in]{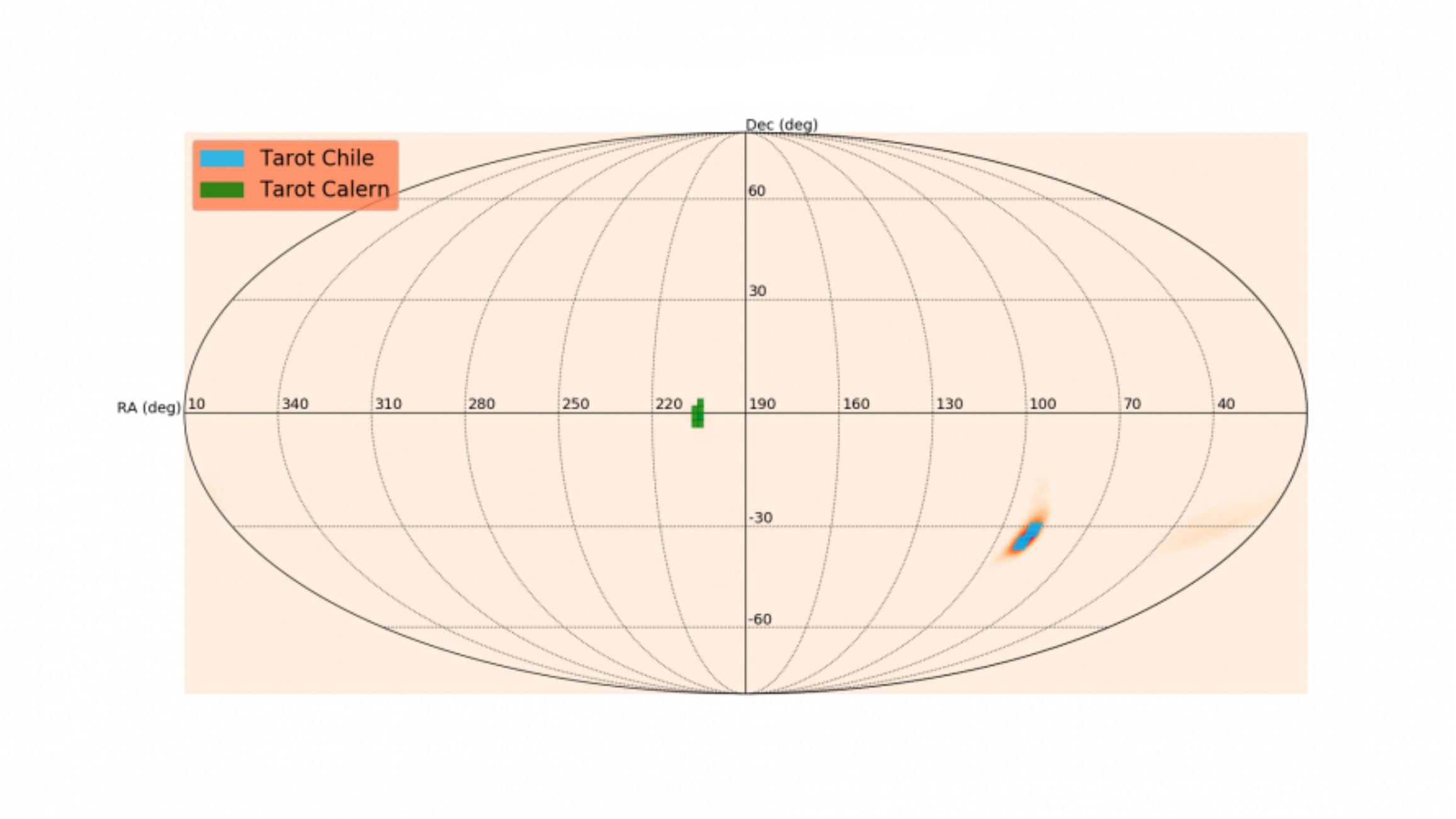}}\\
\subfloat[S190503bf]{\includegraphics[width = 3.5in]{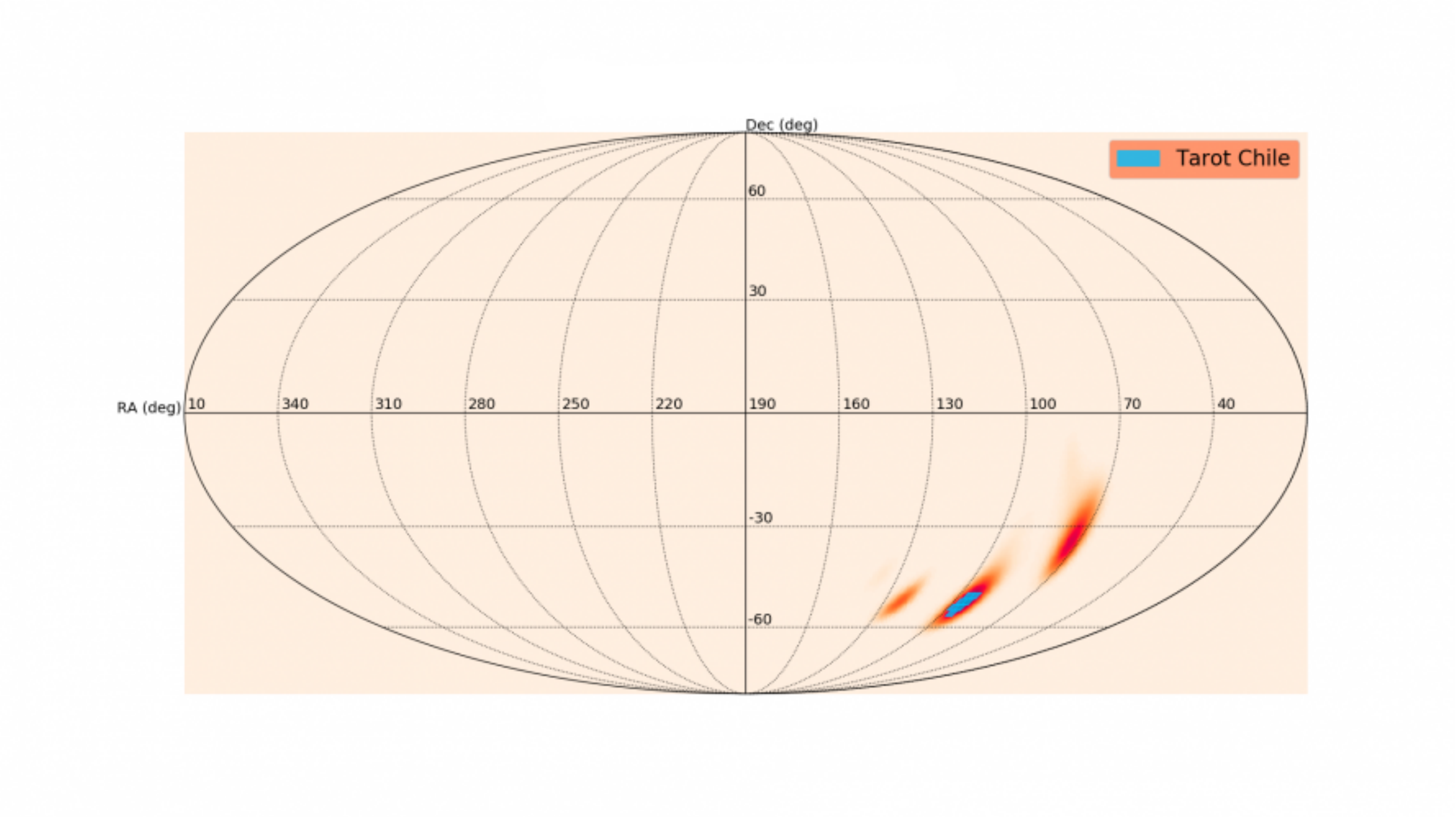}}
\subfloat[S190426c]{\includegraphics[width = 3.5in]{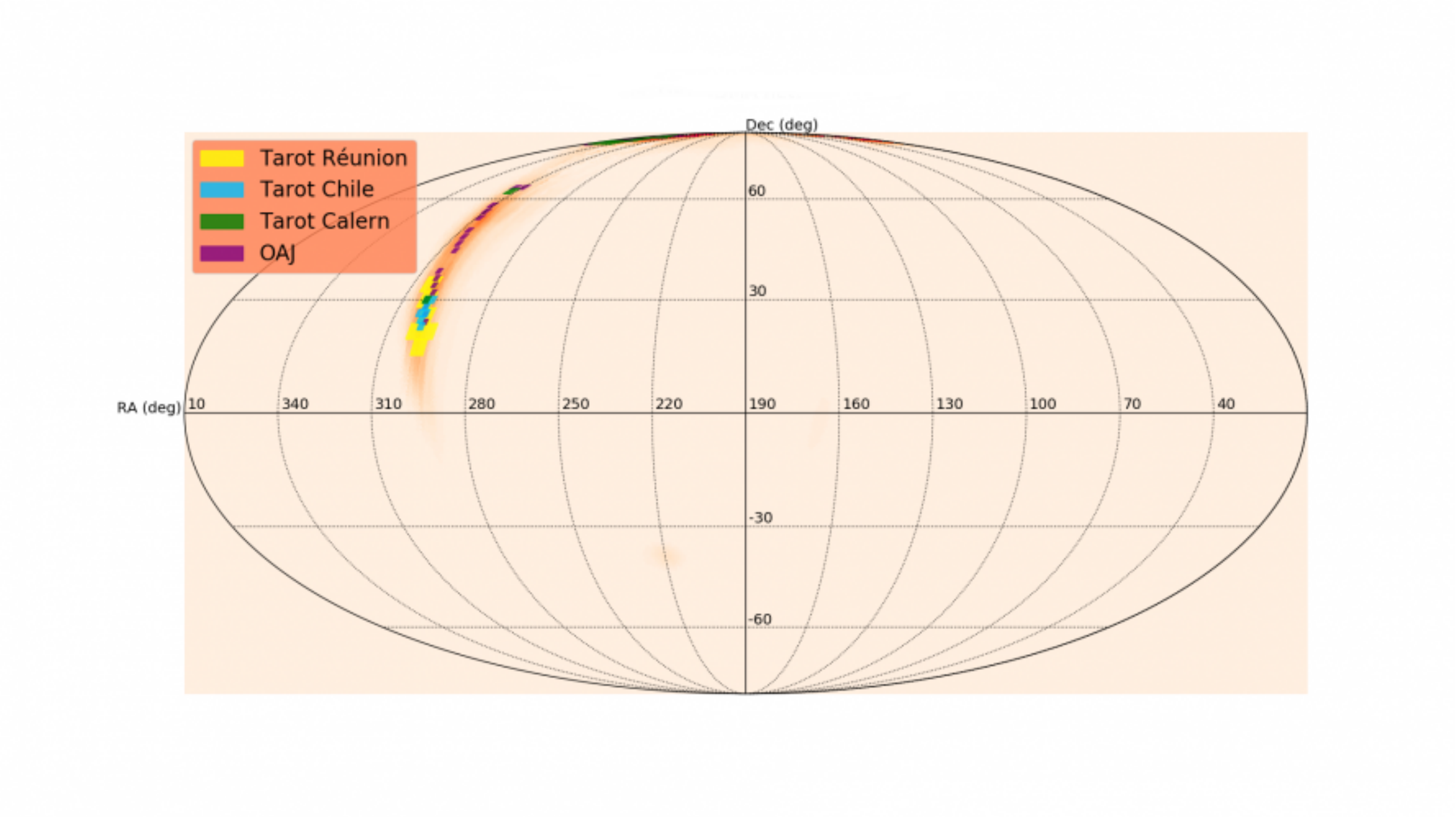}}\\
\contcaption{Sky coverage observations of the GRANDMA consortium during the first six months.}
\end{figure*}

\begin{figure*}
\subfloat[S190425z]{\includegraphics[width = 2.8in]{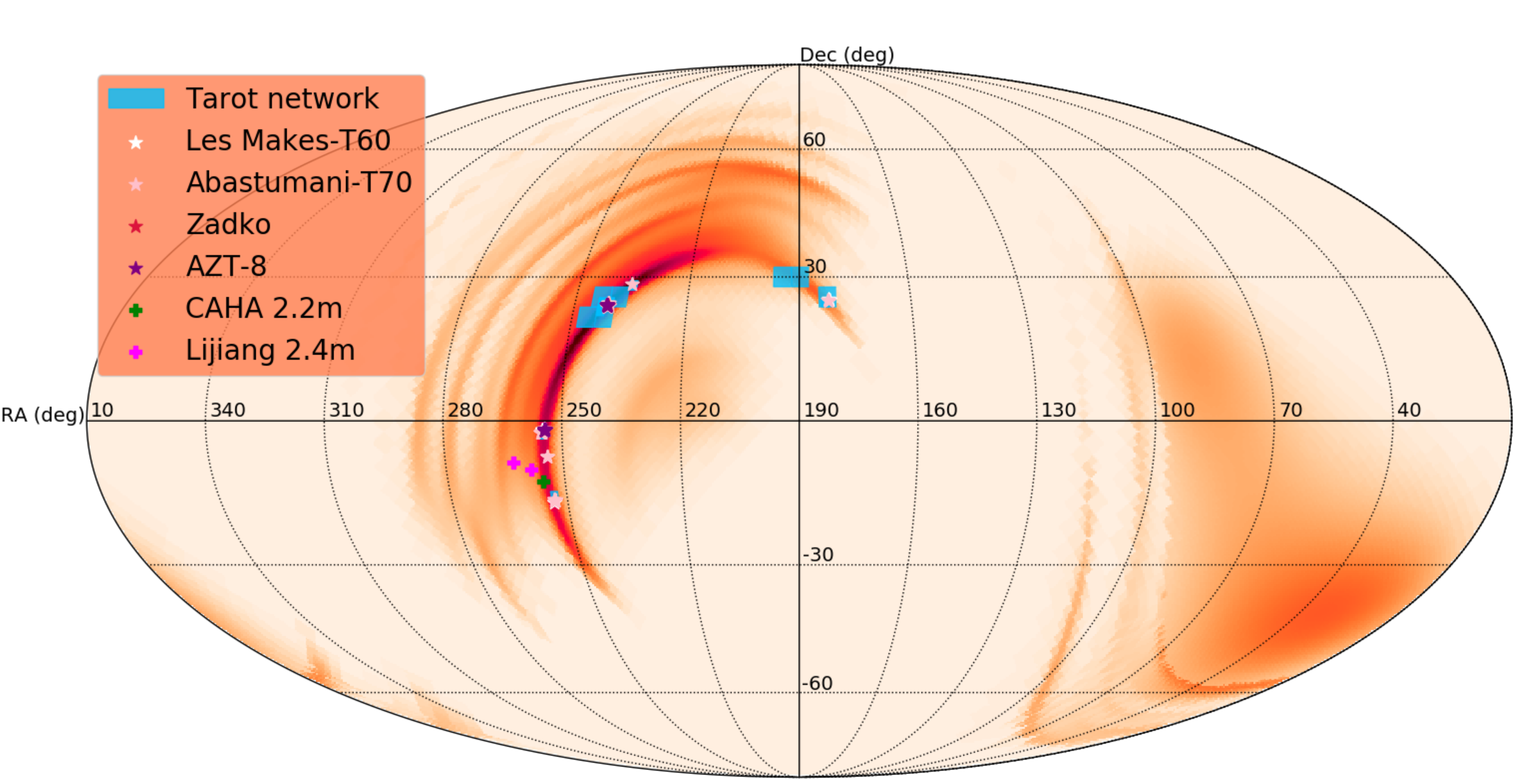}}
\subfloat[S190421ar]{\includegraphics[width = 3.5in]{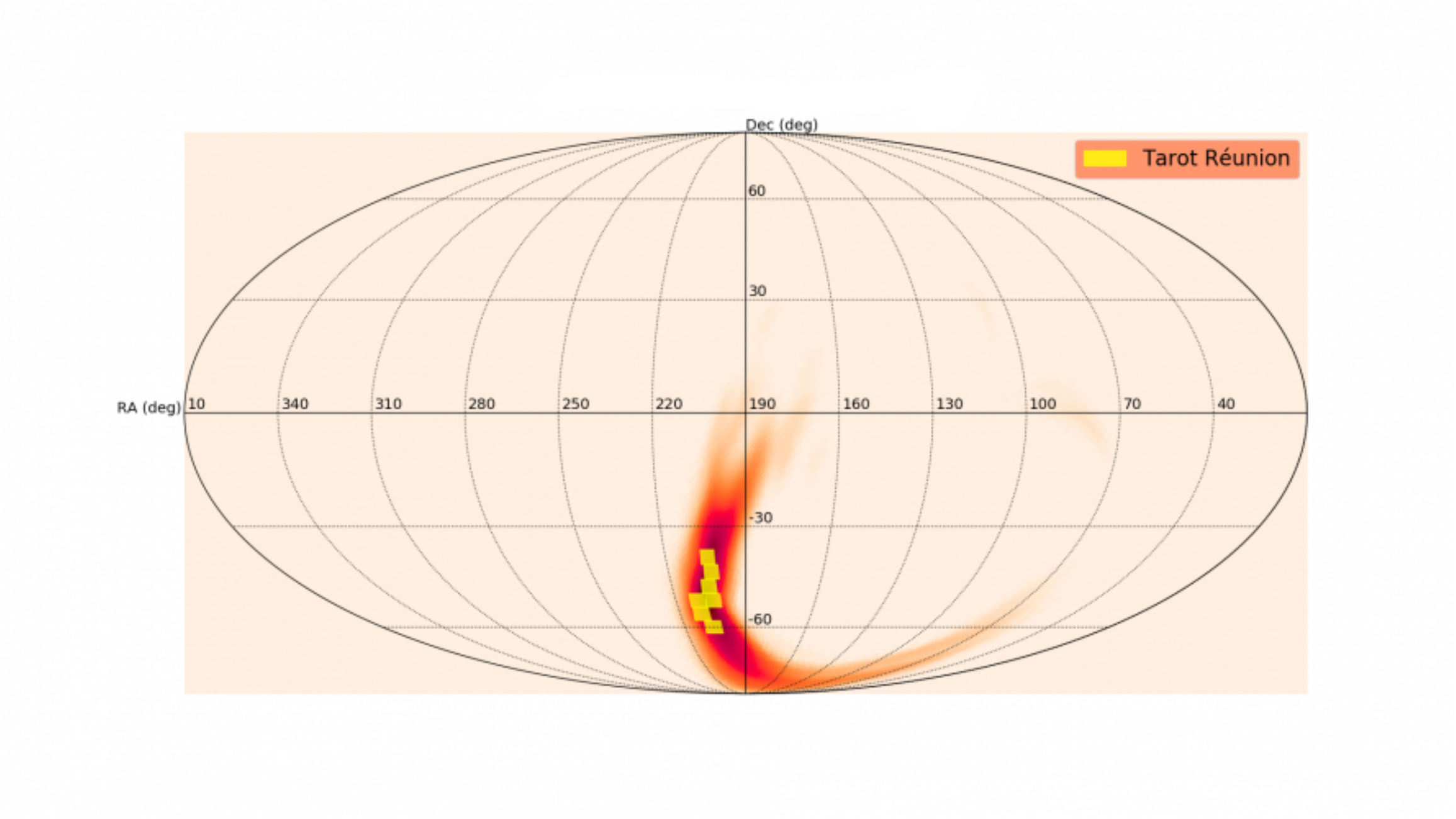}}\\
\subfloat[S190412m]{\includegraphics[width = 3.5in]{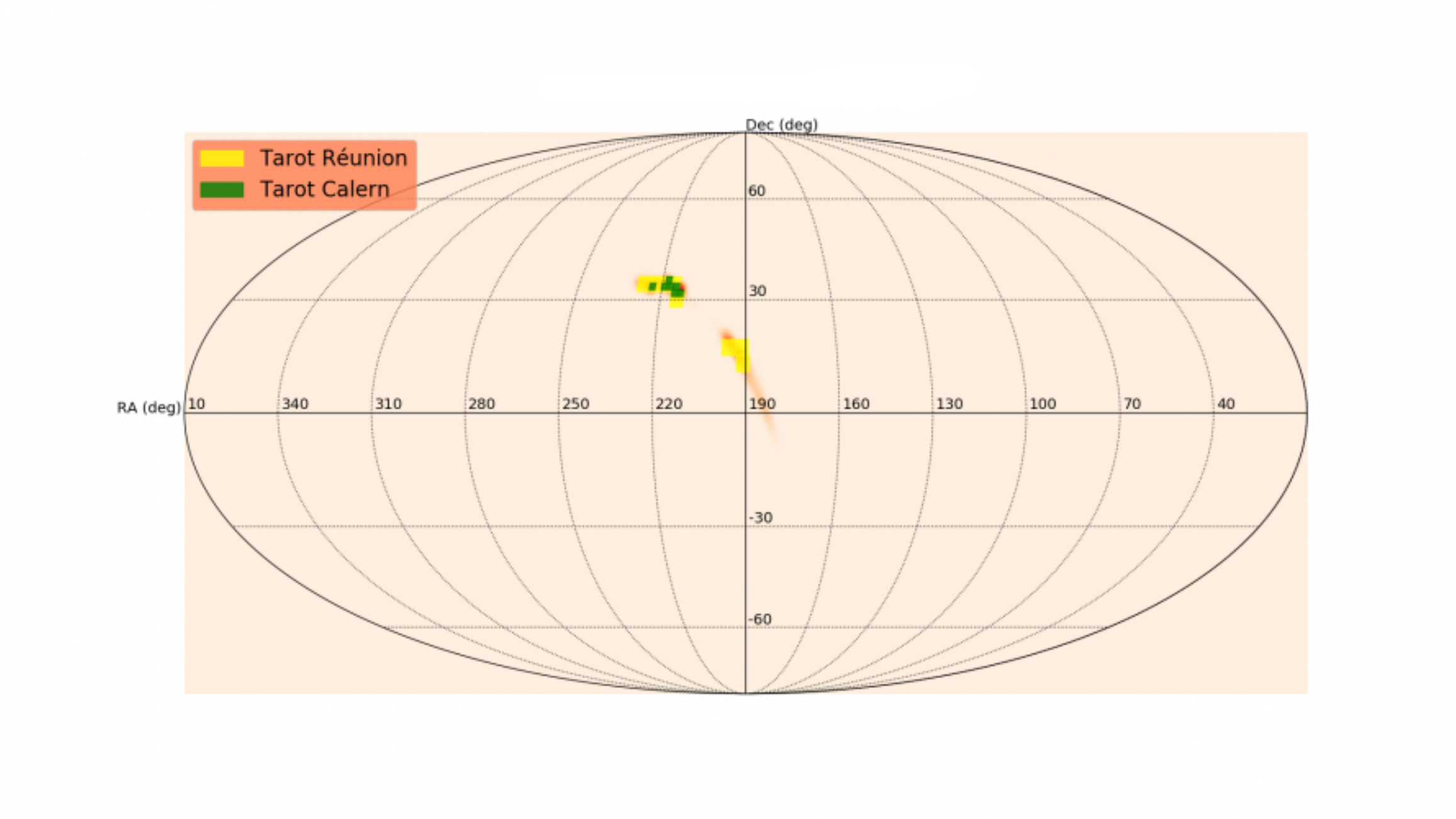}}
\contcaption{Sky coverage observations of the GRANDMA consortium during the first six months.}
\end{figure*}

\section{Full observation logs for S190425z}

\begin{table*}
\caption{List of tiled observations performed by the TAROT network for 190425z. The probability refers to the 2D spatial probability of the LALInference GW skymap
enclosed in a given tile.}
\label{tab:TAROT190425z}
\begin{tabular}{ccccc}
\hline
\multicolumn{2}{c}{} & \multicolumn{2}{c}{}  & \\
\multicolumn{2}{c}{Observation period \tiny{(UTC)}} & \multicolumn{2}{c}{Tile coor. \tiny{(center)}}  & \scriptsize{Coverage}\\
\smallskip
 T$_{\rm{start}}$ & T$_{\rm{end}}$ & RA [$^\circ$] & Dec [$^\circ$] & prob (\%)\\
\hline
\multicolumn{5}{c}{TAROT-TRE}\\
\hline
2019-04-25 14:56:19  & 2019-04-25 22:29:56 & 194.595 & 30.000      &     0.06  \\
2019-04-25 17:03:57 & 2019-04-25 17:55:18 & 182.338 & 25.714  &     0.04 \\
2019-04-25 17:16:58 & 2019-04-25 21:51:14  & 189.73  & 30.000      &     0.07 \\
2019-04-25 18:07:39 & 2019-04-26 01:40:56  & 243.117 & 25.714  &     0.4 \\
2019-04-25 18:20:49 & 2019-04-26 01:09:32 & 246.076 & 21.429  &     0.5 \\
2019-04-25 19:11:40 & 2019-04-25  22:16:40 & 238.442 & 25.714  &     0.8 \\
2019-04-25 19:43:51 & 2019-04-26  00:18:12 & 241.519 & 21.429  &     0.6 \\
\hline
\multicolumn{5}{c}{TAROT-TCA}\\
\hline
2019-04-26 22:25:38 & 2019-04-28 03:32:13  & 255.075 & -2.727  &     0.1 \\
2019-04-26 22:32:16 & 2019-04-28 03:39:05 & 242.609 & 22.727  &     0.2 \\
2019-04-26 23:18:50 & 2019-04-28 03:25:22 & 240.663 & 24.545  &     0.2 \\
2019-04-27 19:51:10 & 2019-04-27 20:58:08 & 242.652 & 24.545  &     0.2 \\
2019-04-27 20:44:47 & 2019-04-27 20:51:17  & 235.227 & 28.182  &     0.2 \\
\hline
\multicolumn{5}{c}{TAROT-TCH}\\
\hline
2019-04-26 06:40:51 & 2019-04-28 01:44:47 & 182.983 & 24.545  &     0.01 \\
2019-04-26 06:45:56 & 2019-04-28 04:51:16 & 235.227 & 28.182  &     0.2 \\
2019-04-26 06:52:44 & 2019-04-28 04:55:49 & 242.652 & 24.545  &     0.2 \\
2019-04-26 06:59:33 & 2019-04-28 03:04:58 & 253.125 & -15.455 &     0.1 \\
2019-04-26 07:18:42 & 2019-04-27 09:38:57 & 242.609 & 22.727  &     0.2 \\
2019-04-26 07:57:48 & 2019-04-28 05:29:50 & 255.075 & -2.727  &     0.1 \\
2019-04-27 03:18:52 & 2019-04-28 06:23:00 & 240.663 & 24.545  &     0.2 \\
\hline
\end{tabular}
\end{table*}

\begin{table*}
\caption{List of galaxies, enclosed in the 3D probability region of S190425z, observed by the T60 telescope at les Makes Observatory, the T70 telescope at Abastumani Observatory, the AZT-8 telescope at Lisnyky Observatory, and the Zadko telescope.}
\label{tab:galaxytarget_S190425z}
\begin{tabular}{cclccc}
\hline
 \multicolumn{2}{c}{} & & \multicolumn{2}{c}{} & \\
\multicolumn{2}{c}{Observation period (UTC)} & $~~~~~~~~$Galaxy name & \multicolumn{2}{c}{Galaxy coordinates}  & Distance\\
 T$_{\rm{start}}$ &T$_{\rm{end}}$ & & RA [$^\circ$] & Dec [$^\circ$] & [Mpc]\\
\hline
\hline
\\
\multicolumn{6}{c}{Les Makes/T60}\\
\\
\hline
2019-04-25 15:18:10 & 2019-04-25 15:33:35 & GWGC [MFB2005]         &  182.081 &  25.233 & 105.2\\
 &   & J120819.37+251357.9         &    &    &  \\
2019-04-25 15:34:40 & 2019-04-25 15:50:05 & SDSSJ120754.97+250102.6   &  181.979 &  25.017 & 106.3\\
2019-04-25 15:34:40 & 2019-04-25 15:50:05 & SDSSJ120754.70+250604.8   &  181.978 &  25.101 & 111 \\
2019-04-25 15:34:40 & 2019-04-25 15:50:05 & HyperLEDA 38516        &  182.038 &  25.079 & 111.7\\
2019-04-25 21:58:35 & 2019-04-26 00:50:20 & HyperLEDA 1109773      &  254.435 &  -1.799 & 136.9\\
2019-04-25 21:58:35 & 2019-04-26 00:50:20 & 2MASS 16574427-0147567 &  254.434 &  -1.799 & 136.4\\
2019-04-25 21:58:35 & 2019-04-25 22:14:00 & 2MASS 16573561-0147457 &  254.398 &  -1.796 & 142.6\\
2019-04-25 21:58:35 & 2019-04-25 22:14:00 & HyperLEDA 1109835      &  254.399 &  -1.796 & 145.1\\
2019-04-25 21:58:35 & 2019-04-25 22:14:00 & 2MASS 16572382-0147498 &  254.349 &  -1.797 & 147.9 \\
2019-04-25 21:58:35 & 2019-04-25 22:14:00 & HyperLEDA 1110982      &  254.423 &  -1.754 & 148.8 \\
2019-04-25 21:58:35 & 2019-04-25 22:14:00 & HyperLEDA 1109807      &  254.349 &  -1.797 & 148.8 \\
2019-04-25 21:58:35 & 2019-04-25 22:14:00 & 2MASS 16574141-0145157 &  254.423 &  -1.754 & 168.5 \\
2019-04-25 22:14:10 & 2019-04-25 22:29:35 & HyperLEDA 1106076      &  254.456 &  -1.934 & 136.3\\
2019-04-25 22:14:10 & 2019-04-25 22:29:35 & 2MASS 16574955-0156027 &  254.456 &  -1.934 & 135.6\\
2019-04-25 22:29:45 & 2019-04-25 22:45:10 & 2MASS 16573281-0138287 &  254.387 &  -1.641 & 157.3\\
2019-04-25 22:45:21 & 2019-04-25 23:00:46 & 2MASS 16570522-0151022 &  254.272 &  -1.851 & 147.2\\
2019-04-25 22:45:21 & 2019-04-25 23:00:46 & HyperLEDA 1107232      &  254.346 &  -1.89  & 148.8\\
2019-04-25 22:45:21 & 2019-04-25 23:00:46 & HyperLEDA 1109171      &  254.296 &  -1.82  & 148.8\\
2019-04-25 22:45:21 & 2019-04-25 23:00:46 & 2MASS 16572303-0153258 &  254.346 &  -1.891 & 155.8\\
2019-04-25 22:45:21 & 2019-04-25 23:00:46 & 2MASS 16571094-0149128 &  254.296 &  -1.82  & 169.9\\
2019-04-25 23:00:56 & 2019-04-26 00:50:20 & 2MASS 16580368-0150074 &  254.515 &  -1.835 & 147.5\\
2019-04-25 23:00:56 & 2019-04-25 23:16:21 & HyperLEDA 1108643      &  254.598 &  -1.839 & 136.9\\
2019-04-25 23:00:56 & 2019-04-26 00:50:20 & 2MASS 16580515-0149544 &  254.521 &  -1.832 & 162.3\\
2019-04-25 23:16:31 & 2019-04-25 23:31:56 & 2MASS 16585823-0202277 &  254.743 &  -2.041 & 132.1\\
2019-04-25 23:16:31 & 2019-04-25 23:31:56 & HyperLEDA 161011       &  254.743 &  -2.041 & 145 \\
2019-04-25 23:16:31 & 2019-04-25 23:31:56 & HyperLEDA 161014       &  254.822 &  -2.074 & 120.5\\
2019-04-25 23:16:31 & 2019-04-25 23:31:56 & 2MASS 16591729-0204281 &  254.822 &  -2.074 & 107.3\\
2019-04-25 23:32:06 & 2019-04-25 23:47:31 & HyperLEDA 161013       &  254.775 &  -1.914 & 133.6\\
2019-04-25 23:32:06 & 2019-04-25 23:47:31& 2MASS 16592042-0153111 &  254.835 &  -1.886 & 147 \\
2019-04-25 23:32:06 & 2019-04-25 23:47:31 & HyperLEDA 161015       &  254.835 &  -1.886 & 148.1\\
2019-04-25 23:32:06 & 2019-04-25 23:47:31 & 2MASS 16590588-0154501 &  254.775 &  -1.914 & 128.1\\
2019-04-25 23:47:49 & 2019-04-26 00:03:14 & 6dFJ1654221-164837      &  253.592 &  -16.81 & 122.3\\
2019-04-25 23:47:49 & 2019-04-26 00:03:14 & 2MASS 16542215-1648370 &  253.592 &  -16.81 & 122.5\\
2019-04-26 00:03:31 & 2019-04-26 00:18:56 & HyperLEDA 1093975      &  255.095 &  -2.372 & 140.4\\
2019-04-26 00:03:31 & 2019-04-26 00:18:56 & HyperLEDA 161019       &  255.096 &  -2.326 & 143.6\\
2019-04-26 00:03:31 & 2019-04-26 00:18:56 & HyperLEDA 1092764      &  255.06 & -2.413 & 145 \\
2019-04-26 00:03:31 & 2019-04-26 00:18:56 & 2MASS 17002062-0225482 &  255.086 &  -2.43  & 145.6\\
2019-04-26 00:03:31 & 2019-04-26 00:18:56 & 2MASS 17002302-0219322 &  255.096 &  -2.326 & 129 \\
2019-04-26 00:03:31 & 2019-04-26 00:18:56 & 2MASS 17001441-0224462 &  255.06 & -2.413 & 147.4\\
2019-04-26 00:03:31 & 2019-04-26 00:18:56 & 2MASS 17002095-0217192 &  255.087 &  -2.289 & 151.5\\
2019-04-26 00:03:31 & 2019-04-26 00:18:56 & 2MASS 17002288-0222202 &  255.095 &  -2.372 & 120.1\\
2019-04-26 00:19:13 & 2019-04-26 00:34:38 & HyperLEDA 899699       &  253.301 &  -16.316 &  118.8\\
2019-04-26 00:19:13 & 2019-04-26 00:34:38 & HyperLEDA 899284       &  253.355 &  -16.349 &  118.8\\
2019-04-26 00:19:13 & 2019-04-26 00:34:38 & HyperLEDA 900018       &  253.27 & -16.291 &  122.8\\
2019-04-26 00:19:13 & 2019-04-26 00:34:38 & 2MASS 16530485-1617273 &  253.27 & -16.291 &  123 \\
2019-04-26 00:19:13 & 2019-04-26 00:34:38 & 2MASS 16532193-1617132 &  253.341 &  -16.287 &  129.8\\
2019-04-26 00:19:13 & 2019-04-26 00:34:38 & 2MASS 16531513-1621223 &  253.313 &  -16.356 &   99.7\\
2019-04-26 00:34:55 & 2019-04-26 00:50:20 & HyperLEDA 1109100      &  254.506 &  -1.823 & 138 \\
2019-04-26 00:34:55 & 2019-04-26 00:50:20 & 2MASS 16580128-0149216 &  254.505 &  -1.823 & 136.4\\
2019-04-26 00:50:34 & 2019-04-26 01:05:59 & HyperLEDA 161003       &  253.86 & -7.206 & 132.3\\
2019-04-26 00:50:34 & 2019-04-26 01:05:59 & 2MASS 16552449-0715255 &  253.852 &  -7.257 & 130.1\\
2019-04-26 00:50:34 & 2019-04-26 01:05:59 & HyperLEDA 161002       &  253.852 &  -7.257 & 130.7\\
\hline
\\
\multicolumn{6}{c}{Abastumani/T70}\\
\\
\hline
2019-04-25 23:08:01 & 2019-04-25 23:15:13 & GWGC UGC07124          &  182.193 &  24.946 & 100.5\\
2019-04-25 23:08:01 & 2019-04-25 23:15:13 & GWGC PGC038615         &  182.325 &  24.968 & 102.7\\
2019-04-25 23:08:01 & 2019-04-25 23:15:13 & GWGC PGC038526         &  182.055 &  24.943 & 105.7\\
\hline
\end{tabular}
\end{table*}

\begin{table*}
\contcaption{List of galaxies, enclosed in the 3D probability region of S190425z, observed by the T60 telescope at les Makes Observatory, the T70 telescope at the Abastumani Observatory and the AZT-8 telescope at the Lisnyky Observatory and the Zadko telescope.}
\begin{tabular}{cclccc}
\hline
 \multicolumn{2}{c}{} & & \multicolumn{2}{c}{} & \\
\multicolumn{2}{c}{Observation period (UTC)} & $~~~~~~~~$Galaxy name & \multicolumn{2}{c}{Galaxy coordinates}  & Distance\\
 T$_{\rm{start}}$ &T$_{\rm{end}}$ & & RA [$^\circ$] & Dec [$^\circ$] & [Mpc]\\
\hline
\hline
2019-04-25 23:22:46 & 2019-04-25 23:29:58 & HyperLEDA 1093975      &  255.095 &  -2.372   140.4\\
2019-04-25 23:22:46 & 2019-04-25 23:29:58 & HyperLEDA 161019       &  255.096 &  -2.326 & 143.6\\
2019-04-25 23:22:46 & 2019-04-25 23:29:58 & HyperLEDA 1092764      &  255.06 & -2.413 & 145 \\
2019-04-25 23:22:46 & 2019-04-25 23:29:58 & 2MASS 17002062-0225482 &  255.086 &  -2.43  & 145.6\\
2019-04-25 23:22:46 & 2019-04-25 23:29:58 & 2MASS 17002302-0219322 &  255.096 &  -2.326 & 129 \\
2019-04-25 23:22:46 & 2019-04-25 23:29:58 & 2MASS 17001441-0224462 &  255.06 & -2.413 & 147.4\\
2019-04-25 23:22:46 & 2019-04-25 23:29:58 & 2MASS 17002095-0217192 &  255.087 &  -2.289 & 151.5\\
2019-04-25 23:22:46 & 2019-04-25 23:29:58 & 2MASS 17002288-0222202 &  255.095 &  -2.372 & 120.1\\
2019-04-25 23:37:08 & 2019-04-25 23:44:20 & HyperLEDA 899699       &  253.301 &  -16.316 &  118.8\\
2019-04-25 23:37:08 & 2019-04-25 23:44:20 & HyperLEDA 899284       &  253.355 &  -16.349 &  118.8\\
2019-04-25 23:37:08 & 2019-04-25 23:44:20 & HyperLEDA 900018       &  253.27 & -16.291 &  122.8\\
2019-04-25 23:37:08 & 2019-04-25 23:44:20 & 2MASS 16530485-1617273 &  253.27 & -16.291 &  123 \\
2019-04-25 23:37:08 & 2019-04-25 23:44:20 & 2MASS 16532193-1617132 &  253.341 &  -16.287 &  129.8\\
2019-04-25 23:37:08 & 2019-04-25 23:44:20 & 2MASS 16531513-1621223 &  253.313 &  -16.356 &   99.7\\
2019-04-25 23:54:01 & 2019-04-26 00:01:13 & 6dFJ1653567-160711  &  253.486 &  -16.12 & 121.9\\
2019-04-25 23:54:01 & 2019-04-26 00:01:13 & 2MASS 16535674-1607110 &  253.486 &  -16.12 & 122.1\\
2019-04-25 23:54:01 & 2019-04-26 00:01:13 & 2MASS 16535385-1614537 &  253.474 &  -16.248 &  109.3\\
2019-04-26 00:04:49 & 2019-04-26 00:12:01 & HyperLEDA 896696       &  253.494 &  -16.552 &  121.4\\
2019-04-26 00:17:30 & 2019-04-26 00:24:42 & 2MASS 17005715-0225086 &  255.238 &  -2.419 & 127 \\
2019-04-26 00:17:30 & 2019-04-26 00:24:42 & 2MASS 17012357-0220366 &  255.348 &  -2.344 & 146.1\\
2019-04-26 00:17:30 & 2019-04-26 00:24:42 & HyperLEDA 1094718      &  255.348 &  -2.344 & 150.5\\
2019-04-26 00:30:30 & 2019-04-26 00:37:42 & HyperLEDA 1697461      &  241.489 &  23.986 & 148.8\\
2019-04-26 00:30:30 & 2019-04-26 00:37:42 & HyperLEDA 1696139      &  241.357 &  23.92  & 136.9\\
2019-04-26 00:42:34 & 2019-04-26 00:49:45 & SDSSJ154308.15+282509.9 &  235.784 &  28.419 & 145 \\
2019-04-26 00:42:34 & 2019-04-26 00:49:45 & HyperLEDA 1835553      &  235.651 &  28.432 & 143.7\\
2019-04-26 00:55:56 & 2019-04-26 01:03:07 & HyperLEDA 1697264      &  240.942 &  23.977 & 157.7\\
2019-04-26 00:55:56 & 2019-04-26 01:03:07 & HyperLEDA 1695306      &  240.892 &  23.875 & 153.6\\
2019-04-26 00:55:56 & 2019-04-26 01:03:07 & HyperLEDA 56887        &  240.925 &  24.095 & 140.9\\
2019-04-26 01:09:56 & 2019-04-26 01:17:08 & HyperLEDA 56963        &  241.153 &  23.663 & 165.5\\
2019-04-26 01:09:56 & 2019-04-26 01:17:08 & HyperLEDA 1691288      &  241.25 & 23.658 & 166.8\\
2019-04-26 01:09:56 & 2019-04-26 01:17:08 & HyperLEDA 1693165      &  241.235 &  23.758 & 170.6\\
2019-04-26 01:09:56 & 2019-04-26 01:17:08 & HyperLEDA 1690684      &  241.15 & 23.625 & 163.5\\
2019-04-26 01:09:56 & 2019-04-26 01:17:08 & HyperLEDA 57021        &  241.26 & 23.669 & 146.9\\
2019-04-26 01:09:56 & 2019-04-26 01:17:08 & HyperLEDA 1690441      &  241.184 &  23.612 & 151.4\\
2019-04-26 01:21:42 & 2019-04-26 01:28:54 & SDSSJ160409.58+241950.7&  241.04 & 24.331 & 168.7\\
2019-04-26 01:21:42 & 2019-04-26 01:28:54 & HyperLEDA 1706506      &  240.988 &  24.402 & 169.4\\
2019-04-26 01:21:42 & 2019-04-26 01:28:54 & HyperLEDA 1706072      &  241.077 &  24.382 & 151.9\\
2019-04-26 01:21:42 & 2019-04-26 01:28:54 & HyperLEDA 1703751      &  241.076 &  24.281 & 147.3\\
\hline
\\
\multicolumn{6}{c}{Lisnyky/AZT-8}\\
\\
\hline
2019-04-25 22:32:38 & 2019-04-25 22:56:24 & HyperLEDA 1107232      &  254.346 &  -1.89  & 148.8\\
2019-04-25 22:32:38 & 2019-04-25 22:56:24 & 2MASS 16572303-0153258 &  254.346 &  -1.891 & 155.8\\
2019-04-25 23:01:18 & 2019-04-25 23:21:22 & 2MASS 16570522-0151022 &  254.272 &  -1.851 &  147.2\\
2019-04-25 23:01:18 & 2019-04-25 23:21:22 & HyperLEDA 1109171      &  254.296 &  -1.82 &    148.8\\
2019-04-25 23:01:18 & 2019-04-25 23:21:22 & 2MASS 16571094-0149128 &  254.296 &  -1.82  & 169.9\\
2019-04-25 23:24:57 & 2019-04-25 23:45:08 & HyperLEDA 1109773      &  254.435 &  -1.799 & 136.9\\
2019-04-25 23:24:57 & 2019-04-25 23:45:08 & 2MASS 16574427-0147567 &  254.434 &  -1.799 & 136.4\\
2019-04-25 23:24:57 & 2019-04-25 23:45:08 & 2MASS 16573561-0147457 &  254.398 &  -1.796 & 142.6\\
2019-04-25 23:24:57 & 2019-04-25 23:45:08 & HyperLEDA 1109835      &  254.399 &  -1.796 & 145.1\\
2019-04-25 23:24:57 & 2019-04-25 23:45:08 & 2MASS 16572382-0147498 &  254.349 &  -1.797 & 147.9\\
2019-04-25 23:24:57 & 2019-04-25 23:45:08 & HyperLEDA 1110982      &  254.423 &  -1.754 & 148.8\\
2019-04-25 23:24:57 & 2019-04-25 23:45:08 & HyperLEDA 1109807      &  254.349 &  -1.797 & 148.8\\
2019-04-25 23:24:57 & 2019-04-25 23:45:08 & 2MASS 16574141-0145157 &  254.423 &  -1.754 & 168.5\\
2019-04-25 23:56:12 & 2019-04-26 00:16:25 & HyperLEDA 1699170      &  241.231 &  24.073 & 165.3\\
2019-04-25 23:56:12 & 2019-04-26 00:16:25 & SDSSJ160504.21+240240.4 &  241.268 &  24.045 & 154.1\\
2019-04-26 00:17:33 & 2019-04-26 00:37:36 & HyperLEDA 140564       &  241.215 &  23.938 & 161.5\\
2019-04-26 00:17:33 & 2019-04-26 00:37:36 & HyperLEDA 1695387      &  241.253 &  23.879 & 154.2\\
2019-04-26 00:17:33 & 2019-04-26 00:37:36 & SDSSJ160454.83+235507.5&  241.229 &  23.919 & 146 \\
2019-04-26 00:17:33 & 2019-04-26 00:37:36 & HyperLEDA 57003        &  241.211 &  23.975 & 176.2\\
2019-04-26 00:17:33 & 2019-04-26 00:37:36 & HyperLEDA NGC6051      &  241.236 &  23.933 & 143.6\\
\hline
\end{tabular}
\end{table*}

\begin{table*}
\contcaption{List of galaxies, enclosed in the 3D probability region of S190425z, observed by the T60 telescope at les Makes Observatory, the T70 telescope at the Abastumani Observatory and the AZT-8 telescope at the Lisnyky Observatory and the Zadko telescope.}
\begin{tabular}{cclccc}
\hline
 \multicolumn{2}{c}{} & & \multicolumn{2}{c}{} & \\
\multicolumn{2}{c}{Observation period (UTC)} & $~~~~~~~~$Galaxy name & \multicolumn{2}{c}{Galaxy coordinates}  & Distance\\
 T$_{\rm{start}}$ &T$_{\rm{end}}$ & & RA [$^\circ$] & Dec [$^\circ$] & [Mpc]\\
\hline
\hline
2019-04-26 00:17:33 & 2019-04-26 00:37:36 & HyperLEDA 57014        &  241.248 &  23.97  & 140.2\\
2019-04-26 00:40:37 & 2019-04-26 01:01:06 & HyperLEDA 1697461      &  241.489 &  23.986 & 148.8\\
2019-04-26 01:03:45 & 2019-04-26 01:24:03 & SDSSJ160519.85+235455.0&  241.333 &  23.915 & 149.7\\
2019-04-26 01:03:45 & 2019-04-26 01:24:03 & SDSSJ160505.95+235227.2&  241.275 &  23.874 & 175.9\\
2019-04-26 01:03:45 & 2019-04-26 01:24:03 & SDSSJ160512.22+235109.2&  241.301 &  23.853 & 136.9\\
2019-04-26 01:03:45 & 2019-04-26 01:24:03 & HyperLEDA 1696139      &  241.357 &  23.92  & 136.9\\
\hline
\\
\multicolumn{6}{c}{Zadko}\\
\\
\hline
2019-04-25 14:56:19 & 2019-04-26 19:00:00 & 2MASS 16580128-0149216 &  254.505 &  -1.822 & 136.4\\
2019-04-25 14:56:19 & 2019-04-26 19:00:00 & 2MASS 16574955-0156027 &  254.456 &  -1.934 & 135.6\\
2019-04-25 14:56:19 & 2019-04-26 19:00:00 & 2MASS 16572303-0153258 &  254.345 &  -1.890 & 155.8\\
2019-04-25 14:56:19 & 2019-04-26 19:00:00 & 2MASS 16580515-0149544 &  254.521 &  -1.831 & 162.3\\
2019-04-25 14:56:19 & 2019-04-26 19:00:00 & 2MASS 16580368-0150074 &  254.515 &  -1.835 & 147.4\\
2019-04-25 14:56:19 & 2019-04-26 19:00:00 & 2MASS 16582363-0154183 &  254.598 &  -1.905 & 186.2\\
2019-04-25 14:56:19 & 2019-04-26 19:00:00 & 2MASS 16574594-0148107 &  254.441 &  -1.802 & 182.0\\
2019-04-25 14:56:19 & 2019-04-26 19:00:00 & 2MASS 16575261-0148017 &  254.469 &  -1.800 & 188.5\\
2019-04-25 14:56:19 & 2019-04-26 19:00:00 & HyperLEDA 1106076 &  254.456 &  -1.934 & 136.3\\
2019-04-25 14:56:19 & 2019-04-26 19:00:00 & HyperLEDA 1107232 &  254.346 &  -1.890 & 148.8\\
2019-04-25 14:56:19 & 2019-04-26 19:00:00 & HyperLEDA 1109100 &  254.506 &  -1.822 & 138.0\\
2019-04-25 14:56:19 & 2019-04-26 19:00:00 & 2MASS 16580128-0149216 &  254.505 &  -1.822 & 136.4\\
2019-04-25 14:56:19 & 2019-04-26 19:00:00 & 2MASS 16574427-0147567 &  254.434 &  -1.799 & 136.4\\
2019-04-25 14:56:19 & 2019-04-26 19:00:00 & 2MASS 16574955-0156027 &  254.456 &  -1.934 & 135.6\\
2019-04-25 14:56:19 & 2019-04-26 19:00:00 & 2MASS 16573561-0147457 &  254.398 &  -1.796 & 142.6\\
2019-04-25 14:56:19 & 2019-04-26 19:00:00 & 2MASS 16572382-0147498 &  254.349 &  -1.797 & 147.9\\
2019-04-25 14:56:19 & 2019-04-26 19:00:00 & 2MASS 16571094-0149128 &  254.295 &  -1.820 & 169.9\\
2019-04-25 14:56:19 & 2019-04-26 19:00:00 & 2MASS 16572303-0153258 &  254.345 &  -1.890 & 155.8\\
2019-04-25 14:56:19 & 2019-04-26 19:00:00 & 2MASS 16580515-0149544 &  254.521 &  -1.831 & 162.3\\
2019-04-25 14:56:19 & 2019-04-26 19:00:00 & 2MASS 16580368-0150074 &  254.515 &  -1.835 & 147.5\\
2019-04-25 14:56:19 & 2019-04-26 19:00:00 & 2MASS 16580581-0143094 &  254.524 &  -1.719 & 113.5\\
2019-04-25 14:56:19 & 2019-04-26 19:00:00 & 2MASS 16575321-0141247 &  254.471 &  -1.690 & 174.9\\
2019-04-25 14:56:19 & 2019-04-26 19:00:00 & 2MASS 16574141-0145157 &  254.422 &  -1.754 & 168.5\\
2019-04-25 14:56:19 & 2019-04-26 19:00:00 & 2MASS 16574594-0148107 &  254.441 &  -1.802 & 182.0\\
2019-04-25 14:56:19 & 2019-04-26 19:00:00 & 2MASS 16575261-0148017 &  254.469 &  -1.800 & 188.5\\
2019-04-25 14:56:19 & 2019-04-26 19:00:00 & 2MASS 16575048-0145247 &  254.460 &  -1.756 & 185.9\\
2019-04-25 14:56:19 & 2019-04-26 19:00:00 & 2MASS 16570522-0151022 &  254.271 &  -1.850 & 147.2\\
2019-04-25 14:56:19 & 2019-04-26 19:00:00 & HyperLEDA 1106076 &  254.456 &  -1.934 & 136.3\\
2019-04-25 14:56:19 & 2019-04-26 19:00:00 & HyperLEDA 1107232 &  254.346 &  -1.890 & 148.8\\
2019-04-25 14:56:19 & 2019-04-26 19:00:00 & HyperLEDA 1109100 &  254.506 &  -1.822 & 138.0\\
2019-04-25 14:56:19 & 2019-04-26 19:00:00 & HyperLEDA 1109171 &  254.296 &  -1.820 & 148.8\\
2019-04-25 14:56:19 & 2019-04-26 19:00:00 & HyperLEDA 1109773 &  254.435 &  -1.799 & 136.9\\
2019-04-25 14:56:19 & 2019-04-26 19:00:00 & HyperLEDA 1109807 &  254.349 &  -1.797 & 148.8\\
2019-04-25 14:56:19 & 2019-04-26 19:00:00 & HyperLEDA 1109835 &  254.399 &  -1.796 & 145.1\\
2019-04-25 14:56:19 & 2019-04-26 19:00:00 & HyperLEDA 1110982 &  254.423 &  -1.754 & 148.8\\
2019-04-25 14:56:19 & 2019-04-26 19:00:00 & 2MASS 16573561-0147457 &  254.398 &  -1.796 & 142.6\\
2019-04-25 14:56:19 & 2019-04-26 19:00:00 & 2MASS 16572382-0147498 &  254.349 &  -1.797 & 147.9\\
2019-04-25 14:56:19 & 2019-04-26 19:00:00 & 2MASS 16564648-0157432 1105380 &  254.193 &  -1.962 & 149.8\\
2019-04-25 14:56:19 & 2019-04-26 19:00:00 & 2MASS 16571094-0149128 &  254.295 &  -1.820 & 169.9\\
2019-04-25 14:56:19 & 2019-04-26 19:00:00 & 2MASS 16572303-0153258 &  254.345 &  -1.890 & 155.8\\
2019-04-25 14:56:19 & 2019-04-26 19:00:00 & 2MASS 16570522-0151022 &  254.271 &  -1.850 & 147.2\\
2019-04-25 14:56:19 & 2019-04-26 19:00:00 & HyperLEDA 1107232 &  254.346 &  -1.890 & 148.8\\
2019-04-25 14:56:19 & 2019-04-26 19:00:00 & HyperLEDA 1109171 &  254.296 &  -1.820 & 148.8\\
2019-04-25 14:56:19 & 2019-04-26 19:00:00 & HyperLEDA 1109807 &  254.349 &  -1.797 & 148.8\\
2019-04-25 14:56:19 & 2019-04-26 19:00:00 & HyperLEDA 1109835 &  254.399 &  -1.796 & 145.1\\
2019-04-25 14:56:19 & 2019-04-26 19:00:00 & 2MASS 16045670+2355583 NGC6051 &  241.236 &  23.932 & 143.6\\
2019-04-25 14:56:19 & 2019-04-26 19:00:00 & 2MASS 16045057+2358303 57003 &  241.210 &  23.975 & 176.2\\
2019-04-25 14:56:19 & 2019-04-26 19:00:00 & 2MASS 16045947+2358123 57014 &  241.247 &  23.970 & 140.2\\
2019-04-25 14:56:19 & 2019-04-26 19:00:00 & 2MASS 16043249+2356393 140561 &  241.135 &  23.944 & 148.2\\
2019-04-25 14:56:19 & 2019-04-26 19:00:00 & 2MASS 16045152+2356153 140564 &  241.214 &  23.937 & 161.5\\
2019-04-25 14:56:19 & 2019-04-26 19:00:00 & HyperLEDA 1699170 & 241.231 &  24.073 & 165.3\\
2019-04-25 14:56:19 & 2019-04-26 19:00:00 & SDSSJ160504.21+240240.4 &  241.268 &  24.044 & 154.1\\
\hline
\end{tabular}
\end{table*}

\begin{table*}
\contcaption{List of galaxies, enclosed in the 3D probability region of S190425z, observed by the T60 telescope at les Makes Observatory, the T70 telescope at the Abastumani Observatory and the AZT-8 telescope at the Lisnyky Observatory and the Zadko telescope.}
\begin{tabular}{cclccc}
\hline
 \multicolumn{2}{c}{} & & \multicolumn{2}{c}{} & \\
\multicolumn{2}{c}{Observation period (UTC)} & $~~~~~~~~$Galaxy name & \multicolumn{2}{c}{Galaxy coordinates}  & Distance\\
 T$_{\rm{start}}$ &T$_{\rm{end}}$ & & RA [$^\circ$] & Dec [$^\circ$] & [Mpc]\\
\hline
\hline
2019-04-25 14:56:19 & 2019-04-26 19:00:00 & SDSSJ160427.52+240639.8 &  241.115 &  24.111 & 151.2\\
2019-04-25 14:56:19 & 2019-04-26 19:00:00 & SDSSJ160454.83+235507.5 &  241.229 &  23.918 & 146.0\\
2019-04-25 14:56:19 & 2019-04-26 19:00:00 & 2MASS 16045670+2355583 NGC6051 &  241.236 &  23.932 & 143.6\\
2019-04-25 14:56:19 & 2019-04-26 19:00:00 & 2MASS 16050427+2355015 57025 &  241.267 &  23.917 & 237.2\\
2019-04-25 14:56:19 & 2019-04-26 19:00:00 & 2MASS 16043346+2350163 56958 &  241.139 &  23.837 & 185.2\\
2019-04-25 14:56:19 & 2019-04-26 19:00:00 & 2MASS 16045057+2358303 57003 &  241.210 &  23.975 & 176.2\\
2019-04-25 14:56:19 & 2019-04-26 19:00:00 & 2MASS 16045947+2358123 57014 &  241.247 &  23.970 & 140.2\\
2019-04-25 14:56:19 & 2019-04-26 19:00:00 & 2MASS 16043249+2356393 140561 &  241.135 &  23.944 & 148.2\\
2019-04-25 14:56:19 & 2019-04-26 19:00:00 & 2MASS 16045152+2356153 140564 &  241.214 &  23.937 & 161.5\\
2019-04-25 14:56:19 & 2019-04-26 19:00:00 & HyperLEDA 1695387 &  241.253 &  23.879 & 154.2\\
2019-04-25 14:56:19 & 2019-04-26 19:00:00 & SDSSJ160454.83+235507.5 &  241.229 &  23.918 & 146.0\\
2019-04-25 14:56:19 & 2019-04-26 19:00:00 & SDSSJ160505.95+235227.2 &  241.275 &  23.874 & 175.9\\
\hline
\end{tabular}
\end{table*}


\section{Affiliations}

$^{1}$APC, Univ. Paris Diderot, CNRS/IN2P3, CEA/lrfu, Obs de Paris, Sorbonne Paris Cit\'e, France\\
$^{2}$N.Tusi Shamakhy astrophysical Observatory\ Azerbaijan National Academy of Sciences, settl.Mamedaliyev, AZ 5626, Shamakhy, Azerbaijan\\
$^{3}$Ilia State University\ E.Kharadze Abastumani Astrophysical Observatory Mt.Kanobili,Abastumani, 0301, Adigeni, Georgia\\
$^{4}$Samtskhe-Javakheti  State  University, Rustaveli Str. 113,  Akhaltsikhe, 0080,  Georgia\\
$^{5}$UPMC-CNRS, UMR7095, Institut d'Astrophysique de Paris, 75014 Paris, France\\
$^{6}$Astronomical Observatory\ Taras Shevshenko National University of Kyiv, Observatorna str. 3, Kyiv, 04053, Ukraine\\
$^{7}$Nuclear Physics Department\ Taras Shevchenko National University of Kyiv, 60 Volodymyrska str., Kyiv, 01601, Ukraine\\
$^{8}$Yunnan Astronomical Observatories/Chinese Academy of Science, Kunming, 650011, China \\
$^{9}$Laboratoire d'Astrophysique de Marseille, UMR 7326, CNRS, Universit\'e d'Aix Marseille, 38, rue Fr\'ed\'eric Joliot-Curie,Marseille, France\\
$^{10}$AGORA observatoire des Makes, AGORA, 18 Rue Georges Bizet, Observatoire des Makes, 97421 La Rivi\`ere, France\\
$^{11}$Instituto de Astrof\'isica de Andaluc\'ia (IAA-CSIC), Glorieta de la Astronom\'ia s/n, 18008 Granada, Spain\\
$^{12}$ARTEMIS UMR 7250 UCA CNRS OCA, boulevard de l'Observatoire, CS 34229, 06304 Nice CEDEX 04, France\\
$^{13}$Ulugh Beg Astronomical Institute, Uzbekistan Academy of Sciences, Astronomy str. 33, Tashkent 100052, Uzbekistan\\
$^{14}$Nikhef, Science Park, 1098 XG Amsterdam, The Netherlands\\
$^{15}$OzGrav-UWA, University of Western Australia, School of Physics, M013, 35 Stirling Highway, Crawley, WA 6009, Australia\\
$^{16}$IRAP, Universit\'e de Toulouse, CNRS, UPS, 14 Avenue Edouard Belin, F-31400 Toulouse, France\\
$^{17}$Universit\'e Paul Sabatier Toulouse III, Universit\'e de Toulouse, 118 route de Narbonne, 31400 Toulouse, France\\
$^{18}$ WuZhou University, WuZhou, GuangXi, 543000, China \\
$^{19}$AIM, CEA, CNRS, Universit\'e Paris-Saclay, Universit\'e Paris Diderot, Sorbonne Paris Cit\'e, F-91191 Gif-sur-Yvette, France\\
$^{20}$LAL, Univ Paris-Sud, CNRS/IN2P3, Orsay, France\\
$^{21}$California Institute of Technology, 1200 East California Blvd, MC 249-17, Pasadena, CA 91125, USA\\
$^{22}$National Astronomical Observatories/Chinese Academy of Science 20A Datun Road, Beijing, 100012, China\\
$^{23}$Physics Department and Astronomy Department, Tsinghua University, Beijing, 100084, China\\
$^{24}$Astronomy and Space Physics Department\ Taras Shevchenko NationalUniversity of Kyiv, 60 Volodymyrska str., Kyiv, 01601, Ukraine\\

\label{lastpage}

\end{document}